\documentclass[universe,review,accept,pdftex,moreauthors]{Definitions/mdpi}
\usepackage{comment} 

\firstpage{1} 
\pubvolume{1}
\issuenum{1}
\articlenumber{0}
\pubyear{2023}
\copyrightyear{2023}
\datereceived{10 March 2023} 
\daterevised{6 June 2023} 
\dateaccepted{14 June 2023} 
\datepublished{22 June 2023} 
\hreflink{https://doi.org/10.3390/universe9070302} 

\newcommand{\prd}{Phys. Rev. D}
\newcommand{\pre}{Phys. Rev. E}
\newcommand{\prl}{Phys. Rev. Lett.}
\newcommand{\pasa}{PASA}
\newcommand{\jcap}{JCAP}
\newcommand{\apjl}{APJL~}
\newcommand{\apj}{APJ~}
\newcommand{\mnras}{Mon.\ Not.\ R.\ Astron.\ Soc.}
\newcommand{\apjs}{Astrophys.\ J.\ Supp.}
\newcommand{\physrep}{Phys. Rep.}
\newcommand{\aap}{A\& A}
\newcommand{\aj}{AJ}
\newcommand{\apss}{Ap\&SS}
\newcommand{\araa}{Annu. Rev. Astron. Astrophys.}
\newcommand{\aapr}{A\&A Rev.}
\newcommand{\pasj}{PASJ}
\newcommand{\nat}{Nature}

\def\ln{{\rm ln}}

\def\bfx{{\bf x}}
\def\bfk{{\bf k}}
\def\bfq{{\bf q}}

\def\bfv{{\bf v}}

\def\calL{{\mathcal L}}

\def\calO{{\mathcal O}}

\renewcommand{\vec}[1]{\boldsymbol{#1}}

\def\bfk{{\bf k}}

\def\bfr{{\bf r}}

\def\calL{{\cal L}}

\def\calO{{\cal O}}

\def\bfx{{\bf x}}

\def\rd{{\rm d}}

\newcommand{\bfs}{\mathbf s}

\newcommand{\hn}{\hat{\vec n}}

\newcommand{\av}[1]{\left\langle{#1}\right\rangle} 

\def\mpch{{h^{-1}\rm{Mpc}}}

\usetikzlibrary{tikzmark}

\usepackage[utf8]{inputenc}
\usepackage{amsmath,amssymb, graphicx, multicol, array, upgreek, natbib, booktabs, multirow, bm, dsfont} 

\usepackage{simplewick}
\usepackage{scalerel}
\usepackage[mathscr]{euscript}
\usepackage{extarrows}
\usepackage{caption}
\usepackage[labelformat=simple]{subcaption}

\DeclareCaptionLabelFormat{subcaptionlabel}{\normalfont(\textbf{#2}\normalfont)}
\usepackage{tikz}
\usepackage{physics}

\newcounter{para}

\Title{Cosmological Probes of Structure Growth and Tests of Gravity}

\TitleCitation{Cosmological Probes of Structure Growth and Tests of Gravity}

\Author{{Jiamin Hou} 
 $^{1,2,}$*$^{,\dagger}$\orcidA{}, Julian Bautista $^{3}$, Maria Berti $^{4,5,6}$\orcidB{}, Carolina {Cuesta-Lazaro} 
 $^{7,8,9}$, C\'{e}sar Hern\'{a}n{dez-Aguayo} $^{10,11}$, {Tilman Tr}{\"o}ster $^{12}$ and Jinglan Zheng $^{13}$}

\AuthorNames{Jiamin Hou, Julian Bautista, Maria Berti, Carolina Cuesta-Lazaro, C\'{e}sar Hern\'{a}ndez-Aguayo, Tilman Tr{\"o}ster and
Jinglan Zheng}

\AuthorCitation{{Hou, J.;} 
 Bautista, J.; Berti, M.; Cuesta-Lazaro, C.; Hern\'{a}ndez-Aguayo, C.; Tr{\"o}ster, T.; Zheng, J.}

\address{%
$^{1}$\quad Department of Astronomy, University of Florida, Gainesville, FL 32611, USA\\
$^{2}$\quad {Max--Planck--Institut f\"ur} 
 Extraterrestische Physik, Postfach 1312, Giessenbachstr., 85748 Garching, Germany\\
$^{3}$\quad Aix Marseille {Univ} 
, CNRS/IN2P3, CPPM, {Marseille} 
, France; {bautista@cppm.in2p3.fr} 
\\
$^{4}$\quad SISSA---International School for Advanced Studies, Via Bonomea 265, 34136 Trieste, Italy; mberti@sissa.it\\
$^{5}$\quad INFN---National Institute for Nuclear Physics, Via Valerio 2, 34127 Trieste, Italy\\
$^{6}$\quad IFPU, Institute for Fundamental Physics of the Universe, via Beirut 2, 34151 Trieste, Italy\\
$^{7}$\quad Center for Astrophysics{,} 
 Harvard \& Smithsonian, 60 Garden St, Cambridge, MA 02138, USA; carolina.cuesta-lazaro@cfa.harvard.edu\\
$^{8}$\quad {The NSF AI Institute for Artificial Intelligence and Fundamental Interactions, 77 Massachusetts Avenue Cambridge, MA 02139, USA} 
 \\
$^{9}$\quad Department of Physics, Massachusetts Institute of Technology, Cambridge, MA 02139, USA \\
$^{10}$\quad Max--Planck--Institut f\"ur Astrophysik, Karl-Schwarzschild-Str. 1, D-85748 Garching, Germany; cesarhdz@mpa-garching.mpg.de\\
$^{11}$\quad Excellence Cluster ORIGINS, Boltzmannstrasse 2, D-85748 Garching, Germany\\
$^{12}$\quad Institute for Particle Physics and Astrophysics, ETH Z\"urich, Wolfgang-Pauli-Strasse 27, \linebreak 8093 Z\"urich, Switzerland; tilmant@phys.ethz.ch\\
$^{13}$\quad Fakult\"at f\"ur Physik, Universit\"at Bielefeld, {Postfach 100131,} 
 33501 Bielefeld, Germany; jzheng@physik.uni-bielefeld.de\\
}

\corres{Correspondence: jiamin.hou@mpe.mpg.de}

\abstract{The current standard cosmological model is constructed within the framework of general relativity with a cosmological constant $\Lambda$, which is often associated with dark energy, and phenomenologically explains the accelerated cosmic expansion. Understanding the nature of dark energy is one of the most appealing questions in achieving a self-consistent physical model at cosmological scales. Modification of general relativity could potentially provide a more natural and physical solution to the accelerated expansion. The growth of the cosmic structure is sensitive in constraining gravity models.
In this paper, we aim to provide a concise introductory review of modified gravity models from an observational point of view. We will discuss various mainstream cosmological observables, and their potential advantages and limitations as probes of gravity models.}

\keyword{modified gravity; large-scale structure; cosmology; observation} 
\begin{document}

\section{Introduction}
\label{sec:intro}
{Our current }
physical description of gravity, Einstein's general relativity (GR), has profound implications for various astrophysical observables.  
The theory precisely predicts the orbital decay of binary pulsars~\citep{Taylor1982:Pulsar}, gravitational time dilation observed in the spectra of white dwarf Sirius B~\citep{Holberg2010:sirusB}, and~gravitational waves (GWs) from merging black holes~\citep{LIGO2016:GW150914} or other compact objects~\citep{LIGO2017:GW170817}.

General relativity is also one of the cornerstones for our understanding of the Universe over billions of years of evolution. However, there still lacks a satisfying explanation for the observed accelerated expansion of our Universe. The~first evidence for cosmic acceleration was found in measurements of type-Ia supernovae~\citep{Perlmutter1999:SNa,Riess1998:SNa} and was later supported by various observations including the cosmic microwave background (CMB), the~3D distribution of large-scale structures, and~GW measurements. To~explain the observed acceleration under GR, an~exotic form of energy with negative pressure has to be introduced in the field equations, the~so-called ``dark energy'' (DE) (see reviews, e.g.,~\cite{Frieman2008:MGreview,Weinberg2013:CCreview}). In~its simplest form, DE is just a cosmological constant $\Lambda$. This additional source of energy could be explained through vacuum energy predicted by quantum field theory. However, this simplest explanation faces various theoretical problems~\cite{Weinberg1989:CC}, mainly driven by the fact that the vacuum energy predicted by quantum field theory disagrees with the observed values by more than a hundred orders of~magnitude.

In addition to DE, an~exotic form of matter that only interacts gravitationally, ``dark matter'', also needs to be introduced to describe, e.g.,~the observed cosmic structure formation and rotation curves of galaxies~\cite{Rubin1980:RotCurve}. The~$\Lambda$CDM model---a model containing DE in the form of a cosmological constant plus cold dark matter---is currently the best fit to most observations. Although~the $\Lambda$CDM model can well explain various astrophysical observations, the~nature of its dark sectors is not well understood: neither DE nor dark matter could be detected so far in lab experiments. Therefore, theoretical physicists around the world have been trying to come up with solutions for the accelerated expansion of our Universe that do not involve introducing new exotic components but rather modify or extend Einstein's theory of~gravity. 

This paper is one of the review series on the cosmological constant problem. The~series will discuss the cosmological constant problem from different perspectives. Here, we will be focusing on testing modified theories of GR, ``Modified Gravity'' (MG) through probes of the growth of structure or the gravitational potential at extragalactic scales $\sim\!\calO(1\,{\mpch})$. For~each probe, we will present their motivations, their promise to detect modifications of GR, as~well as their challenges and~limitations. 

The review organised as follows. We will start with a brief overview of the modified gravity models in {Section} 
~\ref{sec:overview_model}; we will cover two of the most popular parametrizations of the MG models, the~screening mechanisms, and~various parameterised frameworks of gravity. We will proceed with astrophysical probes at different scales in Section~\ref{sec:methods} and tools to distinguish the deviation from GR and list constraints from current surveys. We will discuss the available simulations which assist further tests and our understanding of a few most popular MG parametrizations in Section~\ref{sec:simulations}. In~the next Section~\ref{sec:survey}, we will outline current and next-stage cosmological surveys, including galaxy, CMB, and~radio surveys. Finally, we will conclude the work in Section~\ref{sec:conclusion}. 


\section{An Overview of the Modified Gravity~Models}
\label{sec:overview_model}

\begin{figure}[H]
    \includegraphics[width=0.7\textwidth]{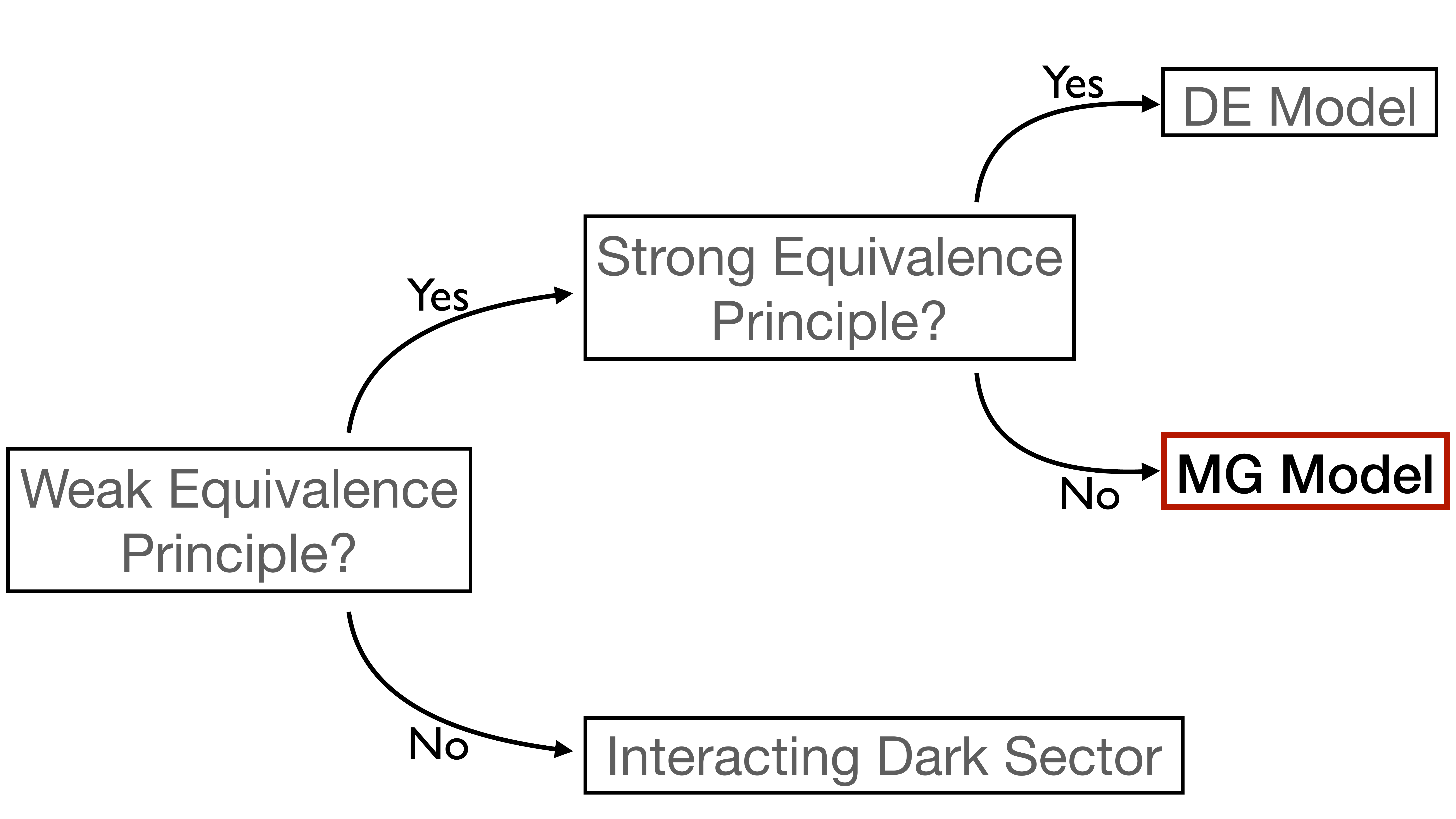}
    \caption{A flowchart distinguishing the DE and modified gravity (MG) models (based on Joyce et al. ~\cite{Joyce2016:DEvsMG}). The~weak equivalence principle (WEP) states that there exists a Jordan frame metric to which matter species are minimally coupled, independent of their composition or structure. Taking scalar--tensor theory as an example, all matter species couple universally to the metric in the presence of a scalar field. The~strong equivalence principle (SEP) further restricts the WEP such that the equivalence principle also applies to objects that exert a self-gravitational force. If~the coupling between matter fields and the metric does not involve any scalar fields, it is termed a DE model  Whereas if the matter fields couple to the metric non-trivially via the scalar field, it is termed a {MG} 
 model (red box) to be considered in this~paper.}
    \label{fig:flowchart_MGvsDE}
\end{figure}

In this section, we briefly overview the most popular MG models. As~discussed in Section~\ref{sec:intro}, although~the cosmological constant provides an excellent explanation for the accelerated cosmic expansion, more natural choices for the cosmological constant can be classified into two categories: (i) modifications of the stress-energy tensor on the right-hand side of Einstein's field equation, leading to DE models; (ii)~modifications of, e.g.,~the Einstein--Hilbert action on the left-hand side of the field equation, leading to MG models. In~practice, there are no clear boundaries between DE or MG models (for a review see, e.g.,~\cite{Joyce2016:DEvsMG}). Nevertheless, it is possible to make certain distinctions based on the strong equivalence principle and observe whether ordinary matter experiences additional forces beyond gravity. Figure~\ref{fig:flowchart_MGvsDE} presents a flowchart distinguishing MG vs. DE following \cite{Joyce2016:DEvsMG}.

In the conformal Newtonian gauge, the~line element is given by
\begin{eqnarray}\label{eq:metric}
    d s^2=a^2\left[-(1+2 \Psi) d \tau^2+(1-2 \Phi) d x^2\right],
\end{eqnarray}
with $\tau$ being the conformal time; the gauge invariant Newtonian potential $\Psi$ and curvature potential $\Phi$ are functions of space and time. In~the case of GR, the~difference between two potentials is negligibly small, and~the $00$ component of the Einstein equation on sub-horizon scales is given by the Poisson equation
\begin{eqnarray}\label{eqn:Poisson}
    \nabla^2 \Psi=4 \pi G a^2 \delta \rho_{\mathrm{m}},
\end{eqnarray}
where $\delta \rho_{\mathrm{m}} \equiv \rho_{\rm m} - \bar \rho_{\rm m}$ is the fluctuation of the matter density. In~the presence of a fifth force, Equation~\eqref{eqn:Poisson} is modified and the structure formed, albeit from the one predicted by GR.
There are different ways to ``modify gravity'', among~which scalar--tensor theories (e.g., Brans and Dicke~\cite{Brans1961:BransDicke}) are probably the most well-studied models. One of the most general scalar--tensor theories with second-order field equations in four dimensions is the Horndeski theory~\cite{Horndeski1974:horndeski}, whose action is constructed as
\begin{eqnarray}
    S=\int \mathrm{d}^4 x \sqrt{-g}\left\{\sum_{i=2}^5 \mathcal{L}_i\left[\phi, g_{\mu \nu}\right] + \mathcal{L}_{\rm m}(g_{\mu\nu}, \psi)\right\}.
    \label{eqn:horndeski_action}
\end{eqnarray}

Each of the four Lagrangian densities in the summation is a function of four parameters $\{\calL_2[K], \calL_3[G_3], \calL_4[G_4],$ and $ \calL_5[G_5]\}$, where $\{K, G_3, G_4, G_5\}$ are arbitrary functions of $(\phi, X)$, with~$\phi$ as a scalar field, $X \equiv-\nabla^\nu \phi \nabla_\nu \phi / 2$. $\mathcal{L}_{\rm m}$ as the matter Lagrangian density, and $\psi$ as the matter~field.

In the following, we will only review two MG models, the~$f(R)$ model and the braneworld model, which can be linked to Horndeski theory by properly choosing the four Horndeski functions. These two models are also the most representative and have been widely tested using different astrophysical probes (see Section~\ref{sec:methods}) and implemented in cosmological simulations (see Section~\ref{sec:simulations}).

\subsection{Conformal Coupling Models: $f(R)$ Gravity}
\label{subsec:fR}
$f(R)$ gravity~\cite{Sotiriou:2008rp,DeFelice:2010aj} is a very popular class of MG models, described by the following gravitational action
\begin{eqnarray}\label{eq:fR_action}
    S = \frac{1}{16\pi G}\int{\rm d}^4x\sqrt{-g}\left[R+f(R)\right] + S_{\rm m}[g_{\mu\nu}, \psi],
\end{eqnarray}
where the cosmological constant $\Lambda$ is replaced by an algebraic function of the Ricci scalar, $f(R)$, and~$S_{\rm m}$ is the action for the matter field $\psi$. 
Various functional forms for $f(R)$ were proposed in early works to explain the cosmic acceleration~\citep{Capozziello2002:quintessence,Capozziello2003:quintessence,Carroll2004:AccCosmo}. However, they do not pass solar system tests of gravity~\citep{Chiba2003:fR,Erickcek2006:SolarConstraint}. The authors of \cite{Hu:2007nk} suggested the functional form of $f(R)$ which is compatible with local gravity tests
\begin{eqnarray}
f(R)=-m^2 \frac{c_1\left(\frac{R}{m^2}\right)^n}{c_2\left(\frac{R}{m^2}\right)^n+1} \approx \rho_{\Lambda, \mathrm{eff}}+f_{R 0}\left(\frac{R}{\bar{R}_0}\right)^{-n}.
\end{eqnarray}

This functional form is chosen such that in the small curvature limit $(\lim_{R\rightarrow 0} f(R) \rightarrow 0)$ we recover $\Lambda$CDM at high redshift and in the large curvature limit ($\lim_{R\rightarrow \infty} f(R) \rightarrow \text{const.}$) it mimics cosmic acceleration at low redshift. $\bar{R}_0$ is the background Ricci scalar today. To~satisfy cosmological and local observations, the~condition $R/m^2\ll 1$ is required, with~$m$ being a free parameter representing a mass scale. The~two free parameters $c_1$ and $c_2$ are adjusted accordingly such that the effective DE density $\rho_{\Lambda, {\rm eff}}$ gives rise to a cosmological constant that matches observations. $f_{R0}\equiv f_R(z=0)$ is the present-day value of the background field, with~the scalar field being $f_R\equiv {df(R)}/{dR}$. This scalar field $f_R$ is of particular importance, and~the impact of $f(R)$ gravity can be viewed in terms of the dynamics of $f_R$. The~larger the amplitude of the scalar field $|f_R|$, the~stronger deviation from~GR.

In varying the action in Equation~\eqref{eq:fR_action} with respect to the metric, we can derive the modified Einstein equations, whose trace can be interpreted as the equation of motion for $f_R$
\begin{eqnarray}\label{eqn:eom_fR}
    \square f_R=\frac{\partial V_{\rm eff}}{\partial f_R}, \qquad V_{\mathrm{eff}}(f_R)=V(f_R)+A(f_R) \bar{\rho}_{\rm m},
\end{eqnarray}
where $\square\equiv \nabla_{\mu}\nabla^{\mu}$ is the d'Alembert operator. The~effective potential $V_{\rm eff}$ consists of two parts; one is a bare function that depends on the scalar field itself; another part knows the external matter field density $\bar{\rho}_{\rm m}$ (see Equation~(20) in Hu and Sawicki~\cite{Hu:2007nk} for the exact form of $V_{\rm eff}$). The~density-dependent term of the effective potential in Equation~\eqref{eqn:eom_fR} is the key that this functional form of $f(R)$ can pass the local gravity tests as its dynamics associate high-density regions with the high curvature of spacetime (see Section~\ref{subsec:screening}).

The interaction range of the scalar field is determined by the Compton wavelength $\lambda_{\rm c}$:
\begin{eqnarray}
    m_{f_R}^2 \equiv \frac{\partial^2 V_{\mathrm{eff}}}{\partial f_R^2}, \qquad \lambda_{\rm c} \equiv m_{f_R}^{-1},
\end{eqnarray}
the larger the scalar field, the~shorter the Compton wavelength, and~the shorter range the fifth force can mediate its interaction.
The presence of the fifth force leads to a different structure formation history, embodied in the modified Poisson equation. Under~the quasi-static\endnote{Quasi-static limit assumes that the time derivative of the scalar field perturbation is negligible compared to the spatial gradient of the scalar field.} and weak field approximations,\endnote{Weak field approximation assumes that the amplitudes of the scalar
field perturbations and gravitational potentials are much smaller than the speed of light squared.} the~Poisson equation takes the following modified form
\begin{eqnarray}
    \boldsymbol{\nabla}^2\Psi &\approx& \frac{16\pi{G}}{3}a^2\delta\rho_{\rm m} - \frac{1}{6}a^2\delta{R}\nonumber\\
    &=& 4\pi{G} a^2 \delta\rho_{\rm m} - \frac{1}{2}\boldsymbol{\nabla}^2 f_R,\label{eq:fR_Poisson_qsa}
\end{eqnarray}
where the first term is the standard Poisson equation, and~the second term represents the fifth force, $F_{\rm 5th} \propto \boldsymbol{\nabla} f_R$, generated by the scalar field $f_R$. Under~the static limit, Equation~\eqref{eqn:eom_fR} can be rewritten as
\begin{eqnarray}
    \boldsymbol{\nabla}^2 f_R =-\frac{a^2}{3}\left[\delta R-8 \pi G \delta \rho_{\mathrm{m}}\right], \label{eq:fR_eom_qsa}
\end{eqnarray}
where $\delta R=R(f_R)-\bar{R}$ and $\delta \rho_{\rm m} = \rho_{\rm m}-\bar{\rho}_{\rm m}$ is the perturbation of non-relativistic matter density. 
The $f(R)$ model can be shown to be equivalent to a scalar--tensor theory in which the scalar field has a universal coupling to different matter species by a conformal transformation~\cite{Barrow1988:ConformalHO,Chiba2003:fR}. At~the same time, the~connection between the $f(R)$ gravity and Horndeski can be seen by setting~\cite{Baker2019NovelProj}
\begin{eqnarray}
    G_4 = \phi = f_R, \quad K= f(R)-Rf_R, \quad G_3 = 0.
\end{eqnarray}

Hereafter, we follow the naming convention for the $f(R)$ model: for a given value of the present-day $f_{R0}$, we take the absolute value of its logarithm and call it F$|\log (f_{R0})|$. 
As an example, $f_{R0}=10^{-5}$ is called the F5~model.

\subsection{Derivative Coupling Models: DGP~Gravity}
\label{subsec:DGP}
In the DGP model~\citep{Dvali:2000hr}, the~Universe is a four-dimensional ``brane'' embedded in a five-dimensional spacetime or bulk. The~total action of the model is
\begin{eqnarray}
S &= \int_{\rm brane} {\rm d}^4x \sqrt{-g} \frac{R}{16\pi G} + \int {\rm d}^5x \sqrt{-g^{(5)}} \frac{R^{(5)}}{16\pi G^{(5)}} +S_{\rm m}(g_{\mu\nu},\psi_i)\,,\label{eq:S_dgp}    
\end{eqnarray}
where $g_{\mu\nu}$, $g$, $R$ and $G$ are the metric tensor, the~determinant of the metric, the~Ricci scalar and the gravitational constant in the 4D brane, respectively, and~$g^{(5)}$, $R^{(5)}$ and $G^{(5)}$ are their equivalents in the 5D bulk. $S_{\rm m}$ is the action of the matter fields $\psi_i$ which are assumed to be confined on the brane. The~transition from 4D to 5D is governed by the crossover scale $r_{\rm c}\equiv G^{(5)}/(2G)$.

There are two branches of solutions for the DGP model. The {\emph self-accelerating branch}~\cite{Deffayet2001:sDGP} can lead to cosmic acceleration purely gravitationally without introducing a cosmological constant. However, this branch of the solution is theoretically unstable~\cite{Luty2003:DGP} and the observed expansion history does not seem to align with the predictions of the self-accelerating DGP model~\cite{Fang2008:sDGP}. The~{\emph normal branch} is theoretically stable, but~it {\emph cannot} lead to an accelerated Hubble expansion; a trivial negative pressure energy stress component~\cite{Sahni2003:braneworld} or an extra DE component must be added to match the observational data~\citep{Schmidt2009:DGPsim}. Nevertheless, the~DGP model remains attractive as a benchmark model with the so-called Vainshtein screening mechanism that will be described in detail below, and~in this paper, we will only discuss the normal branch of the DGP model (nDGP).
The structure formation in the nDGP model is governed by the modified Poisson and scalar equations in the quasi-static and weak field limits~\citep{Koyama:2007ih}:
\begin{eqnarray}\label{eq:poisson_nDGP}
\nabla^2\Psi = 4\pi G a^2 \delta \rho_{\rm m} + \frac{1}{2}\nabla^2\phi\,,
\end{eqnarray}
\begin{eqnarray}\label{eq:dgp_eqn}
\nabla^2 \phi + \frac{r_{\rm c}^2}{3\beta_{\rm dgp} a^2c^2} \left[ (\nabla^2\phi)^2 -\left(\nabla_i\nabla_j\phi\right)^2\right] = \frac{8\pi\,G\,a^2}{3\beta_{\rm dgp}} \delta\rho_{m}\,,\label{eq:phi_dgp}
\end{eqnarray}
where $\phi$ is a scalar degree of freedom related to the bending modes of the brane, i.e.,~it describes the position of the brane in the fifth dimension. The~total modified gravitational potential ${\Psi}$ is given by ${\Psi}={\Psi_{\rm N}}+\phi/2$, with~${\Psi_{\rm N}}$ being the standard Newtonian potential. Again, the~fifth force is proportional to the field's gradient and is given by $F_{\rm 5th}=\nabla \phi/2$. The~parameter $\beta_{\rm dgp}(a)$ is a time-dependent function that depends on the crossover scale $r_{\rm c}$ (see Equation~{2.25} 
 in~\cite{Koyama:2007ih}).


The DGP model gives rise to a cubic interaction $\sim\!(\partial \phi)^2 \square \phi$ in its four-dimensional effective theory~\citep{Luty2003:DGP}, with~$\phi$ being the Galileon~\citep{Nicolis2009:galileon}. The~Galileon is a scalar field with a shift symmetry (in analogy to Galilei transformation in classical mechanics). The~generalized Galileon~\citep{Deffayet2011:galileons} can be mapped to the Horndeski theory~\cite{Kobayashi2011:gInfaltion}.
The cubic Galileon model can be reduced to a Horndeski model with the mapping
\begin{eqnarray}
    G_4 = 1,\quad\, K=-c_2 X, \quad\, G_3 = c_3 X/M^3,
\end{eqnarray}
where $c_3$ and $M$ are free parameters in the Galileon model. Note that it is common to fix $c_2=-1$.

In this paper, we follow the naming convention for the nDGP model and denote the model as N$H_0 r_c$, where the product of the Hubble parameter at $z=0$ and the crossover scale characterises the departure from GR. As~an example, $H_0 r_c=5.0$ is called the N5~model.

\subsection{Screening~Mechanisms}
\label{subsec:screening}

As discussed in Sections~\ref{subsec:fR} and~\ref{subsec:DGP}, MG models lead to a change in the Poisson equation, which inevitably results in an altered equation of motion for the gravitational acceleration. To~pass the constraints imposed within the solar system, such as radio frequency shift~\cite{Bertotti2013Cassini}, the~lunar laser ranging constraints~\cite{Merkowitz2010LLR,Murphy2012apollo}, and~the earth-based torsion balance experiments \cite{Adelberger2003:GravInvSqLaw,Kapner2007:GravInvSqLaw}, one needs solutions such that suppresses the modification of GR within the solar system.
In the examples of early $f(R)$ models, the~problem was that those models introduce an extremely light scalar degree of freedom, producing a long-range fifth force and dissociate the space-time's curvature from the local density. Therefore, the~essence of the screening mechanism is to re-associate the fifth force with the local environment. Suppose we focus on the modified Poisson equation; in this case, the~mechanisms can be summarized as (i) thin-shell screening: adding a non-linear scalar potential such that the effective potential becomes density-dependent via the dynamics of the scalar field itself; (ii) kinetic screening: non-linear generalization of the Laplacian~operator.

\subsubsection{Thin-Shell~Screening}
In the thin-shell models, the~effective potential is usually split into two terms: a bare potential that only depends on the scalar field and an environmental-dependent term $V_{\rm eff} = V(\phi) + \rho_{\rm m} A(\phi)$, with~$\rho_{\rm m}$ being the environmental density (see Equation~(\ref{eqn:eom_fR})). By~adjusting the functions $V(\phi)$ and $A(\phi)$, the~scalar field can acquire a large mass in the high-density regions. Given that the Compton wavelength is inversely proportional to the mass $\lambda_{\rm c}\propto m_{\rm eff}^{-1}$, the~larger the mass, the~shorter the distance the scalar field can mediate the fifth force. There are two widely studied screening mechanisms, the chameleon mechanism and {symmetron mechanism}. In~the following, we will focus on the chameleon mechanism since it is related to the $f(R)$ model, presented in Section~\ref{subsec:fR}. More discussions about the symmetron mechanisms can be found in~\cite{Hinterbichler:2010es,Hinterbichler:2011ca}.

Chameleon screening can be achieved with certain choices of the function $f(R)$~\cite{Khoury:2003aq,Khoury:2003rn,Mota:2006fz,Brax:2008hh}. 
A typical choice of the potential for the chameleon mechanism is
\begin{eqnarray}
    V(\phi)\propto \phi^{-n},\qquad A(\phi)\propto e^{\beta \phi/M_{\rm pl}},
\end{eqnarray}
with $M_{\rm pl}$ as the reduced Planck mass and $\beta$ as the coupling strength of the scalar field to the matter. The~choice of the potential leads to the modified Poisson equation in Equations (\ref{eq:fR_Poisson_qsa}) and (\ref{eq:fR_eom_qsa}), where one can have a quick peek into two opposite regimes of solutions. In~the large-field limit, when $|f_R|$ is relatively large (e.g.,~in the case of large $|f_{R0}|$), the~perturbation $\delta f_R$ is small compared to the background field $|\bar{f}_R|$, and~$|\delta R|\ll 8\pi{G}\delta\rho_{\rm m}$, so that the Poisson Equation~\eqref{eq:fR_Poisson_qsa} can be approximated as
\begin{eqnarray}\label{eqn:GR_Poisson_qsa}
    \boldsymbol{\nabla}^2{\Psi} \approx \frac{16\pi{G}}{3}\delta\rho_{\rm m}a^2.
\end{eqnarray}

Comparing Equation~\eqref{eqn:GR_Poisson_qsa} with the standard Poisson equation in $\Lambda$CDM, the~enhancement of gravity is $4/3$ the strength of the standard Newtonian force (see also Equation (8) in~\cite{Deffayet2001:sDGP}). In~the small-field where $|f_R|$ takes very small values, the~left-hand side of Equation \eqref{eq:fR_eom_qsa} is negligible and therefore we have $\delta{R}\approx8\pi{G}\delta\rho_{\rm m}$. Plugging this into Equation~\eqref{eq:fR_Poisson_qsa} we recover the standard Poisson~equation.

\subsubsection{Kinetic~Screening}
Kinetic screening works by modifying the Laplacian operator in the equation of motion. Two examples of kinetic screenings are the {Vainshtein mechanism} and the {K-mouflage mechanism}. In~the following, we will briefly overview the Vainshtein mechanism since the nDGP model (see Section~\ref{subsec:DGP}) is a representative class of modified gravity models that feature the Vainshtein screening mechanism \cite{Vainshtein:1972sx}. More details for K-mouflage mechanism can be found in~\cite{Babichev:2009ee,Brax:2014wla,Brax:2014yla}.

The difference in thin-shell and kinetic screening can already be seen by comparing Equations~(\ref{eq:fR_Poisson_qsa}) and (\ref{eq:fR_eom_qsa}) and Equations~(\ref{eq:poisson_nDGP}) and (\ref{eq:dgp_eqn}). Instead of adding a scalar potential, the~kinetic screening modifies the Laplacian operator of the equation of motion.
To further illustrate how the Vainshtein mechanism works, let us for simplicity consider solutions in spherical symmetry, defining excess mass enclosed in radius $r$ as $M(r) \equiv 4\pi\int^r_0\delta\rho_{\rm m}(r')r'^2{\rd}r'$. Solving Equation~\eqref{eq:phi_dgp} we find that the fifth force is given by 
\begin{eqnarray}\label{eq:dvarphidr_in}
F_{\rm 5th} = \frac{1}{2}\frac{\rd\phi}{\rd r} = \frac{2}{3\beta_{\rm dgp}}\frac{r^3}{r_{\rm V}^3}\left[\sqrt{1+\frac{r_{\rm V}^3}{r^3}}-1\right]F_{\rm GR},
\end{eqnarray}
where $r_{\rm V}$ is the {Vainshtein radius}
\begin{eqnarray}\label{eq:r_V}
r_{\rm V} \equiv \left[\frac{8r^2_{\rm c} r_{\rm S}}{9\beta_{\rm dgp}^2}\right]^{1/3} = \left[\frac{4GM(R)}{9\beta_{\rm dgp}^2H^2_0\Omega_{\rm rc}}\right]^{1/3}\,,
\end{eqnarray}
with the Schwarzschild radius $r_{\rm S} \equiv2GM(R)/c^2$, $\Omega_{\mathrm{rc}}=1 /\left(4 H_0^2 r_c^2\right)$, and~the total mass of the spherical object $M(R) \equiv 4\pi\int^R_0\delta\rho_{\rm m}(r')r'^2{\rd}r'$.
On scales larger than the Vainshtein radius $r\gg r_{\rm V}$, gravity receives an additional contribution $1/3\beta_{\rm dgp}$ ($\beta_{\rm dgp}>0$ for the nDGP model). On~the other hand, for~$r\ll r_{\rm V}$ the fifth force $F_{\rm 5th}\approx 2(r/r_{\rm V})^{2/3}/(3\beta_{\rm dgp})$ is suppressed.
\subsection{Parameterised Frameworks of~Gravity}
\label{subsec:pms_gravity}
Above we give an overview of specific models of DE/MG. When approaching cosmological observations, it might be useful to first assess in which direction the data constrain gravity by adopting a phenomenological approach. Working with parameterised frameworks of gravity is a powerful and efficient way to test GR with no preference for a given model. In~the following, we present the most popular parameterised~frameworks.

\subsubsection{MG Phenomenological~Functions}
\label{subsubsec:phenomen_func}
Let us consider a Universe described by the metric of Equation~\eqref{eq:metric}, where perturbations are determined by the potentials $\Psi$ and $\Phi$. A~straightforward approach to MG/DE theories is to cast all the modifications to the perturbation evolution into two parameterised functions. One can refer to these as modified growth or MG phenomenological parameters. Usually, one describes how the coupling between gravity and matter density is modified, i.e.,~how it changes the Poisson equation. The~other accounts for the variation among $\Psi$ and $\Phi$. In~the past decade, various choices of parametrisation have been explored. We review some of the most popular~ones.  

Following the notation of~\cite{Planck2016:XIV_DEMG}, possible MG parameters can be introduced by modifying the standard equations in the sub-horizon quasi-static approximation
\begin{eqnarray}
    {\rm (i)}\,\, Q(a, k):&&  -k^2 \Phi \equiv 4 \pi G a^2 Q(a, k) \rho \Delta,\\ \label{eqn:param_Q}
    {\rm (ii)}\,\,\mu(a, k):&&  -k^2 \Psi \equiv 4 \pi G a^2 \mu(a, k) \rho \Delta,\\ \label{eqn:param_mu}
    {\rm (iii)}\,\,\Sigma(a, k):&& -k^2(\Phi+\Psi) \equiv 8 \pi G a^2 \Sigma(a, k),\\ \label{eqn:param_Sigma}
    {\rm (iv)}\,\, \eta(a, k): && \eta(a, k)\equiv \Phi / \Psi, \label{eqn:param_eta}
\end{eqnarray}
with $\Delta\equiv\delta + 3 aHv/k$ as the Gauge-invariant co-moving density perturbation, $\rho$ as the energy density, and~$v$ as the irrotational component of the peculiar velocity. $Q$ and $\mu$ modify the Poisson equation for $\Phi$ and $\Psi$, respectively. $\Sigma$ parametrizes the change in the lensing response to the massless particle in a given matter field, and~$\eta$, the~so-called gravitational slip parameter, reflects the non-zero anisotropic tensor. It is sufficient to choose two from the equations (i--iv) given that they are not independent. {In Table~\ref{tab:EFT_vs_other_models}, we summarise the relation between the gravity models and the phenomenological functions. These functions are identical to ones for GR and the standard DE models (e.g.,~quintessence). For~clustering DE models (e.g.,~$k$-essence~\cite{Armendariz-Picon2000:kessence,Armendariz-Picon2001:kessence}), there is no gravitational slip, but~the Poisson equations can be modified (e.g., see review~\cite{Ishak2019:MGreview}).\endnote{Strictly speaking, $\Psi-\Phi$ can be sourced by an anisotropic pressure and a short-wave correction term~\cite{Sefusatti2011:ClusQuintessence,Hassani2020:kessence} in the $k$-essence model. However, the~absolute difference in the two potentials is shown to be small and can be safely neglected~\cite{Hassani2020:kessence}.} In general, MG models can introduce modifications to the Poisson equation and incorporate anisotropic stress terms with slip parameters that deviate from unity (e.g.,~Brans--Dicke, $f(R)$, and~DGP theories). However, it is important to note that there can be exceptions within specific subclasses of models, such as the generalized cubic covariant Galilean model, which exhibits an equal potential $\Psi=\Phi$~\cite{Frusciante2020:GCCG}. Nevertheless, it is worth mentioning that the opposite signs of $\mu-1$ and $\Sigma-1$ would tend to disfavour all Horndeski models \cite{Pogosian2016:phenom}.
Moving forward, our focus will centre on $(\mu, \Sigma)$ due to their observational significance. Specifically, $\mu$ holds a connection to the Newtonian potential and can be constrained through galaxy clustering via RSD (see Section~\ref{subsubsec:Pk_rsd}). On~the other hand, $\Sigma$ is linked to the Weyl potential and can be probed by the lensing and ISW effects discussed in Section~\ref{subsec:galaxy_lensing} (1, 2 and 4), 
respectively, while the slip parameter $\eta$ often possesses a simple functional form in many models, tending to have weaker constraints from the data~\cite{Zhao2009:GRtest}.}

\begin{table}[H] 
\caption{{Connection} 
 between well-known models of DE/MG and the EFT formalism (from Bloomfield et al.~\cite{bloomfield:2013}, Frusciante and Perenon~\cite{Frusciante:2019}). Here, $0$ means that the EFT function is identically equal to $0$, $\checkmark$ indicates that the function is present, $-$ indicates that the function is not present, while $\star$ indicates that the function is related to other EFT functions. {In addition, we also show the relationship between different gravity models and the phenomenological functions $\mu$, $\Sigma$,and $\eta$. For~each case, it is sufficient to specify two of these functions, as~the remaining functions can be derived from the other two.}}
 \label{tab:EFT_vs_other_models}
 \begin{adjustwidth}{-\extralength}{0cm}
 \newcolumntype{C}{>{\centering\arraybackslash}X}
\begin{tabularx}{\textwidth+\extralength}{p{4cm}cccccccccCCC}
\toprule
 & \boldmath{$\Omega$} & \boldmath{$\Lambda$} & \boldmath{$c$} & \boldmath{$M_2^4$} & \boldmath{$\bar{M}^3_1$} & \boldmath{$\bar{M}^2_2$} & \boldmath{$\bar{M}^2_3$} & \boldmath{$\hat{M}^2$} & \boldmath{$m_2^2$} & \boldmath{$\mu$} & \boldmath{$\Sigma$} & \boldmath{$\eta$} \\ \midrule
$\Lambda$CDM & $0$ & $\checkmark$ & -- & -- & -- & -- & -- & -- & -- & 1 & 1 & 1 \\
$f(R)$ & $\checkmark$ & $\checkmark$ & -- & -- & -- & -- & -- & -- & -- & $\geqslant1$ & $\geqslant1$ & $\leqslant1$ \\
Brans--Dicke & $\checkmark$ & $\checkmark$ & $\checkmark$ & -- & -- & -- & -- & -- & -- & $\geqslant1$ & $\geqslant1$ & $\leqslant1$\\
Quintessence & $0$ & $\checkmark$ & $\checkmark$ & -- & -- & -- & -- & -- & -- & 1 & 1 & 1 \\
$k$-essence & $0$ & $\checkmark$ & $\checkmark$ & $\checkmark$ & -- & -- & -- & -- & -- & $\geqslant 1$ & $\geqslant 1$ & 1 \\
DGP & $\checkmark$ & $\checkmark$ & $\checkmark$ & $\checkmark\star$ & $\checkmark$ & -- & -- & -- & -- & $\neq1$ & $\neq1$ & $\neq1$ \\
Horndeski & $\checkmark$ & $\checkmark$ & $\checkmark$ & $\checkmark$ & $\checkmark$ & $\checkmark$ & $\checkmark\star$ & $\checkmark\star$ & -- & \multicolumn{2}{c}{$(\mu-1)(\Sigma-1) > 0$}  & $\neq 1 \| 1$ \\ \bottomrule
\end{tabularx}
\end{adjustwidth}
\end{table}

\subsubsection{Effective Field Theory of Dark~Energy}
Effective field theory (EFT) is a general theoretical technique, first employed in a cosmological scenario to describe inflation \citep{Creminelli:2006xe,Cheung:2007st,Weinberg:2008hq}. It was later applied to describe DE by means of a unifying and model-independent framework, referred to as EFTofDE~\citep{Gubitosi:2013, bloomfield:2013}. Indeed, the~idea behind EFTofDE is to construct the most general, single scalar field action to be {effective}, i.e.,~to be easily interfaced with observations, and {unifying}, in~the sense that it aims to include the highest possible number of DE--MG models as special~cases.

The recipe to construct the EFTofDE action can be summarized as~follows:
\begin{enumerate}
    \item Usually, the~validity of the weak equivalence principle is assumed a priori. This makes the Jordan frame, where the metric is universally coupled to the matter fields, the~best-suited framework.
    We refer to~\cite{Gubitosi:2013} for details on the Jordan frame and the alternative formulation in the Einstein frame;
        \item The action is constructed within the unitary gauge~\cite{Creminelli:2006xe,Cheung:2007st}. In~practice, this means that the perturbation of the extra scalar degree of freedom $\phi$ representing the DE--MG framework is vanishing. This corresponds to foliate the 4D spacetime in 3D hypersurfaces by breaking the time-translation symmetry and fixing a preferred time slicing; 
    \item The chosen foliation is characterised through the unit vector $n_\mu$ perpendicular to the time slicing
\begin{eqnarray}
        n_\mu \equiv - \frac{\partial_\mu \phi}{\sqrt{-\hat{g}^{\mu\nu} \partial_\mu \phi \partial_\nu \phi}} = - \frac{\delta^0_\mu}{\sqrt{-\hat{g}^{00}}},
    \end{eqnarray}
    where $\hat{g}_{\mu\nu}$ is the Jordan frame metric. From~the unit vector, we can define the extrinsic curvature $K_{\mu\nu}$ as
   $K _{\mu} {}^{\nu} = \nabla _{\mu} n ^{\nu}$;

    \item We construct the action from all the perturbed operators invariant under the residual symmetry of spatial diffeomorphisms, such as the upper time component 
    of the metric $\delta \hat{g}^{00}$, the~Riemann tensor $\delta R_{\mu\nu\alpha\beta}$, the~Ricci tensor $\delta R_{\mu\nu}$ and scalar $\delta R$, the~extrinsic curvature $\delta K ^\mu_\nu$ and its trace $\delta K$;
    \item Due to the broken time-translation symmetry, the~coefficient of the operators in the action are allowed to be time-dependent functions. We call these parameters EFT functions.  
\end{enumerate}

The resulting EFTofDE action in conformal time up to the second order in perturbations is
\begin{eqnarray}\label{eq:EFT_action}
\small
    \begin{split}
        S_{\rm EFT} = &\int {\rm d}^4 x \sqrt{-\hat{g}} \Bigg\{ \frac{M^2_{\rm Pl}}{2}\left[1 + \Omega(\tau)\right]R + \Lambda(\tau) - c(\tau) a^2 \delta \hat{g}^{00} \\
        &+ \frac{M_2^4(\tau)}{2} \left( a^2 \delta \hat{g}^{00}\right)^2 - \frac{\bar{M}^3_1(\tau)}{2} a^2 \delta \hat{g}^{00} \delta K - \frac{\bar{M}^2_2(\tau)}{2} \left(\delta K\right)^2 \\
        & - \frac{\bar{M}^2_3(\tau)}{2} \delta K^\mu_\nu \delta K^\nu_\mu + \frac{\hat{M}^2(\tau)}{2} a^2 \delta \hat{g}^{00} \delta R^{(3)}\\
        &+ m_2^2(\tau)(\hat{g}^{\mu\nu} + n^\mu n^\nu) \partial_\mu \left(a^2 \hat{g}^{00}\right) \partial_\nu \left(a^2 \hat{g}^{00}\right)\\
        &+ \dots \Bigg\} + S_m \left[\hat{\psi}^{(i)}_m,\hat{g}_{\mu\nu}\right],
    \end{split}
\end{eqnarray}
where $S_m$ is the matter action. There are nine time-dependent EFT functions: $\{\Omega,\, \Lambda,\, c \}$ that multiply first-order (in perturbations) operators and affect both the background and the perturbation evolution, and~$\{M_2^4,\, \bar{M}_1^3,\,  \bar{M}_2^2,\, \bar{M}_3^2,\, \hat{M}^2,\, m_2^2\}$ for second-order operators that only enter in the evolution of the perturbations. We recover the GR limit when all the EFT functions vanish, with~the exception of $\Lambda$, and~the EFT action reduces to the standard Einstein--Hilbert~one.

The EFTofDE framework allows for exploring a wide range of models. Within~the so-called {pure} EFT approach, one can test the bare effect of each operator in the action, by~parameterising the evolution of the EFT function. Otherwise, one can link the EFT parameters to well-known DE--MG models, i.e., the {mapping} approach. For~example, $f(R)$ theories of gravity (see Section~\ref{subsec:fR}) are simply connected to the EFT formalism as follows~\citep{Gubitosi:2013}
\begin{eqnarray}\label{eqn:connect_EFT_fR}
    \Omega = f_R, \quad \Lambda = \frac{M^2_{\rm Pl}}{2} (f - Rf_R), \quad c=0,
\end{eqnarray}
with $M_2^4=\bar{M}_1^3=\bar{M}_2^2=\bar{M}_3^2=\hat{M}^2= m_2^2=0$. The~corresponding relations can also be found for the other models considered in this paper. In~Table~\ref{tab:EFT_vs_other_models} we gather a summary of how the EFT formalism is linked to specific models of~DE.

The EFT formalism offers a powerful parameterised framework derived from solid theoretical assumptions. It is widely used in the literature and numerical tools to work within the EFT frameworks are the Einstein--Boltzmann solver EFTCAMB~\citep{Hu:2013twa} and the Monte Carlo Markov chain sampler EFTCosmoMC\endnote{See: \url{http://eftcamb.org/} {(accessed on 17 June 2023).}} 
~\cite{Raveri:2014cka}. For~a complete review of EFTofDE and a summary of the most recent constraints, we refer the reader to~\cite{Frusciante:2019}.

\subsubsection{The $\alpha$-Basis~Parameterisation}
Although the EFTofDE framework allows for exploring a very wide parameter space, it is not always straightforward connecting each EFT function to a specific physical interpretation. An~alternative parameterisation that offers a clearer physical meaning is the $\alpha$-basis. First introduced by~\cite{Bellini:2014fua}, with~the $\alpha$-basis all the modifications of gravity described by Horndeski theories are cast into four independent functions of time: $\alpha_K(t)$, $\alpha_B(t)$, $\alpha_M(t)$ and $\alpha_T(t)$. In~the following we summarize the physical meaning attached to each of these functions~\citep{Bellini:2014fua}:
\begin{itemize}
    \item $\alpha_K$, dubbed as {kineticity}, is connected to the kinetic energy term in the Horndeski Lagrangian (see Equation~(\ref{eqn:horndeski_action})) and depends on the functions $K$, $G_3$, $G_4$ and $G_5$;
    \item $\alpha_B$ quantifies the {braiding}, i.e.,~the mixing between the kinetic terms of the scalar field and the metric. It depends on the $G_3$, $G_4$ and $G_5$ functions;
    \item $\alpha_M$ describes the running of the effective Planck mass and can be easily expressed in terms of Horndeski functions as
\begin{equation}
        HM_{\rm Pl}^2 \alpha_M \equiv \frac{{\rm d} M_{\rm Pl}}{{\rm d}t},
    \end{equation}
    with $H$ being the Hubble parameter and $M_{\rm Pl}^2 \equiv 2\left(G_4-2 X G_{4, X}+X G_{5, \phi}-\dot{\phi} H X G_{5, X}\right)$;
    \item $\alpha_T$ is the {tensor speed excess}, that quantifies the speed excess of GWs with respect to the speed of light. It depends on the Horndeski functions as
\begin{equation}\label{eqn:alpha_T}
        \alpha_T \equiv \frac{2 X}{M_{\rm Pl}^2}\left[2 G_{4, X}-2 G_{5, \phi}-(\ddot{\phi}-\dot{\phi} H) G_{5, X}\right].
    \end{equation}
\end{itemize}

The $\alpha$-basis parametrization is widely used and is implemented in the publicly available Einstein--Boltzmann solver hi-class\endnote{See: \url{http://miguelzuma.github.io/hi_class_public/} ({accessed on}).}~\cite{Zumalacarregui:2016pph,Bellini:2019syt}. For further discussion on this parametrization, including its connection to the EFTofDE formalism, we refer the reader to~\cite{Bellini:2014fua, Frusciante:2019}.

\subsection{Why Do We Still Study $f(R)$ and DGP?}
\label{subsec:gw_constraints}
Given the discussion in Sections~\ref{subsec:fR} and \ref{subsec:DGP},
we may now ask ourselves this question: {why would we further study $f(R)$ and DGP gravity models given that we still need a cosmological constant $\Lambda$ or DE component to explain the cosmic acceleration?} One of the motivations for studying the MG theories was to explain accelerated expansion by replacing the cosmological constant with a more natural model. Now that a large class of self-accelerated models is being ruled out by constraints from large-scale structures and GWs~\cite{Lombriser2016:GWacc,Lombriser2017:GWacc}, MG theories are less attractive from a cosmological~perspective. 

Nevertheless, these two benchmark models remain meaningful in certain ways. First, they are the few survivors from various tests at different astrophysical scales. At~small scales, they can successfully incorporate screening mechanisms to recover Newtonian gravity; at large scales, the~enhancement in structural growth that they predict can be allowed by astrophysical probes when properly choosing the corresponding parameters. Second, we need to bridge observations and theory. These two models, with~well-developed simulations and analytic templates, are illustrative examples of what MG signals could possibly look like for a given observable, especially when  MG-induced features are dominated by systematics of instrumental and astrophysical origin. 
Finally, the~cosmological constant is a phenomenological model that empirically describes cosmic acceleration. It is not surprising that there is room to introduce further parameters to this model, and, in the meantime, maintain this model's validity while numerous attempts teach us what room is left. 

As an alternative to the model-specific approach, the~parametrization framework is another way of searching for deviations from GR and high complementary, especially given that there are many MG models and it is hard to study all models exhaustively. We briefly mentioned this idea in Section~\ref{subsubsec:phenomen_func} and will mention some recent developments~below.


\subsubsection{Survivors from~GWs}
In GR, the~GWs travel at the speed of light. Many MG models alter this prediction. Therefore, GW can put constraints on the MG models. A~simple model-independent parametrization of the speed of the GW is
\begin{eqnarray}
    c^2_{\rm T} = c^2 \left(1+\alpha_{\rm T}(t)\right),
\end{eqnarray}
where $c^2_{\rm T}$ is the speed of the tensor modes, $t$ is the physical time, and $\alpha_{\rm T}$ is given in Equation~\eqref{eqn:alpha_T}. The~difference in the arrival time of the photons and GWs results in a bound of the property function $\alpha_{\rm T}(t)$. The~property function $\alpha_{\rm T}$ can be related to the $G_4$ and $G_5$ functions defined in Equation~\eqref{eqn:horndeski_action}~\cite{Kobayashi2011:gInfaltion,Bellini:2014fua}.
Applying the condition $\alpha_{\rm T}=0$ we have the speed of the GW as the speed of light $c_{\rm T}=c$, and~at the same time, eliminating part of $\calL_4$ and the entire $\calL_5$ in the Horndeski Lagrangian given by Equation~\eqref{eqn:horndeski_action}. Hence, models that involve quartic and quintic Galileans are ruled out as they invoke the full $\calL_4$ and $\calL_5$, while the cubic Galileans can still~survive. 

In the same way, we can ask ourselves which EFT functions survive the GW170817 event~\citep{LIGO2017:GW170817}. Being careful with the adopted notation and formalism, it is possible to map $\alpha_{\rm T}$ into EFT functions. In~the formalism of this paper this relation would be~\citep{bloomfield:2013,Bellini:2014fua,Frusciante:2019}
\begin{eqnarray}
    \alpha_T = - \frac{\Bar{M}^2_2 / M_{\rm Pl}^2}{1 +\Omega + \Bar{M}^2_2 / M_{\rm Pl}^2}.
\end{eqnarray}

If $\alpha_T$ is suppressed by the GW detection, this implies that the $\Bar{M}_2$ function is vanishing.\endnote{Of course one could try to force $\alpha_T = 0 $ with very large $\Omega$. However, this would lead to exotic theories incompatible with the Universe we observe.} 
This simple condition on the EFT theories allows for a better understanding of the implications of GW detection on the surviving theories. In~the $f(R)$ case, $\bar{M}_2$ is zero (see discussions around Equation~(\ref{eqn:connect_EFT_fR})); therefore, GW does not constrain the $f(R)$ model. 


Another impact of MG is through the running Planck mass $\alpha_{\rm M}$, linked to the $\Omega$ parameter in the EFT formalism. In~the tensor sector, this acts as a friction term in the equation of motion of GWs. A~running Planck mass can thus change the GW amplitude and lead to a difference in the luminosity distance by a GW and its electromagnetic counterpart~\cite{Nunes2019:PGWhorndeski}; at the same time, it can also move around the amplitude of the primordial peak of the B-mode spectrum~\cite{Pettorino2015:GW}. As~the tensor/scalar ratio $r$ also shifts the peak amplitude, there is a degeneracy between the $r$ and $\alpha_{\rm M}$. This degeneracy can be lifted by combining with probes sensitive to the scalar sector (e.g.,~see Section~\ref{sec:methods} for lensing or integrated Sachs--Wolfe effect) since a running Planck mass can affect background evolution and structure growth~\cite{Pettorino2015:GW}.
{Ref.} 
\cite{Matos2021:GRfR} studied the GW luminosity distance in $f(R)$ gravity with simulated mock data for an Einstein telescope (ET)~\cite{Sathyaprakash2012:ET}. They found that the ET-like data in the first running decade could only provide constraints for $|f_{R0}|<10^{-2}$ and was not helpful in constraining the $f(R)$ gravity. In~the case of DGP models, the~GW amplitude could ``leak'' into higher dimensions. The~leakage could cause additional GW damping during its propagation. Therefore, constraints can also be obtained by comparing GW and EM luminosity distances. {The authors of Ref.} \cite{Pardo2018:GWdgp} performed a study on the DGP model and found that the model was very poorly constrained by the GW170817 event, where only wavelengths comparable to the cosmic horizon (very low frequency) could leak into extra~dimensions.

\subsubsection{Further Parametrization}

Let us review the idea of parametrizing MG effects.
The parametrization scheme in Equations~(\ref{eqn:param_Q})--(\ref{eqn:param_eta}) is built on cosmological perturbation theories and is applicable only at linear regimes. The authors of \cite{Thomas:2020duj} took the post-Friedmann approach~\cite{Hu2007:PPF,Milillo2015:PF} and developed a parametrization framework which can be applied to all scales. Going beyond the post-Friedmann equation at the leading order, the~authors 
re-grouped the first-order post-Friedmann equation into a form that separates the large- and small-scale limit. By~neglecting some additional terms corresponding to ``intermediate scales'' (we will clarify this term shortly), the~parametrization can be performed analogously to linear~parametrization.

To understand what gravity models could be parameterized by this approach, let us take one step back and examine the GR-$\Lambda$CDM scenario. In~GR-$\Lambda$CDM, on~large scales one can use the cosmological perturbation theory (PT) and on small scales one can use Newtonian gravity; the potential arises from the cosmological PT (Equation (\ref{eq:metric})) and coincides with the Newtonian potential (Equation (\ref{eqn:Poisson})), while there are no ``intermediate scales'' at which neither of the theories apply. The~approach in~\cite{Thomas:2020duj} applies to most GR models that do not have an intermediate scale, including $f(R)$ and cubic Galilean models. In~a companion paper~\cite{Srinivasan:2021gib}, the~authors implemented this idea into $N$-body simulations and demonstrated the framework for weak lensing (WL)~observables.

While the parametrization approach allows one to assess more general classes of models, connections back to physics models can become ambiguous as the phenomenological behaviour of the models can be non-unique. This approach can nevertheless provide a null test of the standard model of cosmology and {a general test of gravity}.

Lastly, an~interesting alternative model-independent approach that does not rely on fixing a parametrization is gravity reconstruction \cite{Zhao:2010dz,Hojjati:2011xd,Casas:2017eob,Raveri:2019mxg,Raveri:2021dbu}. Latest results from \cite{Pogosian:2021mcs} show how it is possible to restrict the theory space directly from cosmological observations, reconstructing the redshift evolution of the MG functions via numerical interpolation. The author of \cite{Pogosian:2021mcs} developed a sophisticated machinery, prompting its use with future high-precision observations. 

\section{Observational Probes of Gravity~Models}
\label{sec:methods} 

Modifying GR can impact the expansion history of the Universe and the process of structure formation, while the expansion history inferred via geometric probes, such as standard rulers from baryon acoustic oscillation~\cite{Peebles1970:BAO,Sunyaev1970:BAO} and {distances of type-Ia supernovae}~\cite{Riess1998:SNa,Perlmutter1999:SNa}, can be indistinguishable from that of standard GR, the~physics behind structure formation can be substantially~modified. 

\begin{figure}[H]
    \includegraphics[width=\textwidth]{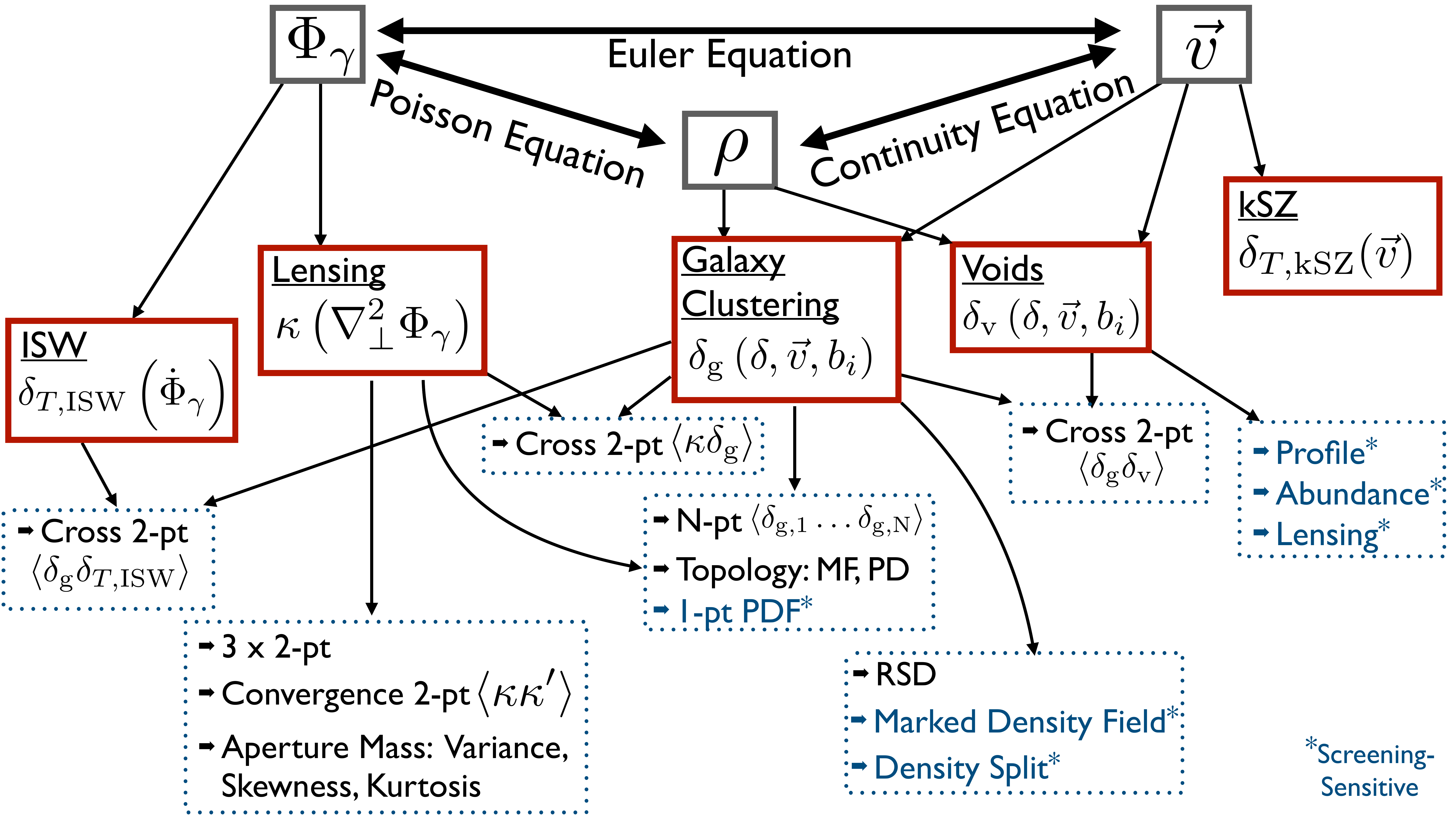}
    \caption{Connections between the basic physical quantities (boxed in grey) and derived observables (boxed in red). The~derived observables can form different combinations as observational probes in terms of summary statistics, topological properties, and~environmental dependence (boxed in dotted blue). We also highlight the probes sensitive to the screening mechanism in blue. They are useful tools in probing deviations from GR. More details see Section~\ref{sec:methods}.}
    \label{fig:probes_tools}
\end{figure}

At mildly non-linear to linear scales ($\calO(10)- \calO(100)\,\mpch$), the~theory of gravity can be described by the following  quantities: the Weyl potential $\Phi_{\gamma}(\bfk,t)$, the~density field $\rho(\bfk,t)$, and~the velocity field $\bfv(\bfk,t)$. As~shown in Figure~\ref{fig:probes_tools}, these three quantities are not fully independent: the potential and density contrast field are connected via the Poisson equation, {the potential and velocity field are linked through the Euler equation, and~the time derivative of the density contrast field is connected to the velocity field through the continuity equation.} Modifications to the gravitational potential can affect the photon trajectories emitted from the CMB or galaxies, create secondary temperature anisotropies in the CMB photons, and~induce new galaxy-clustering patterns. At~non-linear scales ($\sim\!\calO(10^{-2})- \calO(1)\,\mpch$), the~screening mechanisms have further {influence} on the Poisson equations at the characteristic screening scales, and~essentially modify the potentials, matter distribution, and velocity field. Hence, the~screening effects can lead to modified density profiles, environmental-dependent density distributions, and cluster abundance and probability density~distributions.  

Therefore, we divide the astrophysical probes into the following three categories based on basic physical quantities: potential, density, and~velocity. There are derived quantities from each of these categories, such as convergence, density contrast field, and~temperature anisotropies. Further, one can construct summary statistics to extract the information. In~this section, we will start with the relevant observables for the various astrophysics probes derived from the basic physical quantities described above. Then we will further discuss how the different datasets can be summarized to detect modifications of~gravity.

\subsection{Derived~Observables}
\label{subsec:derived_obs}
\unskip
\subsubsection{Potential-Based Derived~Observables}
From the current constraints, we expect the cosmological potential perturbations in Equation~\eqref{eq:metric} to be very small, implying that linear approximations tend to be valid. Here are some simple interpretations of the potentials:  $\Psi$ governs the dynamics of non-relativistic objects, and therefore determines the structural growth. The~Weyl potential, $\Phi_{\gamma} = (\Phi+\Psi)/2$, determines the null geodesics (light propagation), since $\Phi_{\gamma}$ is conformally invariant at linear order.\endnote{A conformal metric transformation is $g_{\mu\nu}\rightarrow e^{2\omega}g_{\mu\nu}$. At~linear order it leads to $\{\Phi,\Psi\} \rightarrow \{\Phi+\omega,\Psi-\omega\}$. Thus, the sum of the potential is conformally invariant.}

\begin{enumerate}
\item [(i)] \textbf{Temperature fluctuation.} {As the CMB }
photons travel between the last scattering surface and the observers, their wavelengths are altered when they travel through the time-varying gravitational potentials. The~{integrated Sachs--Wolfe}(ISW) \cite{Sachs1967:ISW} effect measures the decay of gravitational potential due to cosmic expansion. 
The integrated temperature anisotropies can be expressed as\endnote{We approximate the optical depth $\tau \ll 1$ and $e^{-\tau(z)}\rightarrow 1$.}
\begin{eqnarray}\label{eqn:T_fluctuate}
\left(\frac{\Delta T}{T_{\rm CMB}}\right)_{\rm ISW}(\hn)\approx-\int  \frac{d}{d\tau}({\Phi}+{\Psi})\, d \tau,
\end{eqnarray}
where $T_{\rm CMB}$ is the average temperature of the CMB background. Here, the two potentials $\Phi$ and $\Psi$ are allowed to be different, with~$\tau$ as the conformal~time.

\item [(ii)]\textbf{Deflection of photon trajectories.}\label{para:deflect_photon}
In the presence of MG, the~modified potential can source different inhomogeneous density distributions between the photons sources and the observer compared to the ones given by GR. The~additional inhomogeneity of the density field thus deflects the trajectories of photons. When the detected photons are emitted by the last scattering surface, the~derived observables are {CMB lensing}(see reviews \cite{Lewis2006:ReviewCMBlensing,Hanson2010:ReviewCMBlensing}). Alternatively, when the photons are emitted from distant source galaxies, the~galaxy shapes can be subtly distorted by intervening mass distribution and are hence named {cosmic shear} or {galaxy WL} (see review \cite{Mandelbaum2018:WLreview}). 
\end{enumerate}

The deflection of photons can be expressed in terms of the lensing potential $\psi$, the~integral of the Weyl potential $\Phi_{\gamma}$ along the photon trajectory from its source to the observer
\begin{eqnarray}
    \label{eqn:lensing-potential}
	\psi(\hn) =  \frac{2}{c^2}\int_0^{\chi_{\mathrm{s}}} d \chi\, \frac{\chi_{\mathrm{s}}-\chi}{\chi\chi_{\mathrm{s}}} \Phi_{\gamma}(\chi\hn, \chi)\,,
\end{eqnarray}
where $\chi_{\mathrm{s}}$ is the co-moving radial distance to the photon source. A~spatially flat universe and the Born approximation are assumed.
Observables, such as the deflection angle and the isotropic and anisotropic distortions, can be expressed in terms of derivatives of the lensing potential. 
For example, the~isotropic component of the distortion matrix, the~convergence $\kappa$, is given by
\begin{eqnarray}\label{eqn:kappa}
    \kappa\left({\hn}\right) &=& \frac{1}{2}\nabla^{2}_{\hn}\psi\left( {\hn}\right) \\ \nonumber
    &=&  \frac{1}{c^2}\int_0^{\chi_{\mathrm{s}}} d \chi\, \frac{\chi(\chi_{\mathrm{s}}-\chi)}{\chi_{\mathrm{s}}} \nabla^2_{\perp}\Phi_{\gamma}(\chi\hn, \chi) \,,
\end{eqnarray}
where $\nabla_{\hn} \equiv \chi \nabla_{\perp}$ are the gradients with respect to the angular position $\hm$ and the co-moving transverse distance, respectively. 
If the perturbations in the potential $\Phi_{\gamma}$ are varying on scales much smaller than $\chi_{\mathrm{s}}$, the longitudinal component of the Laplacian $\nabla^2_{\parallel}$ does not contribute to the integral in Equation \eqref{eqn:kappa}. 
This thin-lens approximation allows substitution of the full Laplacian $\nabla^{2}$ for its transversal part $\nabla^2_{\perp}$ in Equation~\eqref{eqn:kappa}, thus {applying} the Poisson equation to cast the convergence as an integral over the mass density contrast, weighted by the lensing efficiency ${\chi(\chi_{\mathrm{s}}-\chi)}/{\chi_{\mathrm{s}}}$ (see review \cite{Bartelmann2010:WLreview}).


\subsubsection{Density-Based Derived~Observables}

\begin{enumerate}
\item [(i)]\textbf{Density contrast field.} As can be seen from the Poisson equation, the~metric perturbations directly source the {density contrast field} given by
\begin{eqnarray}\label{eqn:delta}
\delta(\bfx) \equiv \frac{\rho(\bfx)-\bar{\rho}}{\rho}.
\end{eqnarray}
\end{enumerate}

The density contrast field is a Gaussian random field at the leading order. Hence, its two-point statistics capture most of the information in the field. Late-time gravitational interaction induces non-linearities, and~information in the field thus spreads beyond two-point statistics. However, the~matter field is not a direct observable. Instead, it is traced by luminous tracers such as {{individual} galaxies} or {diffuse gases}. As~galaxies or large-scale structures are only formed when the density crosses a certain threshold, {the contrast of the field of galaxies} and the matter are related via {galaxy bias} (see review by~\cite{Desjacques2018:BiasReview}). Furthermore, due to the galaxies' peculiar velocities, the~observed galaxy positions are distorted by the velocity fields, known as redshift space distortions (see Section~\ref{para:rsd}).

In addition to overdensities, underdense regions of the Universe, $\delta_{\rm v}$, contain complementary information (see Figure~\ref{fig:probes_tools}).
Due to the screening mechanisms, MG effects are only suppressed in high-density regions, while density profiles of underdense regions are modified with respect to GR: the centre will become emptier as more mass outflows towards their high-density surroundings~\citep{Zivick2015:VoidMG,Cai2015:MGvoid}. In~addition, the~lensing signal is also modified due to the modified lensing potential~\citep{Barreira2015:VoidMG,Baker2018:VoidLensingMG}. Moreover, underdense regions have an additional advantage, where properties are largely insensitive to the baryonic and galaxy formation physics~\citep{Alam2021desiMG}. Three-dimensional spherical underdense regions are usually recognized as
{cosmic voids}. One typical way of identifying voids is using a rectangular-grid-based spherical void finder~\citep{Padilla2005:Void}.
Alternatively, there are other approaches to quantify underdense regions, e.g., watershed or Voronoi-based algorithms \citep{Platen2007:WVFvoid,Neyrinck2008:zobov}. In~2D, the~spherical void finder also has its projected version over the line-of-sight (LoS). When underdense regions are identified via circumcircles of triangular Delaunay tessellation cells, they are ``tunnels''~\citep{Cautun2018:VoidMG}; alternatively, if~underdense regions are identified by placing random circles of identical radii in the sky plane, they are ``troughs''~\citep{Gruen2016:WLtroughDES}.

\subsubsection{Velocity-Based Derived~Observables}
\label{subsubsec:vel_observable}
\begin{enumerate}
\item [(i)]\textbf{Redshift space distortions.}
\label{para:rsd}
MG forces leave strong imprints on the matter velocity fields. On~sufficiently large scales, galaxies trace matter velocity fields and there is no velocity bias between the galaxies and matter distribution. The galaxies' peculiar velocities add an additional component to the galaxies' redshifts and lead to an anisotropic clustering pattern, which is known as {redshift space distortion} (RSD). The~observed density contrast field $\delta_g({\bf s})$ is in redshift space, thus receiving a velocity correction {that is parallel to the LoS direction}. In~the linear regime and under the distant observer approximation, the~density contrast field in redshift space is given by
\begin{eqnarray}\label{eqn:delta_rsd}
    \delta_{\rm g}(\bfs) \approx \delta_{\rm g}(\bfr) - \mathcal{H}^{-1}\frac{\partial v_{\parallel}}{\partial r_{\parallel}},
\end{eqnarray}
with the co-moving Hubble scale $\mathcal{H}=aH$, the~{LoS-parallel} velocity $v_{\parallel}=\bfv\cdot \hn$, and~the real and redshift space positions are related as  $\bfs = \bfr + v_{\parallel}\cdot\hn/\mathcal{H}$.

\item [(ii)]\textbf{Temperature~anisotropies in CMB Photons.}
\label{para:ksz}
As CMB photons travel through the Universe, they can be inverse Thompson scattered by hot ionized gas. 
The induced shifts in the photon temperature are referred to as Sunyaev--Zeldovich effects. In~the case of the {kinetic Sunyaev--Zeldovich} (kSZ) effect~\cite{Sunyaev1980:kSZ}, the~shift in photon temperature is caused by the bulk motion of the ionised gas in clusters. The~shift temperature can be expressed as 
\begin{eqnarray}\label{eqn:T_shift}
\left(\frac{\Delta T}{T_{\rm CMB}}\right)_{\rm kSZ}(\hn)=\left(\frac{\sigma_T \rho_{g 0}}{\mu_e m_p}\right) \int_0^{z_{*}} d z \frac{(1+z)^2}{H(z)} \Theta e^{-\tau(z)} \mathbf{q} \cdot \hn,
\end{eqnarray}
where $\sigma_{\rm T}$ is the electron Thomson cross-section, $\rho_{g0}$ is the mean gas density at redshift $z=0$, $\mu_e m_p$ is the mean mass per electron, $z_*$ is the redshift at which reionization ends, $\tau$ is the Thomson optical depth, $\Theta$ is the electron ionization fraction which depends on the primordial helium abundance and the number of ionized helium electrons~\cite{Shaw2012:kSZ,Ma2014:kSZ}, and~$\bfq\equiv (1+\delta)\bfv$ is the density-weighted peculiar velocity of the electron (bulk peculiar velocity of ionized regions or clouds). The~dot product $\bfq\cdot \hn$ in Equation~\eqref{eqn:T_shift} implies that only the LoS-parallel component velocity contributes to the temperature anisotropies.
Given that the velocity is proportional to the Fourier modes $\bfk$ of matter density contrast $\tilde{\bfv}\propto \bfk$ (we use $\sim$ to denote Fourier space quantities) in linear theory, only $\bfk$ modes parallel to the LoS can contribute to the anisotropy. However, since Fourier modes parallel to the LoS direction cancel with each other after the projection, the~angular correlation at smaller scales {mainly} receives contributions from components perpendicular to the LoS direction $\bfq_{\perp}$.

\item [(iii)]\textbf{Direct peculiar velocity measurements.}
\label{para:pecvel} The radial component of the peculiar velocity of a galaxy can be directly estimated when both the redshift and redshift-independent distance estimates are available. Distance estimates can be obtained via well-known correlations of galaxy properties such as the Tully--Fisher relation for spiral galaxies ~\cite{Tully1977:tullyfisher, McGaugh2000:TullyFisher} and the fundamental plane for ellipticals \cite{Djorgovski1987:FundamentalPlane}. These relations provide distances to galaxies with 20\% uncertainties for redshifts reaching $z\sim 0.1$. Another alternative to obtain distances is using type-Ia supernovae, which have smaller intrinsic scatters in luminosity after standardisation and can provide distances with 7\% uncertainties. The~statistics of the peculiar velocities and their cross-correlation with the galaxy density field can yield tighter constraints on the growth rate at $z < 0.1$, where cosmic variance limits the constraints from~RSD. 
\end{enumerate}

\subsection{Two-Point~Statistics}
The lowest-order summary statistic is the two-point statistics. They can capture most information in a mildly non-Gaussian field.
In principle, one can make any combinations $_2C_{\rm N}\equiv N!/(N-2)!/2!$ from the derived {quantities} discussed in Section~\ref{subsec:derived_obs} to form a two-point statistics {in Fourier space}
\begin{eqnarray}
    \av{\tilde{X}(\bfk_1), \tilde{Y}(\bfk_2)}&\equiv&(2\pi)^3 \delta_{\rm D}(\bfk_1-\bfk_2) P_{xy}(k_1) , \\ \nonumber
\end{eqnarray}
or in configuration space
\begin{eqnarray}
    \av{X(\bfx_1), Y(\bfx_2)} &\equiv& \xi_{xy}(\bfx_1 - \bfx_2)  \nonumber
\end{eqnarray}
where $\av{\ldots}$ denotes a statistical ensemble average. One can compute auto-correlation when $X$ and $Y$ are the same tracers and cross-correlation when $X$ and $Y$ are different tracers. In~the following, we will discuss both~cases.

\subsubsection{Redshift Space Power~Spectrum}
\label{subsubsec:Pk_rsd}
From Equation~\eqref{eqn:delta_rsd} we can construct the galaxy power spectrum in redshift space. At~linear order at large scales, the~galaxy power spectrum receives an LoS-dependent boost due to the velocity-dependant term and becomes anisotropic~\citep{Kaiser1987:rsd}
\begin{eqnarray}
P_{\rm g}^s(\mathbf{k})=\left(b_1+f \mu^2\right)^2 P_m(k),
\end{eqnarray}
with $\mu = \hat{k}\cdot\hat{n}$, $b_1$ as the linear bias and {the linear growth rate} as $f=d\,\ln D/d\, \ln a$. In~GR, the~growth rate can be accurately described by $f(z)=\Omega_{\rm m}^{\gamma}$ with $\gamma\approx 0.55$ (with a weak dependence on other parameters)~\citep{Linder2005:growth,Linder2007:growth}. The~overall amplitude of the matter power spectrum in the later epoch depends on the amplitude of primordial scalar power spectrum $A_{\rm s}$.
The combination $f(z)\sigma_8(z)$, {with $\sigma_8$ being the root-mean-square of the linear density field smoothed with a top-hat filter of $8\,\mpch$}, has also been one of the most common methodologies applied in current surveys to constrain deviation from GR (see Section~\ref{subsec:constrain_current_survey} and {Figure} 
~\ref{fig:fs8_compiled}), particularly at large scales ($\sim\calO(20)-\calO(150) \mpch$). Alternatively, this amplitude-induced degeneracy can be re-cast as the ratio between $f$ and the galaxy bias and expressed as $\beta\equiv f/b_1$, with~$b_1$ as the galaxy linear~bias. 

The power spectrum modelling is often performed following a perturbative approach (see Section~\ref{subsec:observable_models}), where most of the perturbation models are developed for GR-only scenarios. It is thus important to validate the GR-based galaxy-clustering models on the mock catalogues. The authors of \cite{Jennings2012:mgRSD} studied the power spectrum of $N$-body simulation with $f(R)$ gravity. The authors of \cite{Barreira2016:mgRSD} applied the BOSS wedge analysis pipeline to the $N$-body simulation with nDGP. The authors of \cite{Hernandez2019:mgRSD} further extended the study to the $f(R)$ and nDGP~models.

\begin{figure}[H]
    \includegraphics[width=0.7\textwidth]{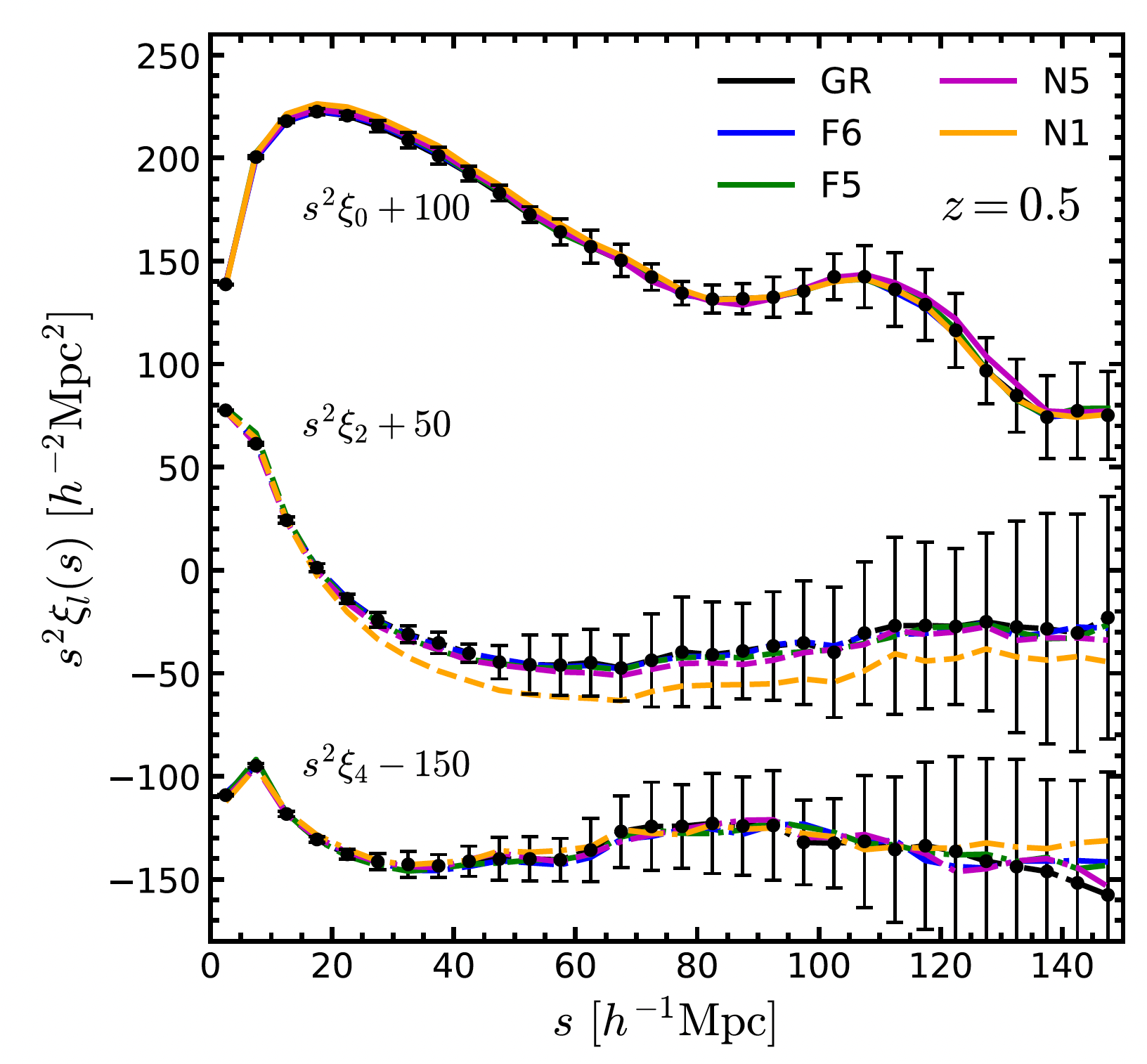}
    \caption{Impact of MG on the galaxy two-point correlation function multipoles at a snapshot $z=0.5$. For~visibility, the~multipole moments $\ell=0,2,4$ are shifted by a constant, respectively. GR is in black with error bars inferred from the standard deviation of five simulations. Coloured curves denote different MG models. A~clear deviation from GR can be seen for the quadrupole moments $\zeta_2$, while other models are consistent within the 1$\sigma$ error bar of GR. (Credit:Hern{\'a}ndez-Aguayo et al.~\cite{Hernandez2019:mgRSD}).}
    \label{fig:2PCF_3mul_rsd}
\end{figure}

However, the~large-scale RSD test requires relatively high statistical precision to detect deviations from GR. For~example, the authors of \cite{Hernandez2019:mgRSD} studied mock simulations for a BOSS-like survey and found significant deviation for DGP models while the potential for $f(R)$ is less promising. Figure~\ref{fig:2PCF_3mul_rsd} shows the impact of different gravity models on the measured galaxy two-point correlation when decomposed into Legendre polynomial basis. The~quadrupole moment, which encodes the leading-order anisotropic information,
is most affected by the MG effects due to the peculiar velocities. There is an obvious deviation of the N1 model from GR, while the rest of the models are all within the $1\sigma$ standard deviation. Future surveys, such as DESI, will improve statistical precision, but~models with smaller deviations from GR would still remain challenging for large-scale RSD~\citep{Hernandez2021:GalFORM}.

Given that MG models, such as $f(R)$, can produce large deviations at small non-linear scales, RSD effects in these regimes are important. {The authors of Ref.}~\cite{He2018:smallscaleRSD} studied the imprints of $f(R)$ gravity for small-scale RSD ($<\calO(10) \mpch$) and found no evidence of $f(R)$ gravity in BOSS data. However, the~analysis needs to assume certain relations between the circular velocity profile and the effective halo mass, while the time-dependent screening mechanism could evade this assumption. Nevertheless, the~small-scale information is valuable and helps to distinguish different gravity~models.

In addition to the galaxy auto-correlation, cosmic voids as large underdense regions of the Universe are also affected by the RSD. Since the underdense regions are less screened, they are expected to be more sensitive to the fifth force~\citep{Hui2009:EPmg,Clampitt2013:voidMG,Lam2015:VoidMG}. One may further expect that the modelling of the galaxy--void cross-correlation is simpler as the dynamics around voids are more linear. Linear RSD models were studied in~\cite{Hamaus2015:VoidRSD,Hamaus2016:VoidDyn,Cai:2016jek} developing a linear RSD model for galaxy--void cross-correlation. The~cross-correlation has been used to infer the growth rate of cosmic structure~\citep{Hamaus2014:VoidGalaxy,Hamaus2015:VoidRSD,Hamaus2020:BOSSvoidRSD,Hawken2020:eBOSSvoid,Nadathur2020:eBOSSvoid}. Yet, the~modelling relies on the following assumptions when mapping between real and redshift space: the void number conservation and the void centre position invariance. Additionally, one also needs to assume that the real space galaxy density and velocity fields around the void centres are isotropic~\citep{Nadathur2019:VoidRSDModel}. The~potential breakdown of these assumptions boils down to the unknown void centre in real space for realistic applications.
Based on these concerns, the authors of \cite{Nadathur2019:VoidRecon} applied the reconstruction methods to restore the galaxies' real space positions before applying the void finders, followed by an extended modelling~\citep{Nadathur2019:VoidRSDModel} as a complementary approach to the non-reconstructive~analysis.

\subsubsection{Angular Power Spectrum for LoS-Integrated~Observables}
\label{subsubsec:angular_Pk}

The angular power spectrum is practical in studying projected quantities. 
The angular power spectrum for LoS-integrated observables at linear order with Limber approximation~\citep{Limber1953:approx} can be written as (see, e.g.,~Equation~(45) in \cite{Joyce2016:DEvsMG}\endnote{The authors of \cite{Joyce2016:DEvsMG} worked in the unit $c=\hbar=1$, and we restored the $c$ in the equation.})
\begin{eqnarray}
    \label{eqn:Cell-general}
    C_{\ell}^{X Y}=\left.\int \frac{\mathrm{d} z}{c} \frac{H(z)}{\chi^2(z)}\left[F_X(k, z) F_Y(k, z) P(k)\right]\right|_{k=\ell / \chi(z)},
\end{eqnarray}
where the kernels $F_i(k,z)$ for tracers $i=X \,\text{or}\, Y$ consist of the scale-dependant growth factor $D(k,z)=\delta(k,z)/\delta(k,0)$. In~this paper, we will show the expressions under linear approximations and specify the power spectrum $P(k)$ for each tracer separately in the following. In~practice, various non-linear effects need to be accounted for in the modelling. The~power spectrum at large scales is sensitive to MG effects, as~screening mechanisms work less efficiently at large scales with low density.  
For potential-induced observables, it is also practical to introduce a scale-dependent variable $G$ that quantifies deviation from GR:
\begin{eqnarray}\label{eqn:G_func}
    G(k, z) \equiv \frac{1}{2}[1+\eta(k, z)] \mu(k, z)(1+z) D(k, z),
\end{eqnarray}
with $\eta$ and $\mu$ defined in Equations (\ref{eqn:param_mu}) and (\ref{eqn:param_eta}). In~the following, we will discuss four types of angular power spectra for LoS-integrated~observables.

\begin{enumerate}
\item [(i)] \textbf{Galaxy lensing.}
\label{subsec:galaxy_lensing}
One advantage of galaxy WL is that it can access smaller-scale information relative to galaxy-clustering analysis.\endnote{Here, we only compare to galaxy clustering focusing on BAO and RSD analysis. There is, in principle, no restriction in applying galaxy-clustering analysis to highly non-linear regimes.}
Contemporary galaxy WL analyses measure two-point statistics of the observed shear field, which is the anisotropic distortion induced by the lensing potential introduced in Equation~\eqref{eqn:lensing-potential} (e.g., \cite{Troxel2018, Hildebrandt2020:KiDS+VIKING-450, Hikage2019:HSC, Hamana2020:HSC-xi, Asgari2021-KiDS1000-CS, Amon2022-DES-Y3-CS, Secco2022-DES-Y3-CS-moddeling}. 
In photometric surveys, source galaxies are binned into tomographic redshift bins, and~the correlation of the shear field within these tomographic bins, as~well as their cross-correlations, are estimated with a selection of two-point statistics. 
Common examples of employed two-point statistics are correlation functions, (pseudo) angular power spectra, or~complete orthogonal sets of E-/B-mode integrals (COSEBIs) \cite{Schneider2010}. 
All these statistics can be related to the angular power spectrum of the shear field through linear transformations. 
For first-order, the~angular power spectrum of the shear field is equivalent to that of the convergence given in Equation~\eqref{eqn:kappa} (see, e.g., Equation~(6.32) in \cite{Bartelmann2010:WLreview}. The~power spectrum is the matter power spectrum at redshift today $P(k)\equiv P_{\rm mm}(k,z=0)$, and the~kernel $F_{\kappa}(k,z)$ that enters the angular power spectrum Equation~\eqref{eqn:Cell-general} can be written as
\begin{eqnarray}\label{eqn:lensing_kernel}
    F_{\kappa_i}(k,z) = G(k, z) \frac{3 H_0^2 \Omega_{m, 0}}{2 a(z)}\chi(z) \int_z dz' n_i(z') \frac{\chi(z')-\chi(z)}{\chi(z')}\,,   
\end{eqnarray}
where $n_i(z)$ is the distribution of source galaxies in tomographic redshift bin $i$.
\end{enumerate}

As both a higher matter fraction and more clustered matter cause an increase in the lensing signal, there is a strong degeneracy between the constraints on the matter fraction $\Omega_{\rm m}$ and the mass fluctuation $\sigma_8$ derived from WL data. 
For galaxy WL, the~parameter combination
\begin{eqnarray}\label{eqn:S8}
    S_8 \equiv \sigma_8 \left({\Omega_{\rm m}}/{0.3}\right)^{0.5}  
\end{eqnarray}
{quantifies the well-constrained direction in the} $\Omega_{\rm m}-\sigma_8$ {space}.

\begin{figure}[H]
    \includegraphics[width=0.7\textwidth]{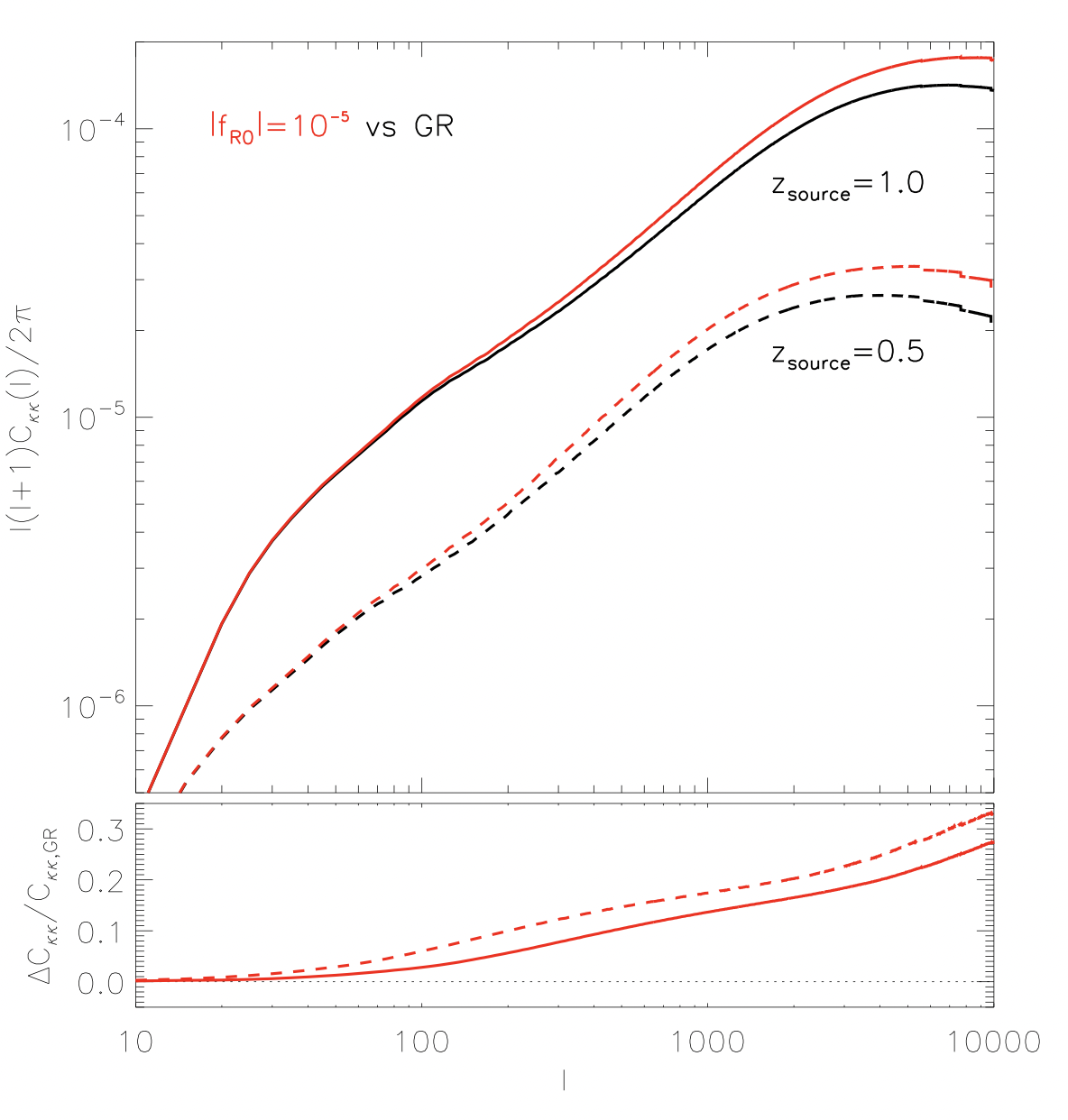}
    \caption{{Upper}: {Convergence} 
 power spectrum (see Equation~(\ref{eqn:Cell-general}) with kernel Equation~(\ref{eqn:lensing_kernel})) for GR (black) and F5 model $|f(R)|=10^{-5}$ at two source redshifts $z_{\rm source}=0.5$ (dashed) and $z_{\rm source}=1.0$ (solid). {Lower}: fractional difference in the convergence power spectrum for GR and F5 model (Credit:~\citep{Li2018:GGLfR}).}
    \label{fig:Convergence_Pk_MG_Li}
\end{figure}

Figure~\ref{fig:Convergence_Pk_MG_Li} shows the lensing convergence power spectrum at two source redshifts $z_{\rm source}=0.5 \,\text{and}\, 1.0$ for the F5 model and in comparison to GR, where all MG models have more power compared to the GR case~\citep{Li2018:GGLfR}. There is a power enhancement of 10--30\% for $10<\ell<10^4$ (corresponds to $\calO(10^{-1})-\calO(10^2) \mpch$) due to MG, with~an enhancement increasing towards smaller scales. 

Current stage 3 surveys measure cosmic shear with high precision: the detection significances are in the range of 20--30$\sigma$, depending on the analysis and employed scale cuts {\cite{Asgari2021-KiDS1000-CS, Amon2022-DES-Y3-CS,Secco2022-DES-Y3-CS-moddeling}}. 
These high-precision measurements require careful mitigation of astrophysical and observational systematics to avoid biases during parameter~inference.

One challenge in galaxy lensing analysis is the contamination by {intrinsic alignment} of galaxies. The~observed quantity in galaxy WL is the ellipticities of source galaxies.
If the orientations of these galaxies are uniformly distributed, then the observed ellipticity of a galaxy, despite being very noisy, is an unbiased
estimate of the shear imparted by gravitational lensing.
Alignment of galaxies with the large-scale structures they reside in, as~well as the structures that cause the gravitational lensing, can induce a coherent correlation of galaxy ellipticities that mimics the effect of gravitational lensing.
This intrinsic alignment of galaxies is one of the leading astrophysical systematics in analysing galaxy WL data; modelling the intrinsic alignment ultimately depends on details of galaxy formation, while direct observation of the effect in the faint population of galaxies used in WL surveys is very~challenging.

Another concern is modelling the non-linear scales and baryon feedback. The~LoS kernels in Equation~\eqref{eqn:lensing_kernel} have very broad supports, reaching from the observer up to the source galaxy.
While the kernel peaks roughly halfway between the source and the observer, this broad support means that in principle, any angular scale receives contributions from all physical scales from the matter power spectrum. In~practice, these contributions can be small but they still need to be accounted for to avoid biases in the modelling. 
This sensitivity to smaller scales, the~historically small areas covered by galaxy lensing surveys, as~well as the observational challenges in analysing WL at larger scales, cause WL analyses to push deep into the non-linear regime.
At these scales, $N$-body simulations, or~phenomenological models based thereon, are required to model the non-linear structure formation. 
While a range of models and emulators are available for the matter power spectrum in GR, this is generally not the case for MG.{\endnote{Recent advances such as the reaction formalism~\cite{Cataneo2019MGpk, Bose2020ReACT} have narrowed this gap (see~Section~\ref{subsec:observable_models}).}} Furthermore, galaxy formation processes, such as the feedback from active galactic nuclei, can redistribute large amounts of gas, significantly impacting the matter power spectrum (e.g., \cite{Chisari2019}). 
Modelling these feedback processes relies again on the details of galaxy formation, while observational data is challenging for the redshift and mass ranges relevant for galaxy WL.

On the observational side, photometric redshift uncertainties are one of the biggest concerns in galaxy WL analyses. The~galaxy samples from which the shear is estimated are typically faint. This makes it challenging to find representative spectroscopic samples from which to estimate the redshift distribution $n_i(z)$ of the $i$th tomographic bin of the photometric source sample, while current surveys use a combination of sophisticated methodologies to address this problem (e.g., \cite{Hildebrandt2021-NZ, Myles2021-DES-Y3-photoz}), the~greater depth and statistical power of future surveys (see Section~\ref{subsec:photo_survey}) still {poses} significant challenges to the accurate estimation and calibration of their photometric~redshifts.

Shape estimation is another challenge.
Galaxy WL relies on accurate estimation of galaxies' ellipticities or shears. However, most galaxies used for WL are faint, noisy, pixelated, and---in the case of ground-based surveys---poorly resolved due to the large atmospheric point spread function (e.g., \cite{Giblin2021, Gatti2021-DES-Y3-shape-cat}). As~the depth of surveys increases, galaxies increasingly overlap with their neighbours. This blending further complicates the accurate estimate of the shape of single~galaxies.

\begin{enumerate}
\item [(ii)]\textbf{CMB lensing.}
\label{subsec:cmb_lensing}
As CMB photons travel along the LoS, they are deflected by the gravitational potential gradients associated with the large-scale structure in the late-time epoch. The~CMB lensing thus indirectly traces the
underlying matter distribution. Although~CMB lensing and galaxy lensing share common features, the~broad CMB lensing kernel guarantees its sensitivity to higher redshifts. Unlike galaxy lensing, the~source redshifts of CMB lensing are almost exactly known; CMB lensing is thus free from photometric redshift~uncertainties.
\end{enumerate}

In the case of CMB lensing, one also inputs the power spectrum of matter (the same as for galaxy lensing), and~the source distribution $n(z)$ in Equation~\eqref{eqn:lensing_kernel} is given by a Dirac delta function at the redshift of the last scattering $z_*$. 
The resulting kernel $F_{\kappa_\mathrm{CMB}}(k,z)$ peaks at a {redshift} of $\sim\!2$, compared to galaxy WL {with $z\!\sim\! 0.5$}, and~is thus less sensitive to the (non-linear) growth of structure at lower redshifts~\cite{Huterer2022:GrowthReview}. 
This difference in redshift sensitivity results in a different degeneracy direction in the $\Omega_{\rm m}-\sigma_8$-plane, and~can be approximated by $\sigma_8 \left({\Omega_{\rm m}}/{0.3}\right)^{0.25}$ \cite{Planck2020:VIIILensing}.
While the CMB lensing measurements are noise-dominated at smaller scales, the~large sky area covered by CMB experiments results in a high detection significance of 40$\sigma$~\cite[][]{Planck2020:VIIILensing}.

One main observational challenge in CMB lensing analysis is foreground contaminations. There are {contaminations by various sources}: thermal Sunyaev--Zeldovich effect (tSZ~\cite{Sunyaev1972:tSZ}), cosmic infrared background (CIB), kSZ effect, and~galactic dust or extinction. These foreground contaminations, which can bias cross-correlation measurements, are particular important for CMB experiments with high angular resolution (e.g., ACT, ~\cite{Darwish2021:ACT}). Among~these foregrounds, tSZ is of particular importance. It originates from hot electron gases in galaxy clusters that scatter off the CMB photons and can correlate with galaxies at low redshifts. CIB could be another source of contamination since it is also peaked at $z\sim 2$ (e.g., \cite{White2022:DESILRGxPlanck}). Various foreground mitigation schemes are adopted in the CMB analysis pipelines (e.g., \cite{Planck2020:VIIILensing,Darwish2021:ACT}).\endnote{Mitigation schemes have been shown to pass various null tests. However, one should keep in mind there could still be redshift-sensitive or scale-sensitive residuals.} 

On the modelling side, CMB and galaxy lensing share similar concerns in non-linear gravitational evolution~\citep{Modi2017:CMBModel} and baryonic effects. In~particular, there are simulations and mitigation schemes for baryonic effects studied in~\cite{Chung2020:CMBModelBaryon,McCarthy2021:CMBModelBaryon,McCarthy2022:CMBModelBaryon,Braganca2021:CMBModelBaryon}. 
In addition to the observational and theoretical challenges, quadratic estimators (QE, \cite{Hu2002:QE,Okamoto2003:QE}, which are widely used to recover CMB lensing potential, could be sub-optimal themselves. QEs are numerically efficient but are only statistically optimal when the lensing effect on the primary CMB power spectrum is small. This is the case at large scales ($\ell<3500$) but breaks down for smaller scales ($\ell \gg 3500$), and~can be sub-optimal for high-sensitivity experiments~\cite{Maniyar2021:QE}. The reader can find more details and alternatives in the works \cite{Hirata2003:CMBrecon,Maniyar2021:QE,Millea2022:CMBLens,Legrand2022:CMBLens,Carron2022:QE}.

\begin{enumerate}
\item [(iii)]\textbf{Combined probes analyses.}
\label{subsec:3x2pt}
The two-point statistics of various combinations of different fields can be combined into {$\mathit{N\!\times\!2}$pt analyses}. 
A common joint analysis consists of the combination of the two-point statistics of the lensing fields, the~galaxy density fields, and~their cross-correlations. 
These {$\mathit{3\!\times\!2}$pt analyses} form the backbone of current and future photometric surveys (e.g., \cite{Abbott2019:desY1,Heymans2021:KiDS-1000,Abbott2022:desY3}). 
Jointly analysing the different probes helps to break parameter degeneracies due to different sensitivities of the respective probes to both cosmological parameters, as~well as astrophysical and observational parameters. 
For example, in~the linear regime, the~two-point statistics of the galaxy density field scale as $\propto b^2\sigma_8^2$, while the cross-correlation of galaxy density and lensing---called galaxy--galaxy lensing---and cosmic shear scale as $\propto b\sigma_8^2$ and $\propto\sigma_8^2$, respectively, since lensing does not depend on the galaxy bias $b$.
This combination of probes can therefore break the degeneracy between galaxy bias and clustering amplitude. 
Similarly, observational systematics in lensing, such as multiplicative shear biases, affect different probe combinations to varying degrees and therefore can be handled more robustly in joint analyses than in a single-probe analysis. 
Combined with the high signal-to-noise of these measurements (e.g., $\!\sim\! 90\sigma$ for DES-Y3, \cite{Abbott2022:desY3}), this allows for tight parameter~constraints.
\end{enumerate}

Such joint {$\mathit{3\!\times\!2}$pt analyses} can be extended with other combinations of fields. 
For example, joint analyses of the two-point statistics of galaxy lensing, galaxy density, CMB lensing, and~their cross-correlations are often referred to as {$\mathit{6\!\times\!2}$pt analyses} (e.g., \cite{DES-Y3-6x2pt}).
Cross-correlating CMB lensing (a projected quantity, tracing matter directly via the potential) and late-time galaxy (3D quantity, tracing matter indirectly via density fluctuation up to a galaxy bias) is very powerful in probing structure growth~\citep{Singh2017:PlkxSDSS,Singh2020:CMBLenxBOSS,Omori2019:DESY1xSPT,Krolewski2021:unWISExPlanck,White2022:DESILRGxPlanck}. The~cross-correlation between CMB lensing and galaxies of different types also helps to break parameter degeneracy between bias and amplitude fluctuations~\citep{Schmittfull2018:gCMBLen}. Further, the~cross-correlation of two independent experiments is also free from any correlation in instrumental noise or~systematics.

An example of combining probes designed to be sensitive to deviations from GR is the $E_\mathrm{G}$ statistic~\citep{Reyes2010:EG}:
\begin{eqnarray}
E_G \equiv \frac{P_{g\nabla^2 \Phi_\gamma}}{P_{g \theta}} \simeq \frac{1}{2}(1+\eta) \mu \frac{\Omega_{m,0}}{f(z)}\stackrel{\text { GR }}{=} \frac{\Omega_{m 0}}{f(z)},
\end{eqnarray}
where $P_{g\nabla^2 \Phi_\gamma}$ is the cross-correlation between the galaxy positions and the lensing map, $P_{g\theta}$ is the cross-correlation between the galaxy and velocity divergence field $\theta\equiv \nabla\cdot \bfv$. At~linear order, galaxy bias drops out by taking the ratio of the cross-power spectrum. The~$E_G$ probe is interesting because it is sensitive to the lensing potential and the RSDs simultaneously at larger scales; the relative lensing amplitude and the assembly rate of large-scale structures will differ in the presence of MG. In~the case of GR, the~effective parameters  $\mu=\eta=1$ (see Equations~(\ref{eqn:param_mu}) and (\ref{eqn:param_eta})).

\begin{figure}[H]
    \includegraphics[width=0.7\textwidth]{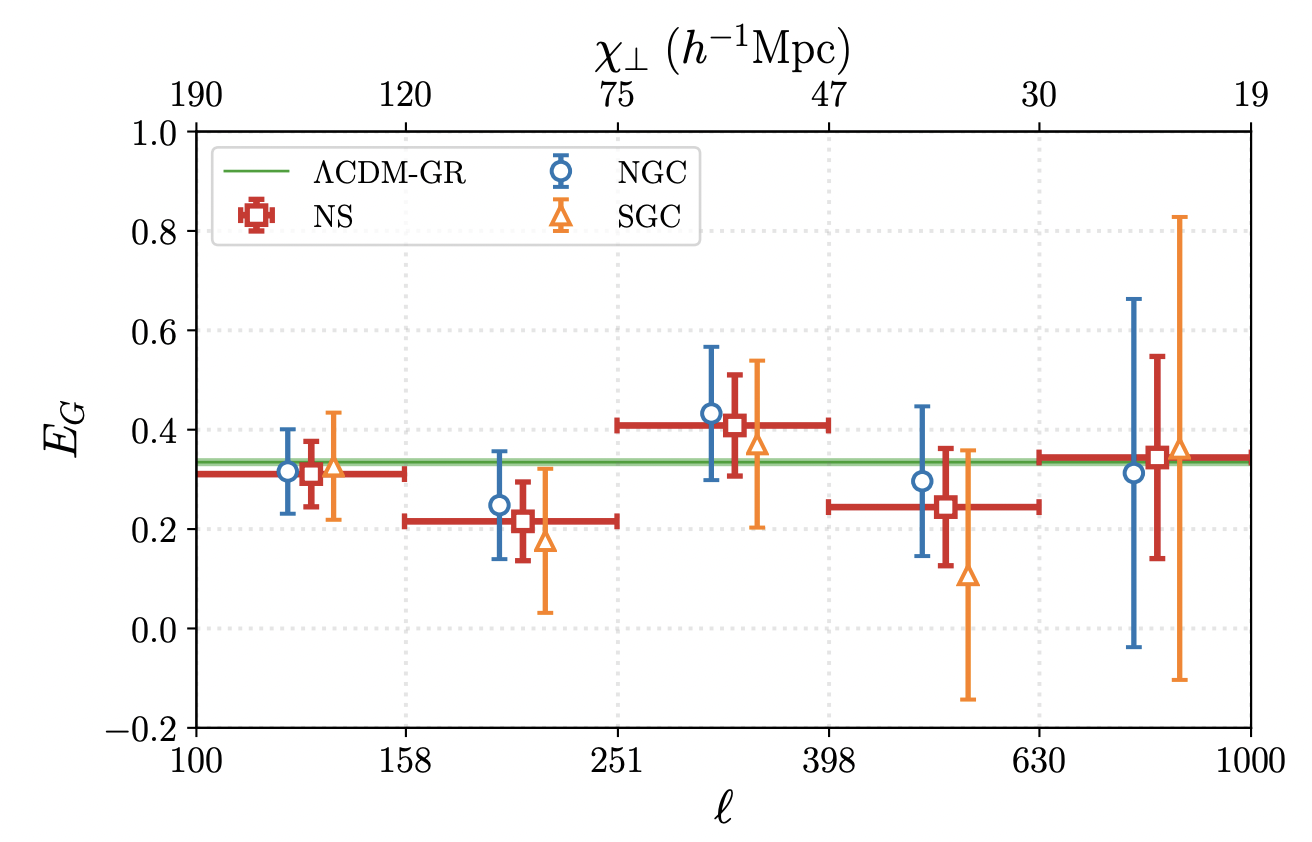}
    \caption{$E_G$ statistic by cross-correlating the Planck CMB lensing map and eBOSS DR16 QSO sample for the Northern Galactic Cap (NGC, blue), Southern Galactic Cap (SGC, orange), and~the combined sky (NS, red). The~green solid line denotes the $\Lambda$CDM using Planck~\citep{Planck2020:VI} with the shaded area being 1$\sigma$ uncertainty from the simulation. The $E_G$ probe is consistent with the scale-independent $\Lambda$CDM prediction across the five scale bins. (Credit:~\cite{Zhang2021:eBOSSQSOxPlanck}).}
    \label{fig:EG_eBOSSQSO}
\end{figure}

The $E_G$ statistic was first applied to galaxy WL~\cite{Reyes2010:EG,Blake2016:RCSLenS,delaTorre2017:VIMOS,Amon2018:KiDS2dFLenSGAMA,Blake2020:KiDsBOSS2dFLenSmg}. The authors of \cite{Pullen2015:CMBLensEG} proposed to apply $E_G$ to CMB lensing and forecasted the cross-correlation with the BOSS sample.
The authors of \cite{Zhang2021:eBOSSQSOxPlanck} applied $E_G$ statistics to the cross-correlation between the Planck CMB lensing map~\cite{Planck2020:VIIILensing} and eBOSS DR16 QSO sample~\cite{Ross2020:eBOSScatalog,Hou2021:eBOSSQSO}. 
Figure~\ref{fig:EG_eBOSSQSO} shows the consistency of the $E_G$ statistic with the scale-independent $\Lambda$CDM prediction across the five scale~bins. 

\begin{enumerate}
\item [(iv)]\textbf{ISW and cross-correlation with galaxy.}
\label{para:gISW}
The ISW effect is sensitive to the time-dependent Weyl potential and can thus be used to constrain $\Sigma$ in Equation~\eqref{eqn:param_Sigma}.
During the expansion of the Universe, gravitational potential decays. At~the same time, the~$f(R)$ gravity enhances structure growth below the Compton wavelength, and~photons thus become colder (decrease in angular power spectrum). Therefore, the~ISW effect affects the larger scales by suppressing the power for low-multipole moments. The~sensitivity at larger scales also implies that the ISW effect is cosmic variance-limited due to the limited number of modes~\citep{Dupe2011:ISW}.
While a direct measurement of the ISW effect from the CMB spectrum is very small~\citep{Ferraro2015:ISW}, expected to be measurable through the cross-correlation with the large-scale structure.
{The authors of \cite{Crittenden1996:RS} proposed the correlation of the X-ray survey and the CMB anisotropy measurements, e.g., the authors of \cite{Giannantonio2008:ISW} performed a combined analysis for galaxy surveys, radio survey, and~hard X-ray counts and found a $4.5\sigma$ detection significance of the ISW signal.}
In the ISW--galaxy cross-correlation case, the~power spectrum is given by the matter clustering $P(k)\equiv P_{\rm mm}(k,z=0)$, and~the ISW kernel is given by
\begin{eqnarray}
    \label{equ:isw-kernel}
    F_\mathrm{ISW}(k,z)=3T_{\rm CMB}\, { H_0^2 \Omega_{m, 0}}\,{k^{-2}} \,{\partial}G(k, z)/{\partial z}\,,
\end{eqnarray}
while the galaxy kernel is
\begin{eqnarray}
    \label{equ:galaxy-kernel}
    F_\mathrm{g}(k,z) = b(z) \Pi(z) D(k, z)
\end{eqnarray}
with $b(z)$ as the galaxy bias and $\Pi(z)$ as a normalized galaxy selection function. Alternatively, expressing the ISW kernel in terms of the ratio between the Newtonian and curvature potential can be used to test the $\eta$ function (see Equation~(\ref{eqn:param_eta})). {The authors of} \cite{Giannantonio2008:ISW,Ferraro2015:ISW} used the ISW--galaxy cross-correlation to constrain $\Sigma$.  
\end{enumerate}

\begin{figure}[H]
\captionsetup[subfigure]{justification=centering}
    \begin{subfigure}[b]{0.49\textwidth}
    \includegraphics[width=.95\textwidth]{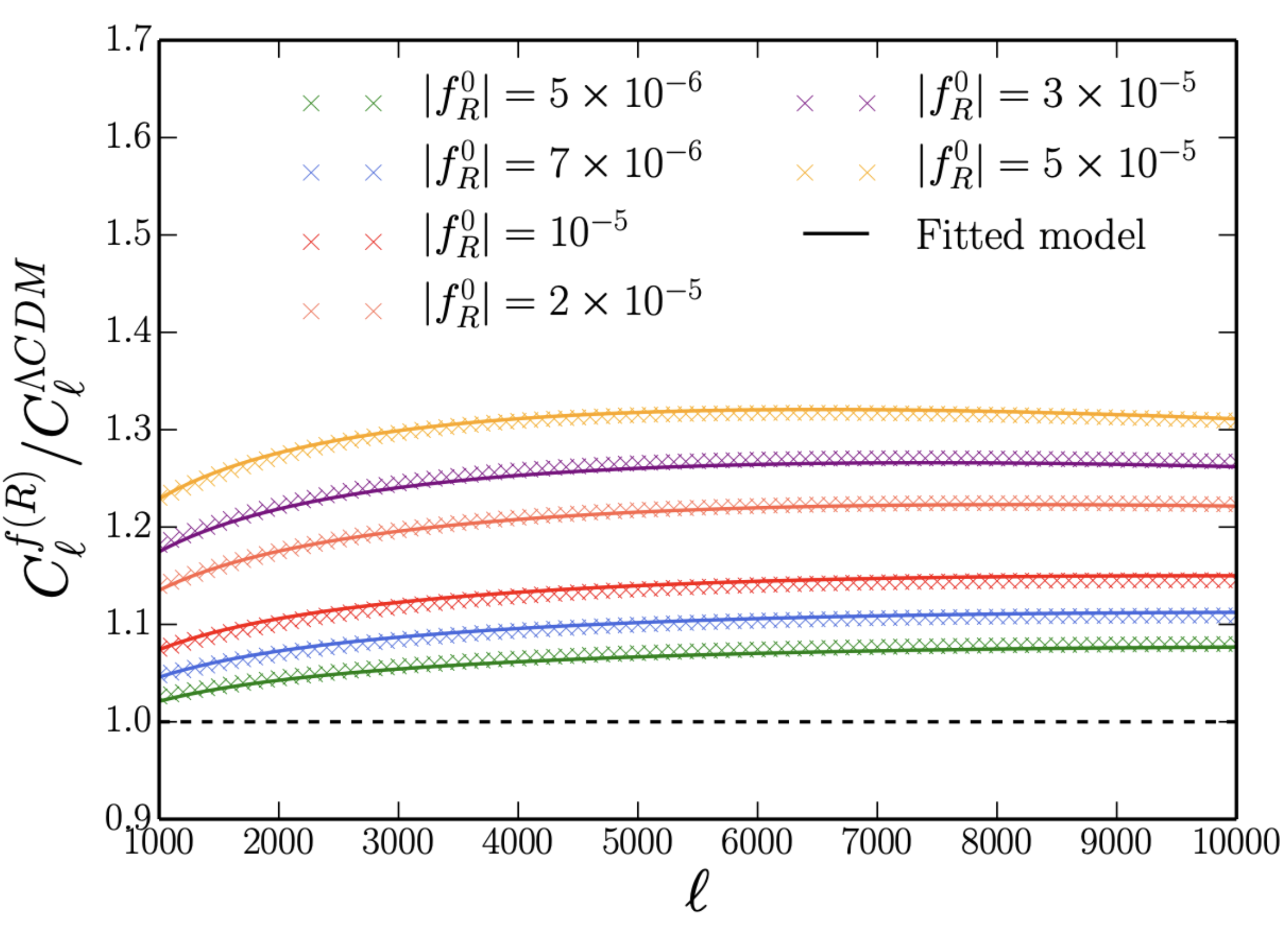}
    \caption{}
    \label{fig:kSZ_MG_fR}
    \end{subfigure}
    \begin{subfigure}[b]{0.49\textwidth}
    \includegraphics[width=.99\textwidth]{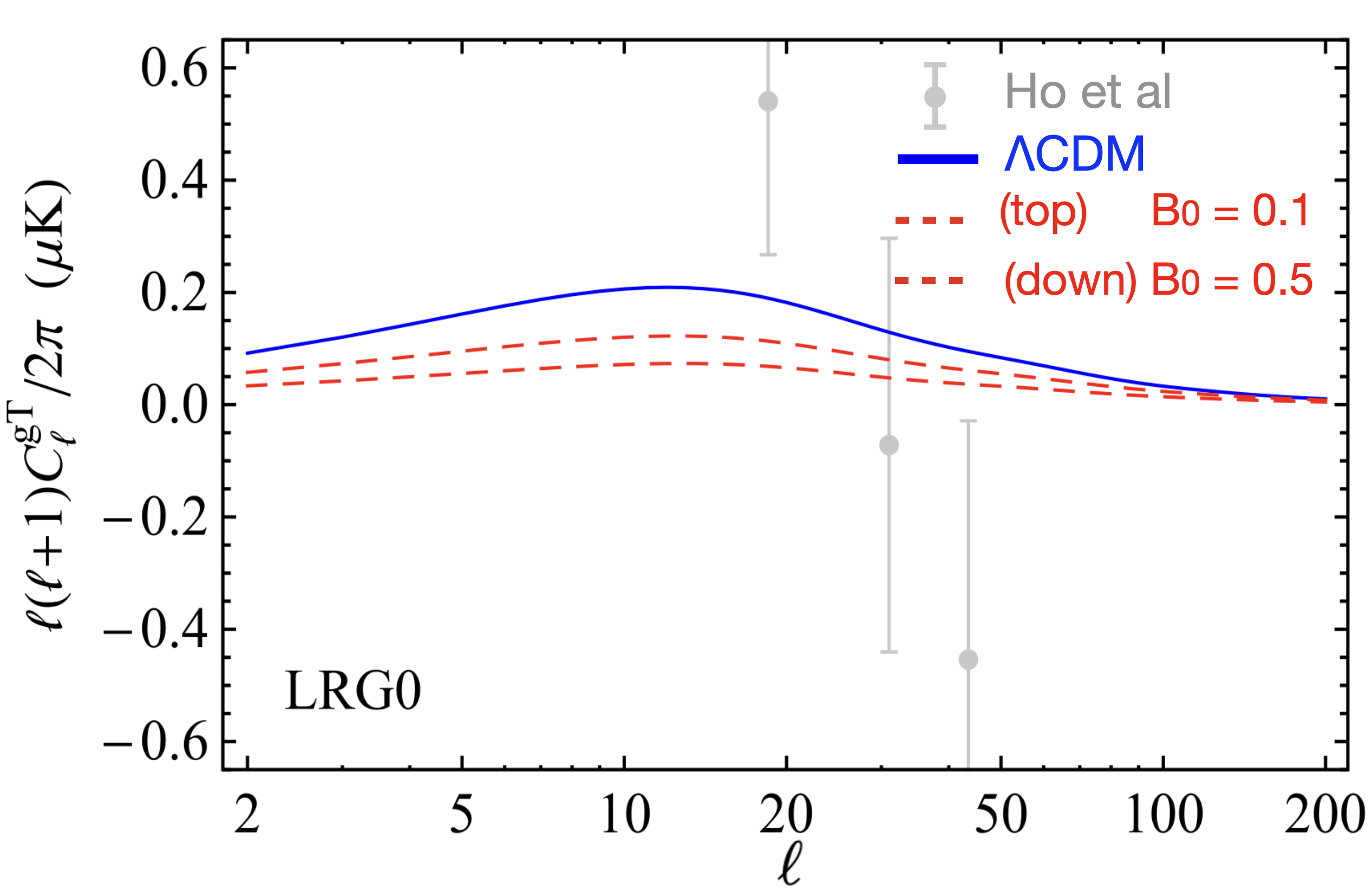}
    \caption{}
    \label{fig:ISW_fR}
    \end{subfigure}
    \caption{(\textbf{a}): Ratio of the kSZ angular power spectrum in the presence of MG compared to GR. The~angular power spectrum computed using MGCAMB~\cite{Zhao2009:mgcamb} and MGHaloFit~\cite{Zhao2014:mghalofit} for $|f^0_R|\in [5\times 10^{-6}, 5\times 10^{-5}]$ (crosses), while a fitted model is the solid curve (see Equations~(16) and (17) and Figure~4 in \cite{Bianchini2016:SZ}). (Credit:~\cite{Bianchini2016:SZ}). (\textbf{b}): Impact of $f(R)$ gravity on the ISW--galaxy cross-correlation function. The~$\Lambda$CDM prediction is given in blue, and~the modified gravity models with different Compton wavelength parameters $B_0=0.1$ and $B_0=0.5$ are in red dashed curves. The~grey data points are from the SDSS Luminous Red Galaxy (LRG) sample~\cite{Ho2008:gISW} (see~\cite{Lombriser2012:ISWfR} for data points from other SDSS samples). The~ISW effect in CMB is suppressed in the presence of MG (reproduced from~\cite{Lombriser2012:ISWfR}).}\label{fig6}
\end{figure}

The ISW--galaxy cross-correlation has been used to test different gravity and DE models~\citep{Lombriser2009:ISWdgp,Song2007:ISWfR,Lombriser2012:ISWfR,Renk2017:ISWGalileon,Krolewski2022:ISWDE}, and~was forecasted to have constraining power for sources selected from radio continuum surveys \citep{Raccanelli2012:Radio}. 
Figure~\ref{fig6}b shows the impact of the $f(R)$ model on the ISW--galaxy cross-correlation power spectrum for different Compton wavelength parameters $B_0\equiv B|_{z=0}$ (see Equation (17) in \cite{Hu:2007nk}). The~data points are from one galaxy subsample of SDSS (see~Section\ref{sec:survey} for more details on the survey). Comparing the $\Lambda$CDM prediction (blue), and~two parametrizations of the $f(R)$ model (red), the~difference is most visible at larger scales with low-multipole moments $\ell \lesssim 100$. The authors of \citep{Hang2021:ISWdesi} used the ISW effect to constrain models which attempt to replace DE with local-inhomogeneity-sourced backreactions \citep{Buchert2015:Backreaction,Racz2017:Inhomogeneity}. 

{While the ISW amplitude can be used to probe the MG effects, the~amplitude degenerates with the galaxy bias.} A robust measurement thus relies on careful calibration of the galaxy bias and contamination fraction of the sources~\citep{Ferraro2015:ISW}. When cross-correlating the ISW effect with galaxies, one can still be affected by galaxy shot noise and imperfect overlap between the galaxy and ISW kernels in redshift~\citep{Foreman2019:ISW}.
Additionally, ISW is a linear probe at larger scales and is not sensitive to screening~effects.

\begin{enumerate}
\item [(v)]\textbf{kSZ power spectrum.}
The amplitude of the kSZ angular power spectrum strongly depends on the normalization of matter perturbations and is thus sensitive to the growth of structure. As~mentioned in Section~\ref{para:ksz}, the photon temperature is shifted by the bulk motion of the ionized gas, and thus the~kSZ effect is also sensitive to the velocity field. The~kSZ angular power is given by taking the expectation value {of} Equation~\eqref{eqn:T_shift}. As~discussed previously, the~angular power spectrum only receives contribution components perpendicular to the LoS direction as the LoS-component averages out due to the integral at linear scales. The~power spectrum $P$ is given by the Vishniac power spectrum \citep{Vishniac1987:kSZ}, which requires modelling the power spectrum of the density field, velocity field, as~well as the cross-correlation of the density--velocity fields \citep{Jaffe1998:ksz,Dodelson1995:ksz,Ma2002:ksz}. The~kernel is given by
\begin{eqnarray}
    F_\mathrm{kSZ} = \frac{\sigma_{\rm T}\rho_{g0}\Theta}{\mu_{e} m_{p}} (1+z)^2 e^{-\tau(z)} \frac{\dot{D}D}{D_0},
\end{eqnarray}
with dots denoting the derivative with respect to the proper time. 
Compared to other secondary anisotropies of the CMB, the~kSZ effect is weak, with~detections of the kSZ power spectrum and cross-correlations with galaxies only in the 3--5$\sigma$ range \cite{Reichardt2021-ksz,Gorce2022-ksz,Calafut2021-ksz-sdss,Tanimura2021-ksz-clusters,Chen2022-ksz-desi-clusters}.
\end{enumerate}

{The authors of \citep{Kosowsky2009:kSZ} pointed out that galaxy cluster velocities can be used as gravity tests via the kSZ effect.} Figure~\ref{fig6}a shows the angular power spectrum model for different $f(R)$ gravities, where the linear growth rate is from MGCAMB~\citep{Zhao2009:mgcamb} and the matter power spectrum is computed using MGHalofit~\citep{Zhao2014:mghalofit}. A~fitted model is shown as the solid curve on top of the model curve (shown as crosses). As~pointed out in~\cite{Bianchini2016:SZ}, the~amplitude of the kSZ power
spectrum depends on the redshift of the reionization epoch. 
Further, the~power spectrum scales as the square of the ionization fraction (see $\Theta$ in Equation~(\ref{eqn:T_shift})), and~underestimating the helium ionization fraction can lead to underestimation of the power spectrum. Moreover, the~baryon physics can reduce the gas density in halos and counteract the MG effects~\citep{Shaw2012:kSZ}. 
The authors of \cite{Mitchell2021:SZ} studied the kSZ effect for both the effect of the nDGP and $f(R)$ models using simulations with full physics, such as star
formation, cooling, stellar and black hole feedback, and~realistic galaxy populations. Using these simulations, they found that both $f(R)$ and nDGP models can enhance power by boosting the abundance and peculiar velocity of large-scale structures. At~the same time, the~sub-grid baryonic physics can suppress the electron transverse momentum (see \mbox{Figures 6--8} in~\cite{Mitchell2021:SZ}).

\subsubsection{Power Spectrum of Line Intensity~Mapping}
Line intensity mapping (LIM or IM)~\cite{Kovetz:2017agg} has recently become a competitive cosmological probe. It has the potential to fill the gap between CMB and galaxy-clustering observations. LIM measures the integrated emission of spectral lines from multiple galaxies and the intergalactic medium in exceptionally large 3D volumes. A~strong point of LIM is its tomographic nature, which allows probing the matter distribution from the present time up to the {epoch of reionization} (EoR, $z\! \sim\! 10$). Therefore, LIM observations may be able to explore the Universe at unprecedented redshifts not accessible by other LSS~probes.

The LIM signal is a biased tracer of the underlying matter distribution at different redshifts. The~observable is the power spectrum of the brightness temperature ($T_b$) perturbations. At~a given redshift, this can be expressed as~\citep{Kovetz:2017agg}
\begin{equation}
    P_{\rm LIM}(k,z) = \Bar{T}_b^2(z) b(z) P_{\rm m}(k,z) + P_{\rm SN}(z),
\end{equation}
where $\Bar{T}_b$ is the mean brightness temperature of the line (see, e.g.,~Equation (10) in \cite{Furlanetto:2006jb}), $b$ is the bias of the considered probe, and $P_{\rm m}$ and $P_{\rm SN}$ are the matter and shot noise power spectra, respectively. Both the bias and the shot-noise term are dependent on the line luminosity function. In the literature, both the angular power spectrum and power spectrum multipoles are widely used. We refer to~\citet{Bernal:2019} for a review on how to model the LIM~signal.

\begin{figure}[H]
    \includegraphics[width=0.6\textwidth]{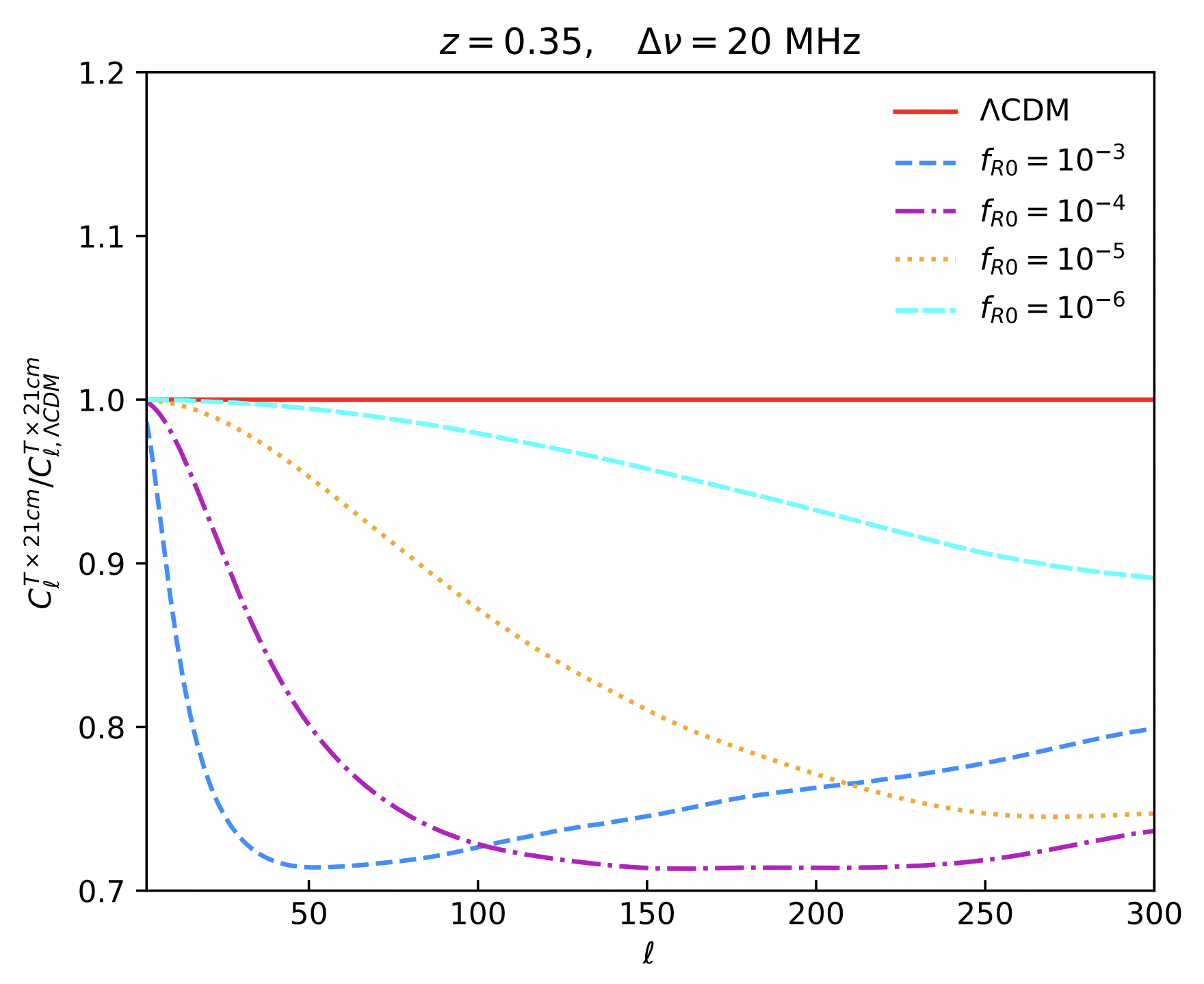}
    \caption{Cross-correlation between the 21 cm and CMB temperature for different gravity models relative to the $\Lambda$CDM model. The~ratio between the $\Lambda$CDM model to itself is the solid line, while the ratio of $f(R)$ models are shown in dashed curves (reproduced from:~\cite{Wang2021:21cmMG}).}
    \label{fig:C21xT_fR}
\end{figure}

The most studied line is the 21 cm signal arising from the spin--flip transition in the neutral hydrogen ground state \citep{Furlanetto:2006jb,Pritchardreview,Ansari_2012,Bull:2014,Villaescusa-Navarro:2018}. During~the reionization epoch at high redshifts, the~Compton wavelength of the scalar-field perturbations can still be considered in the linear regime. Thus, the fifth force is yet unsuppressed before the screening effects undertake the extra structure growth, where the modified growth rate results in modified brightness temperature~\citep{Brax2013:LIM21,Hall2013:LIM21}. Hence, 21 cm
cosmology offers a possibility of probing MG effects at high redshifts. At~lower redshifts, neutral hydrogen HI-dominated galaxies also emit 21 cm (e.g., \cite{Chowdhury2020:21HIgal}). Figure~\ref{fig:C21xT_fR} shows theory-predicted cross power spectra between CMB temperature and 21 cm in response to different gravity models at $z=0.35$, demonstrating changes of 10--25\%~\cite{Wang2021:21cmMG}. Rotational lines of carbon monoxide \mbox{(CO)~\cite{Lidz2011:LIMCO,Breysse2014:LIMCO,Li2016:LIMCO},} fine structure lines of ionized carbon (CII)~\cite{Silva2015:LIMCII,Pullen2018:LIMCII}, as~well as Lyman-$\alpha$, H$\alpha$, and~$H\beta$ lines~\cite{Silva2013:LIMLya,Pullen2014:LIMLya,Gong2017:LIMHabOII,Silva2018:LIMHalpha} are also interesting spectral features that complement the 21~cm line. 

Observationally, LIM suffers from strong continuum foreground contamination, including diffuse galactic synchrotron emission, bright point sources, and~atmospheric turbulence, {which can potentially degrade the predicted constraining power}. Several foreground removal techniques are currently being proposed and tested~\citep{Alonso2015,Wolz2016,Carucci2020,Matshawule:2021,Irfan2021,Soares2021GPR,Spinelli:2021emp,Pourtsidou:2022gsb}. {Overall, LIM is suited to probe gravity at higher and lower redshifts. It additionally allows to constrain BAO feature evolution through a wide range of time, thus also constraining the background evolution~\citep{Kovetz:2017agg}.
The LIM power spectrum is expected to be sensitive to variations in the background equation of state~\citep{Dinda:2018uwm}. Although~only a few detections are available so far~\citep{Masui2013, Anderson:2017ert,Pullen2018:LIMCII,Croft:2018rwv,BOSS:2015ids,Keating:2016pka,Cunnington:2022uzo}, there is a plethora of ongoing and planned experiments (see Section~\ref{subsec:radio_surveys}).} 

\unskip


\subsection{Higher-Order Statistics and Morphology of the~Structure}
In this section, we focus on the gravitational non-linearities induced by MG. In~general, higher-order statistics are expected to be more sensitive to the MG effects than two-point statistics as they are more sensitive to non-linear effects. Higher-order statistics are also helpful when combined with two-point statistics in~breaking galaxy bias degeneracies in models beyond~GR. 

\subsubsection{Galaxy NPCFs/Polyspectra}
Non-linear gravitational evolution induces coupling between different modes and leads to non-Gaussianity. Information is thus spread beyond the two-point statistics. Higher-order statistics thus help to break the parameter degeneracies and can be more sensitive to non-Gaussianity when gravity is enhanced. As~a generalization of two-point statistics, polyspectra or $N$-point correlation functions (NPCFs) measure the probability excess of finding a galaxy $N$-tuplet over a uniform distribution (see, e.g., \cite{Peebles1980:lss}). The~joint probability $d P_{N}$ of finding a galaxy in both volume elements $dV_i$ for $i=1,\ldots,N$ is given as
\begin{eqnarray}
    d P_{N}=\bar{\rho}^n\left(1+\xi^{(N)}\right) d V_1 \ldots d V_N.
\end{eqnarray}

Taking $N=3$ as an example, the~total 3PCF $\xi^{(3)}$ is composed of the reducible 2PCF contribution and a connected contribution, {where the connected contribution (reduced 3PCF) is given by}
\begin{eqnarray}
    \zeta\left(\mathbf{r}_{12}, \mathbf{r}_{13}, \mathbf{r}_{23}\right)=\av{\delta(\bfr_1) \delta(\bfr_2) \delta(\bfr_3)}.
\end{eqnarray}

Despite the numerous studies that have presented 3PCF measurements~\citep{Peebles1975:3pcf,Kayo2004:SDSS3pcf,Nichol2006:SDSS3pcf,McBride2011:SDSS3pcfLC,McBride2011:SDSS3pcfMB,Guo2015:SDSSIII3pcf}, a~brute force measurement of the NPCF poses computational challenges. {For $N_g$ galaxies, while a brute force approach has the complexity that scales as $\mathcal{O}(N_g^N)$, an~algorithm with the complexity of $\mathcal{O}(N_g^2)$ was proposed for the 3PCF~\cite{Slepian2015Algorithm3pcf} and was extended to the arbitrary integer $N\!\geq\! 4$~\cite{Cahn2020:isofunc,Philcox2022:encore,Slepian:Candenza}}. The~algorithm relies on the radial-angular decomposition and can either be applied to discrete data points or gridded data. In~the latter case, the~computation can alternatively be carried out by a fast Fourier transform (FFT) algorithm for the 3PCF~\citep{Portillo2018FFT3pcf}, which has complexity $\mathcal{O}(N_{\rm m}\log N_{\rm m})$, with~$N_{\rm m}$ being the number of mesh grids. An~extended FFT version was implemented in~\cite{Sunseri2022:sarabande}, including the projected 3- and 4PCF. {These improvements drastically accelerated the speed of the NPCF estimators in configuration space and also opened up new avenues for probing fundamental physics~\citep{Cahn2021:ParityIdea,Philcox2022:ParityBOSS,Hou2023:ParityBOSS}.}



Polyspectra are Fourier counterparts of NPCFs. When applied to surveys with realistic geometry and observational effects, they encode complementary information to each other. The~bispectrum as the Fourier counterpart of the connected 3PCF is defined as
\begin{eqnarray}
    \av{\delta\left(\mathbf{k}_1\right) \delta\left(\mathbf{k}_2\right) \delta\left(\mathbf{k}_3\right)} =(2 \pi)^3 \delta_{\mathrm{D}}(\bfk_{123}) B\left(\mathbf{k}_1, \mathbf{k}_2, \mathbf{k}_3\right),
\end{eqnarray}
with $\bfk_{123}\equiv \bfk_1+\bfk_2+\bfk_3$. In Fourier space, most early studies on bispectrum~\citep{Scoccimarro2001:Bk,Feldman2001:Bk,Verde2002:Bk,Gilmarin2015:Bk1,Gilmarin2015:Bk2} were limited to certain choices of triangular configurations due to computational challenge. {Efficient FFT-based algorithms were developed to make use of all $N$-tuplet configurations~\cite{Scoccimarro2015:Bk,Sugiyama2019:BkRSD}.} 

\begin{figure}[H]
    \includegraphics[width=0.8\textwidth]{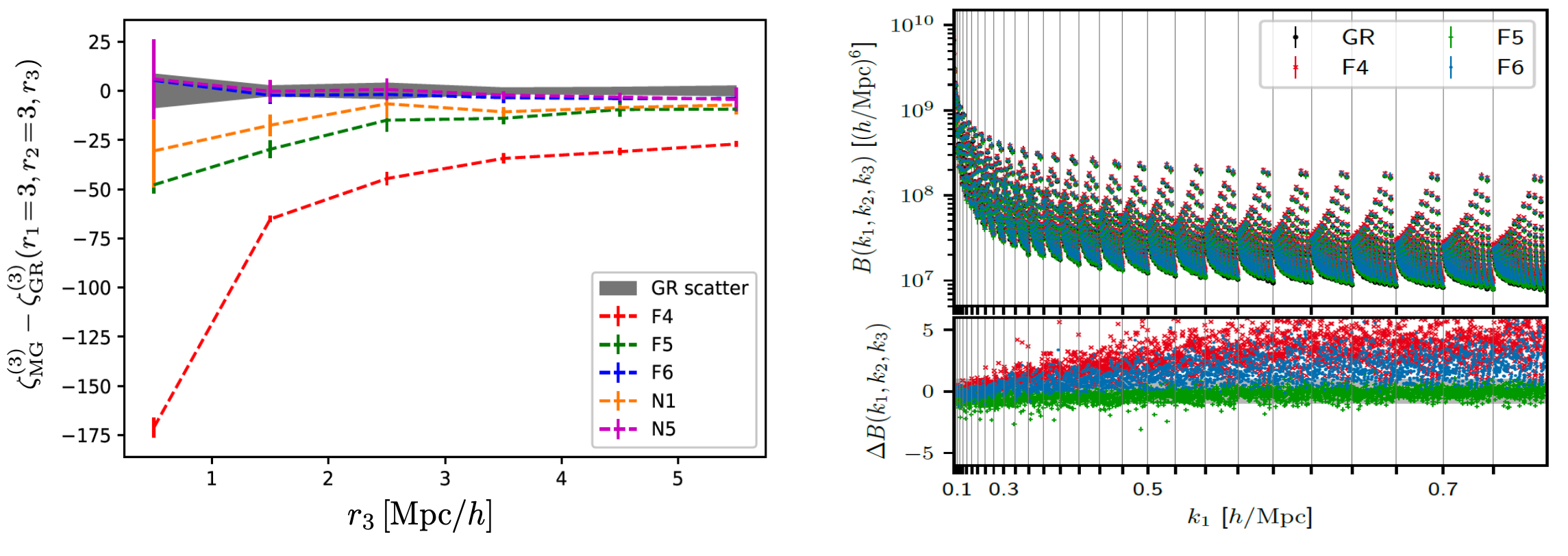}
    \caption{Impact of MG on three-point statistics. {Left:} difference in the 3PCF between GR (black dashed curve) and different MG models (coloured dashed curve) by fixing $r_1=r_2=3\,\mpch$ only. The~grey-shaded region represents the standard deviation of the five HOD GR catalogues. {Right:} upper panel shows the real-space bispectrum for GR (black) and the $f(R)$ model (coloured). Vertical lines correspond to the $k$-bin interval $\Delta k=0.025 \, {\rm Mpc}^{-1 }h$. The~lower panel shows the difference between GR and the MG models normalized by a combined standard deviation $\sigma_{\rm MG+GR}$, computed as a root-sum-squared error of the two models (reproduced from: Alam et al.~\cite{Alam2021desiMG}).}
    \label{fig:3PCF_Bk_MG}
\end{figure}

The authors of \cite{GilMarin2011:fRbk} studied the matter power spectrum and bispectrum for $f(R)$ simulations. They found that the MG-sensitivity of the bispectrum depends on the setup of the initial power spectrum. Nevertheless, the~bispectrum helps break degeneracies in bias and MG models.
The authors of \cite{Alam2021desiMG} measured three-point statistics of HOD catalogues built from MG simulations run with $N$-body codes (see Section~\ref{sec:simulations}). 
Figure~\ref{fig:3PCF_Bk_MG} shows the impact of $f(R)$ and nDGP models on the 3PCF and real-space bispectrum. The~left panel shows the difference of 3PCF between GR (black dashed curve) and the MG models by fixing $r_1=r_2=3\,\mpch$ only. Apart from F6 and N5, all other models show significant deviations from GR at these smaller scales. Moreover, the~results are insensitive to the HOD parameters. However, the~deviations decrease towards larger scales $\sim\!5 \,\mpch$. The~right panel shows the comparison of GR and the $f(R)$ model for the bispectrum monopole. The~difference between the MG models and GR is quantified by $\Delta B\equiv \left(B_{\rm MG}-B_{\rm GR}\right)/\sqrt{\sigma^2_{\rm MG}+\sigma^2_{\rm GR}}$. For~F4 and F6, $\Delta B\sim 3 \,{\rm and}\, 1.6$, respectively. Interestingly, the~difference between F5 and GR is even smaller than F6, which is likely driven by the~HODs.

Higher-order statistics are shown as a promising tool for probing MG effects. They are more sensitive to MG signals compared to the two-point statistics towards smaller scales. However, modelling non-linear scales poses challenges (see Section~\ref{subsec:observable_models}).
At larger scales, higher-order statistics usually have a lower signal-to-noise ratio compared to the two-point statistics. Observational effects such as treatment of non-trivial survey geometry are {also} more complicated compared to the two-point~statistics.

\subsubsection{Convergence-Derived Quantities: Moments of Aperture Mass and Peak~Counting}
\label{subsubsec:apmass_peakcount}

From the convergence (see Equation~(\ref{eqn:kappa})) one can construct various quantities to be applied in conventional lensing analyses (see Section~\ref{subsubsec:angular_Pk}). Apart from these quantities, one can construct the aperture mass $M_{\rm ap}(\vartheta)$~\citep{Schneider1996:WL,Schneider1998:WL}
\begin{eqnarray}
M_{\mathrm{ap}}(\boldsymbol{\theta} ; {\vartheta})=\int \mathrm{d}^2 \theta^{\prime} U_{\vartheta}\left(\left|\boldsymbol{\theta}^{\prime}-\boldsymbol{\theta}\right|\right) \kappa\left(\boldsymbol{\theta}^{\prime}\right),
\end{eqnarray}
where $\boldsymbol{\theta}$ is a 2D vector, and $U_{\vartheta}(|\boldsymbol{\theta}|)$ is a cylindrical filter function with aperture radius $\vartheta$.
The statistics of aperture mass and its higher-order moments are sensitive to non-Gaussianities in the WL signals in low~redshifts. 

The variance of the aperture mass $\av{M^2_{\rm ap}(\vartheta)}$ is given by the integral over the convergence power spectrum.\endnote{The lensing power spectrum is constructed from the angular power spectrum of the convergence $P_{\kappa\kappa}(k)=\chi^2_1 C_{\kappa\kappa}(k\chi_1)$, with~$\chi_1$ as an effective lens distance in co-moving coordinates.} The~third- and fourth-order moment of the aperture mass map, the~{skewness} $\av{M^3_{\rm ap}(\vartheta)}$ and {kurtosis} $\av{M^4_{\rm ap}(\vartheta)}$ ~\citep{Giocoli2015:WLmg} are complementary to the variance of the aperture mass, which are more sensitive to non-Gaussian information. Skewness quantifies the asymmetry of a random variable's probability distribution around its mean, and~kurtosis describes the ``tailedness'' of the distribution with respect to a Gaussian one. In~theory, they are more sensitive to the deviation from GR than the lower moment counterpart. However, it is also hard for these higher-order moments of aperture mass to achieve a higher discrimination efficiency (the ability to distinguish between different models), in~particular in the presence of massive~neutrinos.

\begin{figure}[H]
    \includegraphics[width=0.6\textwidth]{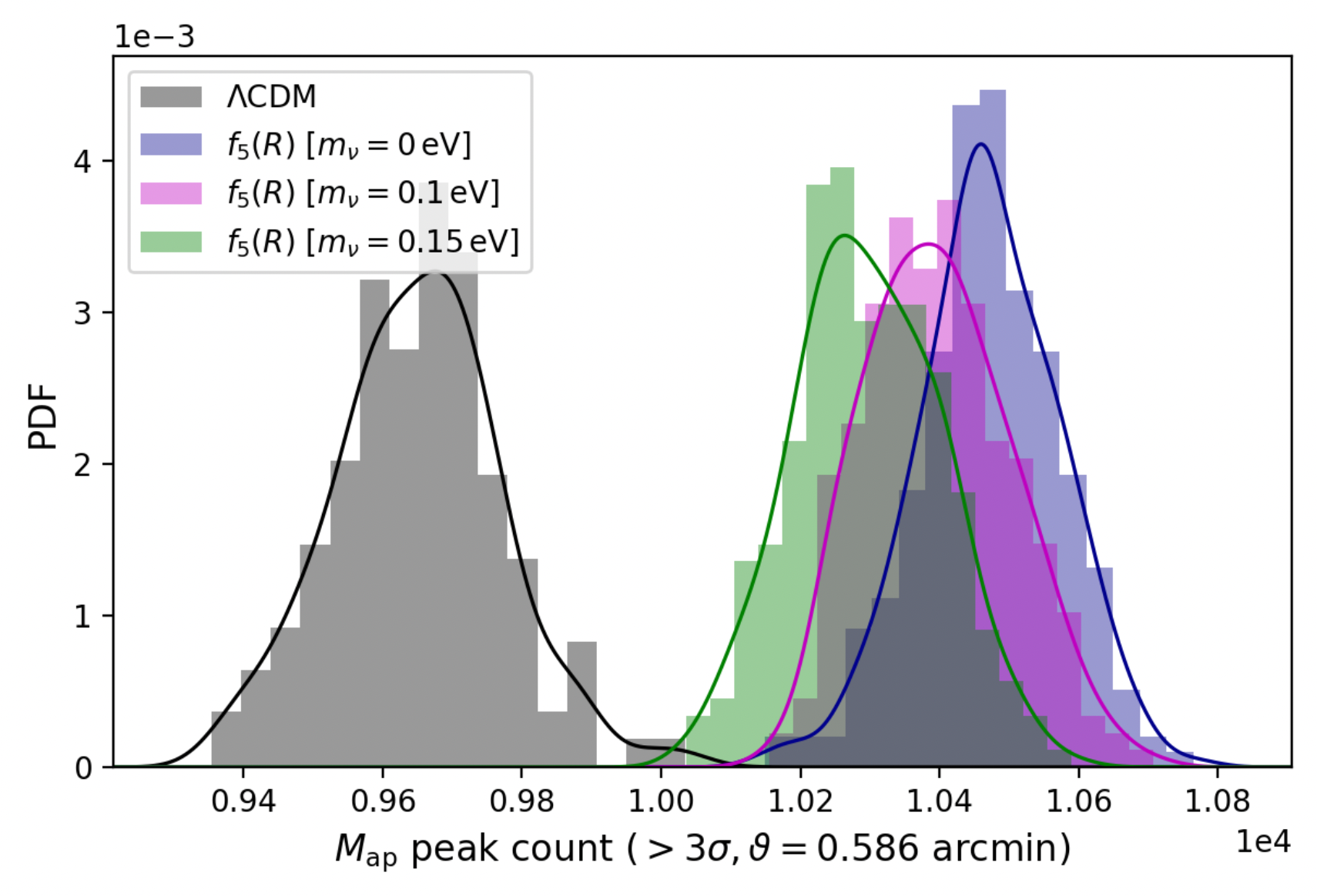}
    \caption{{Peak count} 
 of aperture mass. The~PDF has a clean separation for the GR and MG cases, even in the presence of massive neutrinos (reproduced from~\cite{Peel2018:LensingMG}).}
    \label{fig:Lensing_PeakCount_MG}
\end{figure}

Peak counting is another powerful statistic in probing non-Gaussianities~\citep{Kruse1999:StatWL}. Lensing peaks represent local regions of high convergence, a useful proxy for halo abundance (see Section~\ref{subsubsec:gal_cluster}). A~peak is defined as a pixel with a larger amplitude than its eight neighbouring pixels and exceeds the threshold $n_{\rm thres}\equiv M_{\rm ap}(\vartheta)/\sigma(\vartheta)$, with~$\sigma(\vartheta)$ as the root-mean-square of a given angular scale $\vartheta$. The~peaks trace the largest halos; they probe the tails of the mass function (see Figure~8 and 9 of~\cite{Hagstotz2019:HMFmg}). Figure~\ref{fig:Lensing_PeakCount_MG} shows the impact of the F5 model on the peak count of the aperture mass probability distribution function (PDF) with a threshold $n_{\rm thres} = 3$, where the peak counts appear to be insensitive to neutrino effects~\citep{Peel2018:LensingMG}. However, this conclusion depends on the choice of the $|f_{R0}|$ value, and degeneracy with neutrinos and $\sigma_8$ persists for a different MG parameter choice. The authors of \cite{Shirasaki2017:MFs} also pointed out that peak {counting displays degeneracy} between $|f_{R0}|$ and $\sigma_8$. Although~one can not recover GR by only varying $\sigma_8$, it demonstrates complicated cosmological parameter dependencies with MG effects. Therefore, understanding modelling at non-linear scales is equally important here. Further, when peak counts are applied to the lensing convergence map, the~peak positions can be affected by shape noise; source--lens clustering (correlation between source galaxies and lensing potentials; \cite{Hamana2002:SourceLens}) and intrinsic alignment (see galaxy lensing analysis in Section~\ref{subsubsec:angular_Pk}) are also important systematic effects that need to be studied in the~future. 

\subsubsection{Wavelet Scattering~Transforms}

$N$-point functions provide natural extensions to two-point functions to extract information from the non-Gaussian density field. However, modelling $N$-point functions is theoretically challenging (see Section~\ref{subsec:observable_models}). Moreover, $N$-point functions involve averages over powers of the density field, which increase the variance of their estimators and turns them into less informative summaries of the density~field.

As an alternative to $N$-point functions, {wavelet scattering transforms} (WSTs) have been introduced in cosmology to analyse both WL data \citep{Cheng2020:ScatterTransform,Cheng_2021} and clustering from spectroscopic surveys \citep{Valogiannis_2022}. The authors of \cite{Eickenberg2022:WST} used the Fisher forecast framework and assessed the performance of WSTs. They found that WSTs can improve the cosmological parameter constraints by several factors relative to the power spectrum. 
The 3D WST stacks a series of operations: convolutions of an input field and its modulus. The~convolution operation selects scales, and~the modulus operation converts field fluctuations to their local amplitude. 
The transformation takes an input field, $I_0(\bf x)$, and~produces a first-order output field $I_1 (\bf x)$. Then it takes the field $I_1 (\bf x)$ from the last step as an input to output a second-order field $I_2(\bfx)$. This process can be iterated to generate the $n$th-order field
\begin{eqnarray}
\label{eq:wavelet}
I_1 (\bfx) &=& \left| I_0(\bfx) \ast \psi_{j_1, l_1} \right|,\\ \nonumber
I_2 (\bfx) &=& \left| I_1(\bfx) \ast \psi_{j_2, l_2} \right|,\\
&&\ldots \nonumber
\end{eqnarray}
where the convolution kernel $\psi_{j, l}$ is defined by a family of wavelets labelled by indices $j$ and $l$, where $j$ defines the scale of the wavelet and $l$ its orientation. 
A family of wavelet $\psi_{j,l}$ can be found by dilating and rotating the mother~wavelet. 

The expected {values} of the fields, $S_n$, are called $n$th-order scattering coefficients, which summarize the 3D information contained in the input field
\begin{eqnarray}
    S_n = \av{I_n},
\end{eqnarray}
where $\av{\ldots}$ represents the spatial average of the field. Due to its iterative nature, wavelets and $N$-point statistics are highly tied with each other. For~example, the~first-order scattering coefficient is similar to the power spectrum, while the second-order scattering coefficient can be interpreted as ``clustering of clusters''~\citep{Cheng_2021}, thus encoding part of the four-point statistics~information.

In~\cite{Cheng_2021}, the~authors found that wavelet coefficients can obtain constraints very similar to those of the bispectrum on Rubin-like WL data. For~2D data, the authors of \cite{Mallet2018:WPH} introduced wavelet phase harmonics where the harmonic phase operator only acts on the complex phase of a field while keeping the amplitude of the field unchanged. The authors of \citep{Allys2020:WPH} applied the low-dimensional wavelet phase harmonics to LSS data and numerically forecasted that they provide more stringent constraints on cosmological parameters relative to the joint constraints by power spectrum and~bispectrum. 

Given that wavelet convolutions are defined locally, this summary statistic has the potential to constrain deviations from gravity that incorporate screening mechanisms on smaller scales. Still, there are currently {no applications} of the WST to gravity theories other than GR. The~major theoretical challenge regarding the WST is that it is hard to analytically predict the WST coefficient due to the modulus operation in the estimator (Equation~(\ref{eq:wavelet})), and~parameter inference relies on simulation-based~approaches.

\subsubsection{Topological~Tools}

Topological data analysis (TDA; see review, e.g., \cite{Wasserman2016:TDA} and \cite{Edelsbrunner2002:TDA,Ghrist2008:TDA}) refers to a collection of statistical methods that extract information from the structure in data. Often TDA only refers to a particular method, called {persistent homology}, while a broader definition of TDA includes a large class of data analysis methods that uses notions of shape and connectivity~\citep{Wasserman2016:TDA}. In~this section, we will discuss {Minkowski functionals (MFs)} and {persistent homology} as two representative applications of characterising LSS with topological information. They provide complementary information to the $N$-point statistics.\endnote{An animation demonstrating percolating cosmic structure: \url{https://wwwmpa.mpa-garching.mpg.de/paper/singlestream2017/percolation.html} ({accessed on})} 

These topological methods can be applied to continuous or discretized cosmological datasets, both in 2D and 3D. When analysing discretized galaxy catalogues, one can construct kernel density or apply tessellation tools to convert the catalogue into a continuous density field. Alternatively, one can also expand the radii of circles (2D) or spheres (3D) centred on each galaxy such that the circles form different topologies as the radii~changes. 

MFs, sometimes called genus statistics, are a set of statistics that characterize the topological properties of a collection of data points (see {early applications to galaxy samples~\cite{Gott1989:TDA, Mecke1994:MF}}). It converts the density contrast map into isodensity contours and {includes} quantities such as surface area, volume, curvature, and~the Euler characteristic.\endnote{Euler characteristic is also called genus. It is a topological invariant (a property of a topological space when transformed under a bijective and continuous function---homeomorphisms).}

\begin{figure}[H]
    \includegraphics[width=\textwidth]{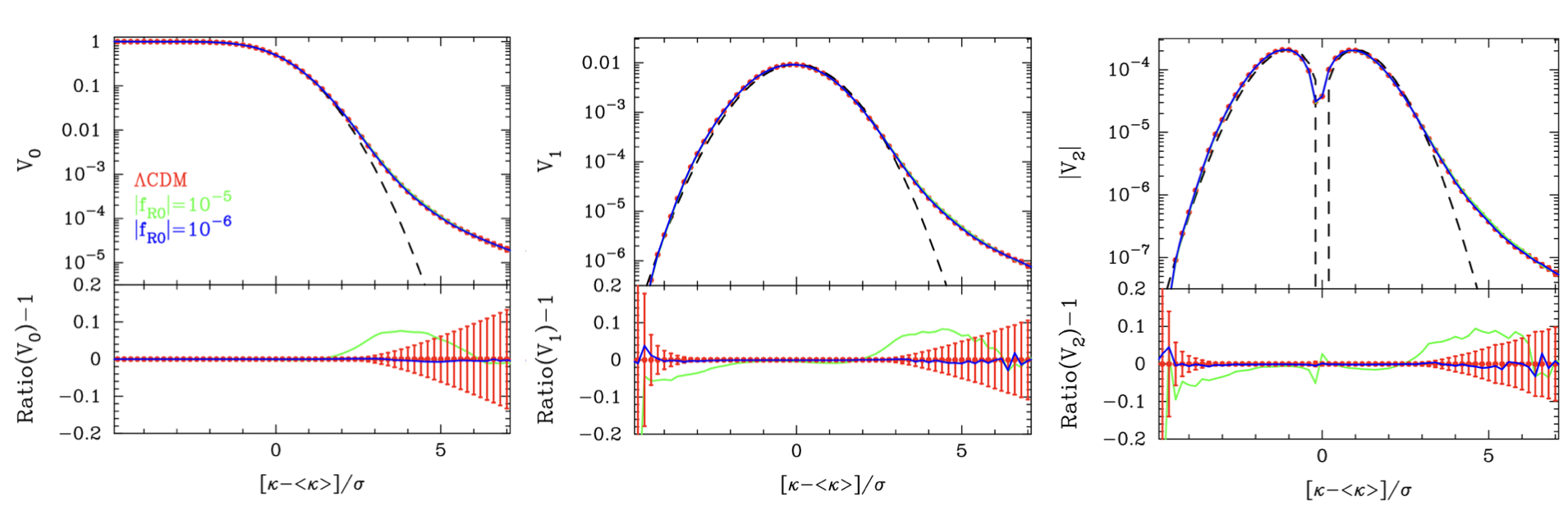}
    \caption{Impact of $f(R)$ gravity on the Minkowski functionals (MFs) of the convergence field $\mathcal{K}$ for $V_0$ (area), $V_1$ (total boundary length), and~$V_2$ (integral of geodesic curvature along the contour). The $x$-axis is the signal-to-noise ratio of the smoothed convergence field. The~MFs (coloured dots) are the average of 100 simulations with the error bars being the standard error of the mean $\sigma_{\rm std}/\sqrt{100}$. The~analytic prediction for a Gaussian random field (GRF) is shown in the black dashed curve (reproduced from~\cite{Shirasaki2017:MFs}).}
    \label{fig:MF_MG}
\end{figure}

At linear scales, since the pattern of matter distribution is scale-dependent and does not change in co-moving {coordinates}, the~genus curve of LSS can be used as a standard ruler as it is insensitive to gravitational evolution, galaxy biasing, and~RSDs \citep{Park2010:LSStda}. When including smaller scales, MFs can identify MG models in which structures grow with different rates on different scales~\citep{Wang2012:topologyLSSmg,Fang2017:MFmg}. The authors of \citep{Shirasaki2017:MFs} applied MFs to the convergence field. The~three MFs in 2D are $V_0$ (area), $V_1$ (total boundary length), and~$V_2$ (integral of geodesic curvature along the contour), where $V_2$ is equivalent to the genus statistics and is tightly connected to the peak counts (see Section~\ref{subsubsec:apmass_peakcount}).  Figure~\ref{fig:MF_MG} shows the response of the three MFs $V_0$, $V_1$, and~$V_2$ to the $f(R)$ gravity~\citep{Shirasaki2017:MFs}. In~the case of the F5 model, {there can be a maximum deviation by $\sim\!10\%$ from $\Lambda$CDM}.

While MFs are mathematically well defined, they are inherently non-local due to the integrated quantities, thus posing challenges when handling overlapping or percolating objects~\citep{Busch2020:reionization}. At~the same time, MFs are tightly connected to the rich framework studied in TDA, the~persistent homology\endnote{An intriguing connection between the Euler characteristic and Betti number (represents the ``dimension'' of a homology group) is discussed in \citet{Bobrowski2020:percolationEC}. The reader can find more information about persistent homology in~\citet{Edelsbrunner2002:TDA,Wasserman2016:TDA} and~\citet{Koplik:TDA} for a visualized explanation.}. When applying persistent homology to LSS, we are interested in one question: {as one varies the radius centred at one galaxy, what changes can we observe {in the representations of a dataset at different scales?}}. To~answer this question, a~central tool is the persistence diagram, which records the creation (birth) or destruction (death) of a connected component (zero-dimensional), loops (one-dimensional), or~voids (two-dimensional) as the scale parameter changes. The~persistence diagram can also be considered an extension of the widely used peak count~statistics.

The persistence diagram has been applied to a subsample of eBOSS DR14 quasars to detect BAO signals~\cite{Kono2020:TDAbao}. It has also been used to constrain the structure growth parameter $S_8$ and the intrinsic alignment parameter $A$ on the cosmic shear data in DES-Y1~\cite{Heydenreich2022:PersistentDESY1}. For further application of persistence diagrams and TDA on cosmic structures, see \citep{Xu2019:VoidTDA,Wilding2021:PersistentHomologyI,vandeWeygaert2011:VGS,vandeWeygaert2011:DEBettiNumber}. 

Topological analysis encodes complementary information into $N$-point statistics. However, modelling topological characteristics is highly non-trivial. In~the case of MFs, while it is possible to obtain analytic predictions beyond GRF approximations, practical challenges {require} further investigations. {For example, the authors of \citep{Appleby2022:MFsys} explored the impact of galaxy weights and survey geometries, there they found the analysis is relative stable with respect to these variations, while the impact of RSDs is at percent level on quasi-linear scales.} In the case of PDs, there are no analytic predictions for PDs to date (to our knowledge). Moreover, there has been no application of persistence diagrams to MG effects so far. Nevertheless, persistence diagrams can have considerably more information than Euler characteristics, since PDs additionally offer information on how long the specific structures~persist.

\subsubsection{One-Point Probability Density~Distribution}
The PDF of the smoothed 3D matter density field \citep{Uhlemann:2019gni}, also known as the counts-in-cells statistics, describes the likelihood of a given environment density as the outcome of the hierarchical structure formation process. 
For the density field in the early Universe, this PDF can be described by a Gaussian distribution that is characterized by the variance of the field at a smoothing scale used to define an environment, directly related to the matter power spectrum. However, gravitational evolution introduces non-Gaussianities in the late-time PDF. Contrary to its name, the~one-point PDF of the density field encodes information about beyond the two-point statistics.

For the late-time density field, the~matter PDF has been shown to be a highly complementary probe of the cosmological parameters to the matter power spectrum within $\Lambda$CDM \citep{Uhlemann:2019gni}, where the authors also showed that analytical models of the matter PDF obtain sub-percent accurate predictions for smoothing scales of the order of tens of~megaparsecs. 

\begin{figure}[H]
    \includegraphics[width=0.6\textwidth]{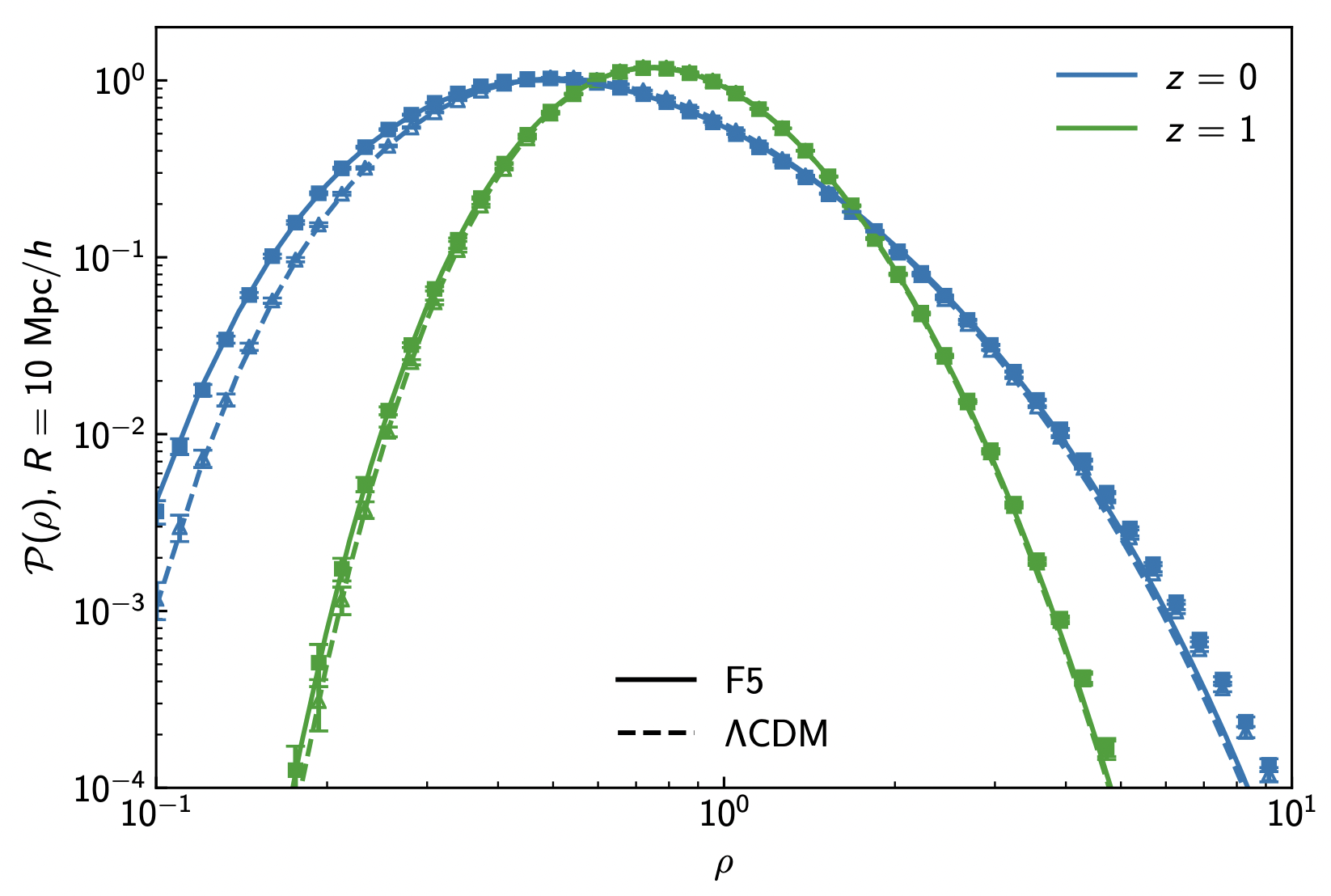}
    \caption{Global shape of the matter PDF for F5 (triangle) and $\Lambda$CDM (square) at redshift $z=0$ (blue) and $z=1$ (green) . The~measurement from the simulation is shown as the dashed curve and the model prediction is the solid curve. The~$f(R)$ gravity modifies the skewness of the distribution and leads to an asymmetric enhancement over $\Lambda$CDM (reproduced from~\citet{Cataneo:2021xlx}).}
    \label{fig:1p-PDF_MG}
\end{figure}

Recently, the authors of \cite{Cataneo:2021xlx} included the effects of MG in the analytical predictions, achieving a similar accuracy to those obtained in $\Lambda$CDM. The~authors showed that chameleon theories mainly impact the skewness of the distribution (see Figure~\ref{fig:1p-PDF_MG}), given that different density fluctuations evolve under varying gravity conditions. Alternatively, they found that in Vainshtein screening theories the enhanced linear structure growth increases the variance of the distribution while producing more under- and overdense structures (heavier tails), compared to the standard cosmology. In \cite{Cataneo:2021xlx}, the authors showed through a Fisher forecast the potential of this statistic to increase the significance of a MG signal, with~improvements of up to six times compared to the power spectrum~alone.

However, this statistic has not yet been used to constrain deviations from GR using real data. In~practice, the~PDF can be estimated from either the 3D galaxy field or 2D WL convergence maps. For~modelling gravitational non-linearity, the~framework of large-deviation statistics \citep{Bernardeau2014:LDT,Uhlemann2016:LDT,Uhlemann:2019gni} has been applied to study matter density PDFs and achieved promising results. However, modelling galaxy biasing, galaxy stochasticity, and~RSDs still remains a challenge. At~the same time, one-point PDFs can be highly complementary to density split statistics (see next paragraph) that match the galaxy density PDF and matter density PDF~quantile-by-quantile.


\subsubsection{Nearest Neighbour~Distributions}
The $k$-nearest neighbour cumulative distribution function ($k$NN-CDF) was introduced as an informative summary statistic to constrain cosmology in~\cite{10.1093/mnras/staa3604}. It is defined as the cumulative distribution function of distances from a set of volume-filling random points to the $k$-nearest tracers, while the two-point correlation function is dominated by overdense regions, the~$k$NN-CDF is sensitive to all environmental densities since the random points have been chosen to be volume-filling instead of~mass-weighted.

In~\cite{10.1093/mnras/staa3604}, the~authors showed that the $k$NN-CDF of a set of tracers is related to integrals of its $N$-point correlation functions, which implies that the information content of $k$NN is related to that of all orders of the correlation function. In~\cite{Banerjee:2021cmi}, the~authors found that combining the large-scale power spectrum with $k$NN measurements to summarize clustering on smaller scales can improve the constraints by a factor of $3$ for $\sigma_8$ and $~60\%$ for $\Omega_m$ over the power spectrum only analysis. To~assess the information content of $k$NNs on smaller scales without an accurate theoretical model, the authors of \citep{10.1093/mnras/staa3604, 10.1093/mnras/stab961} performed a Fisher analysis and found that $k$NNs can improve the constraints on the cosmological parameters compared to the monopole of the two-point correlation function, by~breaking degeneracies present in two-point~clustering.

In \cite{Banerjee:2021cmi}, the authors developed a theoretical model to predict $k$NN-CDF statistics using hybrid effective field theory (HEFT) to fit two-point correlation functions and $k$NN statistics on scales larger than 20 h$^{-1}$ Mpc, based on averages of the underlying field smoothed on different scales.  On~larger scales, all CDFs tend to one, and~therefore the information from the BAO scale is unfortunately lost. Currently, there are no extensions of this modelling approach to MG theories.

\subsection{Environment-Sensitive~Estimators}
As described in Section~\ref{subsec:screening}, modifications to GR in the solar neighbourhood are strongly constrained. For~an alternative theory of gravity to be viable on larger scales, a so-called screening mechanism needs to be incorporated to recover GR in high-density (high-curvature) regions. To~test the existence of screening mechanisms we focus on the effect that the fifth force has in unscreened regions. In~this section, we will describe different probes of gravity that leverage the environmental dependence of screening mechanisms to expose unscreened structures and set constraints on the magnitude of the fifth~force.

\subsubsection{Density Split~Statistics}
Given that extreme environments have shown promise at constraining MG theories that incorporate environment-based screening mechanisms, recent works have developed summary statistics that can extract information simultaneously from a range of environments from both lensing \citep{Friedrich_2018, Gruen_2018, Burger2023} and clustering data \citep{Paillas:2021oli,Paillas:2022wob}.

In~\cite{Friedrich_2018, Gruen_2018, Burger2023}, the~authors used a foreground (low redshift) population of galaxies to divide the sky into patches of equal size but distinct galaxy densities. Then, the~background (high redshift) population of galaxies was used to measure the tangential shear around the patches with different densities. The~stacked shear signals are tracers of the average profile of density contrast around different environment densities. More recently, this idea was extended to 3D clustering studies~\cite{Paillas:2021oli,Paillas:2022wob}, where environmental densities were estimated using the 3D positions of galaxies. 
To estimate the density contrast, the~galaxy field is smoothed with a tophat or Gaussian filter and then sampled at randomly selected positions. These positions are then ranked according to the filtered density contrast, and~split into $n$ bins. For~$n=5$, each bin is also denoted as a ``quintile''. The~random points on each quintile are then cross-correlated with the redshift space positions of the galaxies. The~cross-correlation between the random points belonging to the lowest-density quintile is similar to the void--galaxy cross-correlations, whilst those on the higher end have similar properties to the cluster--galaxy~cross-correlations.

By modelling the clustering dependence on the environment and exploiting the effect of RSDs, it was shown that the density split statistics can constrain the cosmological parameters in a $\Lambda$CDM Universe with much higher accuracy than the two-point correlation function~\cite{Paillas:2022wob}. In~particular, they found that the constraints on the sum of neutrino masses can increase a factor of eight relative to the two-point~statistics. 

To date, there are no studies of the constraining power of density split RSDs in MG scenarios, but~we expect that the environmental dependence would make it a strong test of gravity.
One challenge in using the density split technique is in modelling these statistics. First, the~definition of quintiles requires smoothing the density field, which can potentially mix the small- and large-scale information.
Second, analogous to void identification (see Section~\ref{subsubsec:void}), RSDs can produce systematic offsets in identifying different environments, leading to an ambiguity in the quintile definition. Third, as~in the case of other two-point statistics, the~standard weights are calibrated with respect to the two-point statistics, and~these weights can in fact be non-optimal for the density split method. It can generally be hard to model the scale mixing, RSD-induced quintile definition, and~residual systematics, while other non-analytic approaches such as simulation-based methods (see Section~\ref{subsec:sbi}) could potentially be interesting to investigate further. 

\subsubsection{Galaxy~Clusters}
\label{subsubsec:gal_cluster}
Galaxy clusters form from the largest peaks in the initial overdensity field (see, e.g.,~\cite{Allen2011:ClusterReview}). The~fifth force enhances gravity and faster growth and collapse of dark matter halos at non-linear scales. Therefore, MG leaves imprints on galaxy clusters' abundance, density profile, and~internal gas distribution below the Compton wavelength, below~which the mediation of the fifth force is allowed. We will discuss the three aspects in the~following.

\begin{figure}[H]
    \includegraphics[width=0.8\textwidth]{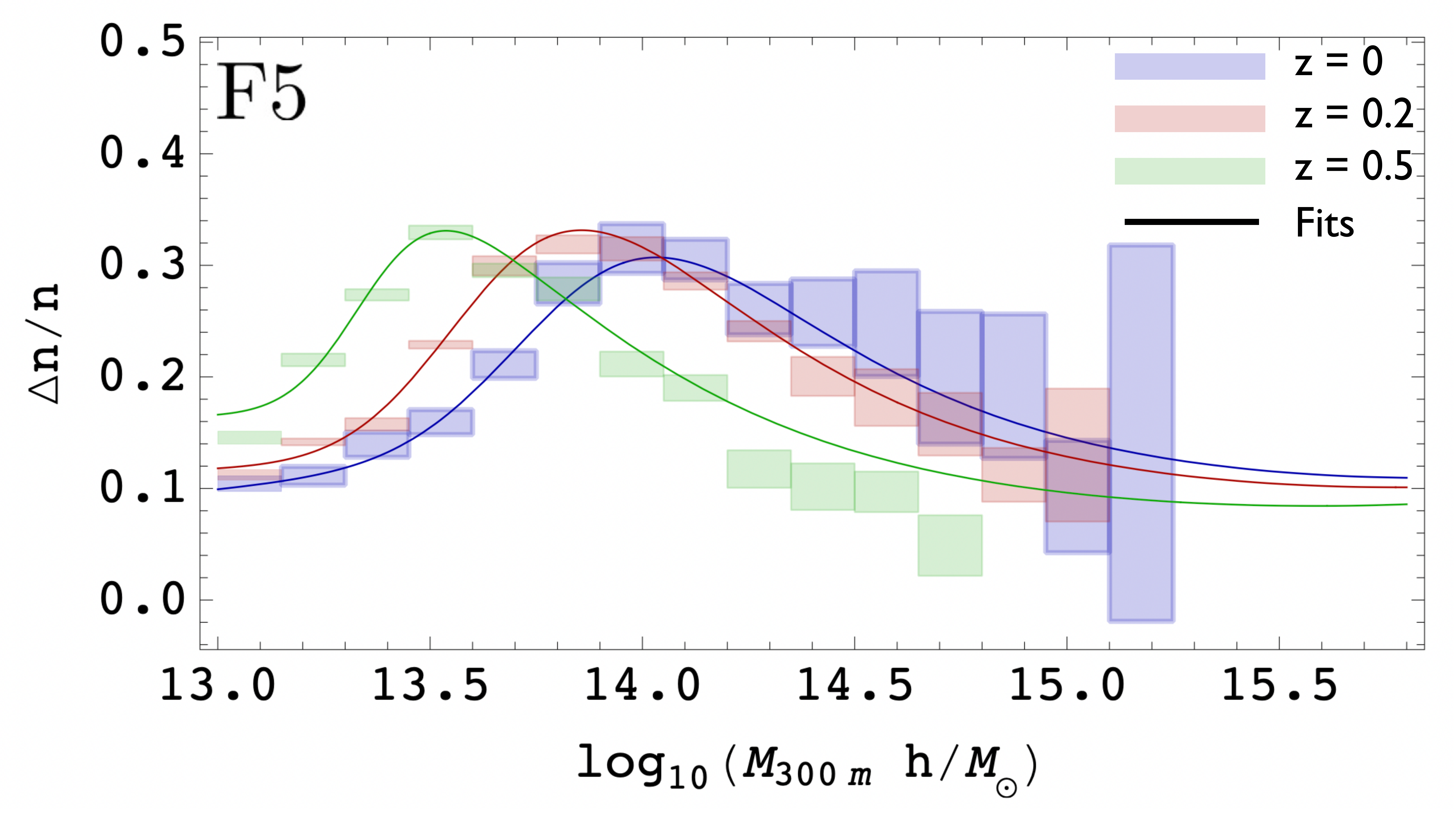}
    \caption{Relative enhancement of the halo mass function for the F5 model at $z=0$ (purple), $z=0.2$ (red), and~$z=0.5$ (green) measured from simulation (rectangles) and fits (solid curves); compared with the full simulation (blue squares) and simulations without chameleon (red triangles, displaced horizontally for visibility) (reproduced from~\cite{Cataneo:2016iav}).}
    \label{fig:cluster_abundance_HMF_fR_Cateneo16}
\end{figure}

\textbf{Abundance}: {The co-moving} 
 halo number density per virial mass $M_\nu$ interval in logarithmic binning $n_{{\rm ln} M_{\nu}}\equiv d\; {\rm ln}n/d\; {\rm ln}M_{\nu}$ can be predicted by a spherical collapse model~\citep{Gunn1972:SPC}. Structures collapse when the density field exceeds a given threshold $\delta_{\rm c}$, and~a generic prediction is that the massive halo abundance is enhanced in the presence of a fifth force. Figure~\ref{fig:cluster_abundance_HMF_fR_Cateneo16} shows the fractional enhancement of the halo mass function in F5 gravity at three redshift bins~\citep{Cataneo:2016iav}. In~the case of $f(R)$ chameleon gravity, when the scalar field value is larger compared to the gravitational potential ($|f_{R0}| \gtrsim 10^{-5}$, {large field limit}), the~generic prediction holds and the $\Lambda$CDM-predicted $\delta_{\rm c}$ is a good description to the ones for the $f(R)$ gravity. In~contrast, when the scalar field value is smaller compared to the gravitational potential ($|f_{R0}| \lesssim 10^{-5}$, {small field limit}), the~Compton wavelength shrinks so that the fifth force mediation range becomes shorter, the~chameleon mechanism modifies the collapsing threshold $\delta_{\rm c}$, and~reduces the deviation at the high-mass end~\citep{Schmidt2009:fRHaloStat}. Given the non-linear threshold $\delta_{\rm c}$, cluster counts also contain information about the higher-order moments of the density field (see Section~\ref{subsubsec:apmass_peakcount}).

\textbf{Density profile}: The impact of the MG effect on the cluster density profile can be split into three regimes: (i) inside the virial radius of halos $\simeq\!\mathcal{O}(1) \, \mpch$; (ii) a few virial radii $\sim\!\mathcal{O}(1-10) \, \mpch$; (iii) splashback radius $\sim\!\mathcal{O}(100) \, \mpch$.
In regime (i) dark matter distribution in the presence of $f(R)$ gravity is described by the Navarro--Frenk--White~ (NFW,~\cite{Navarro1996:NFW}) profile. The~halo concentration is mass- and environmental-dependent and can be modified relative to the concentration of $\Lambda$CDM halos. The authors of \cite{Mitchell2019:ClusterMG} found that there exists a universal fitting formula for the halo concentration of different redshifts and halo masses. In~regime (ii) the halo density profile is enhanced compared to GR. The~density profile can be probed by measuring, e.g.,~the projected mass distribution of cluster--galaxy lensing~\citep{Lombriser2012:ClusterMG}. In~regime (iii) for the scales approaching the initial infall stage of the clusters, the~splashback radii~\citep{Diemer2014:SplashBack} of the galaxy cluster probes the transition from the screened to the unscreened regime. Accreted material from the unscreened regimes can carry a memory of the fifth force and different dynamics~\citep{Adhikari2018:SplashbackMG}.

\textbf{Internal gas}: There are two important observables for clusters, the~{lensing mass} $M_{\rm L}\propto r^2 \nabla \Phi_\gamma(r)$ determined by the gravitational lensing and the {dynamical mass} $M_{\rm D}\propto r^2 \nabla \Psi(r)$ determined by the temperature and pressure of gases measured through X-ray or SZ effects. An~inequality between the two metric potentials $\Phi_\gamma$ and $\Psi$ indicates MG. Comparing the lensing and dynamic mass, one can place constraints for chameleon $f(R)$ gravity~\citep{Terukina2014:fRCluster}, Galilean gravity~\citep{Terukina2015:GalileonCluster}, and~beyond Horndeski theories~\citep{Sakstein2016:BeyondHorndeskiCluster}.

There are a few challenging problems regarding galaxy clusters. The~galaxy number counts rely on the mass-observable scaling relation to estimate the clusters' masses, where the latter is not directly observable. This mass-calibration problem thus involves calibrating the scaling relation and a model that transforms the theoretical mass distribution to the distribution of the clusters~\citep{Ettori2004:ScalingXrayTM,Arnaud2005:scalingTM,Vikhlinin2006:ChandraTM,Stanek2006:scalingLM}. These calibration steps can have a strong impact on the cosmological parameters, including the growth rate~\citep{Salvati2020:Cluster}. Systematics such as mass profile reconstruction can induce spurious MG detection~\citep{Pizzuti2020:ClusterMG}. Generally, disentangling astrophysical processes at cluster scales from gravitational effects is~non-trivial.

\subsubsection{Properties of 2D and 3D Underdense~Regions}
\label{subsubsec:void}
In underdense regions, the~fifth force may be totally or partially unscreened. Precisely, the~fifth force counteracts Newtonian gravity and drives underdensities to expand faster and grow larger~\citep{Cai2015:MGvoid}. Therefore, basic properties of underdense regions, such as their {abundances} and {density profiles} can be powerful in constraining gravity models. In~Section~\ref{subsubsec:Pk_rsd} we discussed {how} the 3D spherical underdense regions can be used to infer the growth rate. In~this section, we will further discuss the basic properties of underdense regions identified with more flexible~definitions. 

Void abundance distribution (number of voids) is usually plotted as a function of void size. The~void abundance distribution can be predicted through the excursion set formalism~\citep{Sheth:2003py}, which has been extended to the chameleon and symmetron models of gravity~\citep{Clampitt2013:voidMG, Lam2015:VoidMG,Voivodic2017:VoidAbundanceMG}. Although~intuitively there should be more voids in the presence of the fifth force, halos are more likely to form in underdense regions, and~the abundance distribution can thus not be a unique indicator of the fifth force. The~abundance distribution can be combined with other void~properties.

Void density profiles depict the density field as a function of radial distance towards the voids centre\endnote{The radial distance is often normalized by voids' radii given that the void density profiles are almost universal in terms of the void size.}. The~density contrast field for voids approaches $\delta\rightarrow-1$ towards the void centre but increases towards the void radii. For~smaller voids, there are usually ridges at the void radii which are associated with the surrounding filaments and walls. For~larger voids, the~ridges are less pronounced. At~radii much larger than the size of the voids, the~density fluctuations again approach zero. Overall, the~density profiles of these underdense regions are modified with respect to GR: the centre will become emptier as more mass outflows towards their high-density surroundings~\citep{Zivick2015:VoidMG,Cai2015:MGvoid}.

Since the convergence of the WL map is tightly related to the projected matter density (see Equation~(\ref{eqn:kappa})), the~void density profile also manifests itself in the properties related to the {void lensing}, such as convergence profile, the~tangential shear profile~\citep{Davies2019:VoidLensMG}, as~well as the mean excess surface density profile around each void \citep{Barreira2015:VoidMG,Cautun2018:VoidMG,Paillas2019:VoidMG}.

Compared to the 3D underdense regions, the~2D regions have a higher signal-to-noise ratio~\citep{Gruen2016:WLtroughDES,Cautun2018:VoidMG}. The authors of \cite{Alam2021desiMG} forecasted the signal-to-noise ratio for the $f(R)$ and the nDGP model for a DESI-like survey with an LSST-like imaging survey. The~forecast used 3D voids as well as 2D tunnels; they found that the tunnels' tangential shear signal has a higher amplitude and is also more sensitive to the MG models (due to the project effect, the~interior area of tunnels has lower density compared to 3D voids), 2D tunneld can thus better distinguish MG from GR compared to the 3D~voids.

Although voids are promising probes in constraining gravity, there are a few caveats and challenges. First, the~void property can be dependent on the void-finding algorithm, and~the optimal void finder is model- and void-statistics-dependent~\citep{Cautun2018:VoidMG,Paillas2019:VoidMG}. Next, conclusions drawn from simulations can also be dependent on the void-identification: The authors of \cite{Barreira2015:VoidMG} compared voids identified using both dark matter field and dark matter halo. The~abundance of larger DM-identified voids is boosted due to the fifth force, while the effect is less pronounced for halo-identified voids. The authors of \cite{Voivodic2017:VoidAbundanceMG} studied a more realistic scenario by populating a simulated dark matter halo with galaxies and found the statistical precision is not enough to distinguish between MG models and GR. The authors of \cite{Wilson2022:void} showed that the void statistics are degenerate with HOD parameters~\citep{Wilson2022:void}, e.g.,~the void size function and velocity-related properties. Furthermore, galaxies are often used to identify voids observationally, where galaxies of different types populate halos differently. Since halo masses can have a different impact on clustering properties, the~resulting void
catalogues, which depend on the minimum halo mass cut, can have different abundances and void profiles~\citep{Cai2015:MGvoid,Nadathur2015:MGvoid}. As~a complementary approach, peaks of the void lensing map can also be used to identify voids, also resulting in a higher signal-to-noise ratio relative to galaxy-identified voids~\citep{Davies2018:VoidLensMG,Davies2019:VoidLensMG}. Finally, baryon effects~\citep{Osato2015:WLbaryon,Weiss2019:PeakBaryon} and massive neutrinos also need to be taken into account~\citep{Fong2019:WLbaryonNeutrino}.

\subsubsection{Marked Correlation~Function}

Due to the screening mechanism, MG can induce a non-standard density dependence in the GR scenario. 
Therefore, up- or down-weighting the density according to their environments can enhance the estimators' environmental sensitivity. A~marked correlation function (mCF) is a generalized 2PCF. It requires weighting each object by an environmental-dependent ``mark'' $m$~\citep{Sheth2005:mark}:
\begin{eqnarray}
\mathcal{M}(r) \equiv \frac{1}{n(r) \bar{m}^2} \sum_{i j} m_i m_j=\frac{1+W(r)}{1+\xi(r)},
\end{eqnarray}
where $n(r)$ is the galaxy number density, $\bar{m}$ is the mean of the entire galaxy sample marker, $W(r)$ is the weighted 2PCF, and~$\xi(r)$ is the standard 2PCF.
In the large-scale limit, the~marked correlation function $M(r)\rightarrow 1$.

The galaxy number density can be used to define a ``mark''~\citep{White2016:mark} and a density field is transformed as
\begin{eqnarray}
m(\delta)=\left(\frac{\rho_*+1}{\rho_*+\rho}\right)^p=\left(\frac{\rho_*+1}{\rho_*+(\delta+1)}\right)^p,
\end{eqnarray}
with $\rho_*$ and $p$ being two free parameters, $\rho$ is a smoothed estimation of the galaxy number density.
Other choices are log transformations~\citep{Neyrinck2009:LogTransf} and Gaussian transformations of the density field~\citep{Llinares:2017ykn}.

\begin{figure}[H]
    \includegraphics[width=0.5\textwidth]{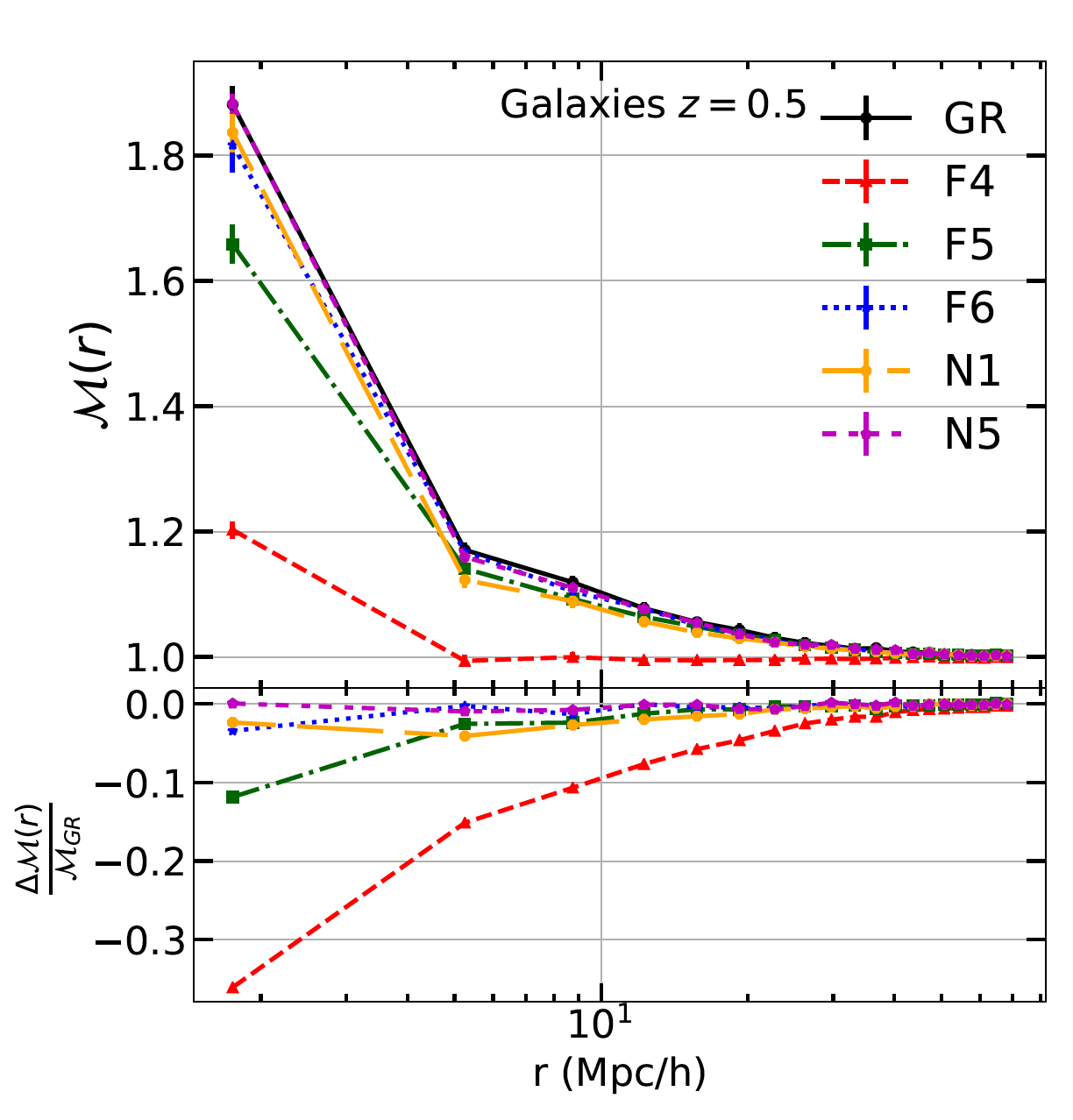}
    \caption{Gravitational potential marked correlation function. (Credit: Alam et al.~\cite{Alam2021desiMG}).}
    \label{fig:mCF_MG_GravPotential}
\end{figure}

Alternatively, gravitational potential $\Phi_{\rm N}$ can also be used to define a mark (see, e.g.,~\cite{Hernandez-Aguayo:2018yrp}):
\begin{eqnarray}
m(\Phi_{\rm N})=\frac{1}{\sqrt{2 \pi} \sigma_{\Phi}} \exp \left[-\frac{\left(\log _{10}\left(\left|\Phi_{\rm N}\right|\right)-\Phi_*\right)^2}{2 \sigma_{\Phi}^2}\right],
\end{eqnarray}
where $\sigma_{\Phi}$ and $\Phi_*$ are free parameters that characterise the width and mean of the Gaussian, respectively. Using such a mark, galaxies hosted by a certain halo mass are up-weighted (see Figure~\ref{fig:mCF_MG_GravPotential}).

The theoretical modelling of the marked correlation function in MG theories is under development (see, e.g., \cite{Aviles:2019fli}). For~instance, the authors of \cite{Aviles:2019fli} modelled mCF using a PT approach to up-weight low-density regions in the Universe. The~authors combined the convolution Lagrangian PT with the $f(R)$ gravity and nDGP models to build a theoretical framework of the mCF for those models. They tested their analytical model against $N$-body simulations (dark matter halos and HOD galaxies) and found good agreement between the theory and simulations for scales $>30\,h^{-1}{\rm Mpc}$.

Marked clustering statistics are a promising tool to test theories of gravity because the screening mechanisms are environmental-dependent. However, there are challenges due to the freedom of selecting the marks or weights. Furthermore, the~definitions of the environment are {manifold} in the literature. Hence, finding an appropriate combination of an environment and mark that increases the MG signal is still a topic that needs further~exploration.

\subsection{Modelling the~Observables}
\label{subsec:observable_models}
Dynamics of density fluctuation fields $\delta$ become non-linear when approaching smaller scales of order a few tenths of $\mpch$. For~most observables discussed in Section~\ref{sec:methods}, PT (see review\cite{Bernardeau2002spt}) is one of the most common approaches for modelling observables involving large-scale structure (see also developments \cite{Crocce2006:RPT,Crocce2008:RPTbao,Matsubara2008:LPT,Bernardeau2008regPT,Baumann2012:EFTofLSS,Carrasco2012:EFTofLSS}). By~perturbatively solving the Vlasov--Poisson system, which governs the dynamics of the density fluctuation field $\delta$, one can analytically achieve the correlator of $N$ density fluctuation fields.
In addition to the matter field, galaxies as biased tracers of the matter field are affected by the RSD. The~PT approach can also be extended to incorporate galaxy biasing (see review\cite{Desjacques2018:BiasReview}) and RSD~\citep{Scoccimarro2004rsdStreaming,Matsubara2008:LPT2,Taruya2010:TNS,Senatore2014:EFTrsd,Lewandowski2018:EFTrsd}. 


The perturbation models mentioned above are all GR-based; there are developments in the perturbation models with MG extensions.
The authors of \cite{Koyama2009mgSPT} presented a calculation for the matter power spectrum using the standard PT in real space. Given that RSD is a sensitive probe of cosmic structure growth, perturbative MG formalism is extended to redshift space~\citep{Bose2016SPTrsdMG} with improved series expansion convergence~\mbox{\cite{Taruya2014SPTrsdMG,Taruya2014regPTmg}.} The authors of \cite{Aviles2017LPTMG} presented a Lagrangian PT calculation for the power spectrum in Lagrangian coordinates, helping to improve the integral convergence. Their calculation can also be applied to a wide range of MG models classified as Horndeski theory. The authors of \cite{Valogiannis:2019nfz} combined convolution Lagrangian PT and a local Lagrangian bias schema to obtain accurate predictions of the redshift space correlation functions in the $f(R)$ and nDGP gravity~models.

The accuracy of PT-based methods decreases rapidly from linear/mildly non-linear scales to non-linear scales ($k > 0.2\, \mpch$ or $r < 20 \,\mpch$). A~different approach is constructed to predict the reaction of the halo model-derived~\citep{Cooray2002:HM} power spectrum in the presence of MG effects. Loosely speaking, the non-linear power spectrum can be obtained from the linear power spectrum via a non-linear mapping operator $\mathcal{K}$. MG effects can change the Poisson equation, halo mass function, and~halo concentration, which all go into the operator and modify the mapping relation such that $\mathcal{K}_{\rm GR} \rightarrow \mathcal{K}_{\rm MG}$. The~operator itself can be determined by comparing the power spectrum from simulation with or without MG effects~\citep{Cataneo2019MGpk}. Using this halo model reaction framework, the authors of \cite{Bose2022nonlinearMG} provided a model-independent prescription for the non-linear matter power spectrum by parametrizing perturbations using EFTofDE.


For higher-order statistics, the authors of \cite{Bose2018BkOneloop} calculated a one-loop matter bispectrum based on PT, while the authors of \cite{Aviles2023:3PCFmg} performed the calculation for the 3PCF for biased tracers. The authors of \cite{Bose2020BkMGsmallscale} assessed various theoretical modelling approaches, including PT, halo modelling, and~fitting approaches for the matter bispectrum in the non-linear regime. However, they found that all the tested models were unlikely to be accurate enough for next-generation lensing surveys, while a corrected formula based on the halo model outperformed the~others.

\section{Simulations for MG}
\label{sec:simulations}
Cosmological simulations are crucial tools in understanding structure formation at smaller scales \citep{Vogelsberger:2019ynw,Angulo:2021kes}. They are even more important in understanding the gravitational dynamics in the presence of the fifth force (see, e.g., \cite{Winther:MGcodes}). Among~many gravity models, two models with screening mechanisms $f(R)$ and nDGP (see Section~\ref{sec:overview_model}) are the most representative, both of which feature modified Poisson equations. Modern cosmological simulations rely on $N$-body codes that solve the modified Poisson equation. Under~the weak-field and quasi-static limit, it reads:
\begin{eqnarray}~\label{eqn:modified_poisson}
    \nabla^2 \Phi=\underbrace{4 \pi G \delta \rho_{\rm m}}_{\substack{\text { Standard } \\ \text { Poisson }}}+\underbrace{f(\phi, \nabla \phi, \nabla^2 \phi, \cdots)}_{\text {Extra field }},
\end{eqnarray}
in the absence of the second term, Equation~\eqref{eqn:modified_poisson} is the standard Poisson equation, whereas the modification is given by the second term due to the extra scalar field $\phi$. The~dynamics of the scalar field are governed by
\begin{eqnarray}\label{eqn:phi_dyn}
    \mathcal{L}[\phi ; \delta \rho_{\rm m}]=\mathcal{J}(\delta \rho_{\rm m}),
\end{eqnarray}
where $\mathcal{L}$ is a non-linear derivative operator and $\mathcal{J}$ is the source field for the scalar field (cf. Equations~(\ref{eq:fR_Poisson_qsa}) and (\ref{eq:phi_dgp})). To~solve Equations~(\ref{eqn:modified_poisson}) and (\ref{eqn:phi_dyn}) a typical method is to represent the extra scalar field on an adaptive grid and solve for their values via relaxation methods\endnote{In {most of the MG simulations with the $N$-body algorithms}, the~field is discretized and solved using a finite-difference method; the relaxation method is usually applied to iteratively solve a system of differential equations. The~goal of each iteration is to update the discretized field with a new value until the equality of the left- and right-hand sides of Equation~\eqref{eqn:phi_dyn} converges.}~\cite{Brandt1977:multigrid,Trottenberg2001:Multigrid,Li-book}.

Given the complexity of the MG models (i.e., the~non-linearity of the equations of motion that governs the evolution of the new scalar degree of freedom, cf. Section~\ref{sec:overview_model}), just a few numerical codes have been used and adapted to evolve dark matter structures through cosmic time in such cosmologies. For~instance, MG $N$-body simulation codes include ECOSMOG~\citep{Li2011:ECOSMOG,Li2013:VainshteinARM,Brax2012:SimMG,Brax2013:SimMG} and ISIS~\cite{Llinares2014:ISISSim}, which are extensions of the AMR code RAMSES \citep{Teyssier:2001cp}. MG-GADGET \citep{Puchwein2013:MGGadget} and MG-AREPO \citep{Arnold2019:MGArepo,Hernandez2021:MGArepo} are based on the TreePM GADGET-2 \citep{Springel:2005mi} and the moving-mesh AREPO \citep{Arepo} codes, respectively. MG-GLAM \citep{Hernandez-Aguayo:MGglam,Ruan:MGglam} is based on the fast full $N$-body code GLAM \citep{Klypin2018:GLAM}. 

Due to the non-linear equation of motion, an~MG simulation takes 2 to 10 times the computational time of its GR ($\Lambda$CDM) counterpart. For~this reason, the~$N$-body gravity solver can be accelerated upon improving the relaxation method (e.g., for ECOSMOG;~ \mbox{\cite{Barreira2015:SimMGVainshtein,Bose2017:SimMGfR})} or FFT-based methods (e.g.,~MG-GLAM; \cite{Hernandez-Aguayo:MGglam,Ruan:MGglam}). 

Acceleration can also be achieved with approximate methods such as modifications to the co-moving Lagrangian acceleration (COLA;~\cite{Tassev2013:COLA}) method to achieve a faster time evolution. These codes, as~suggested by their name, compute the dark matter particles' position in the Lagrangian coordinate and then map to the Eulerian coordinate according to the Lagrangian PT to the $n$th order (e.g.,~MG-COLA; \cite{Winther2017:MGCOLA,Valogiannis2017:MGCOLA}). 





\subsection{Simulations for~Galaxies}


The dark matter-only simulations can be post-processed to obtain galaxy catalogues. Typically halo finders are used to define halos from the dark matter simulations (e.g., Subfind and Rockstar;~\cite{Subfind,Rockstar}). Empirical methods, such as halo occupation distribution (HOD;~\cite{Berlind2002}) or sub-halo abundance matching (SHAM;~\cite{Conroy2006}), can be applied to build the halo--galaxy connection in MG scenarios \citep{Alam2021desiMG}. For~example, the extended lensing physics using analytic ray tracing (ELEPHANT; \cite{Cautun2018:VoidMG}) simulations were run with the ECOSMOG code \citep{Li2011:ECOSMOG,Li2013:VainshteinARM}. The~ELEPHANT simulations consist of five independent realizations of the GR model, three variants of the $f(R)$ gravity ($f_{R0} = -10^{-6}\,-10^{-5}$, and~$-10^{-4}$) and two nDGP ($H_0r_c = 5$ and $H_0r_c = 1$) models. They started at $z=49$ in a box with size $L=1024\,h^{-1}\rm{Mpc}$ and $1024^3$ particles. Galaxy catalogues were generated with the HOD method where the HOD parameters were tuned to match the number density and projected galaxy clustering of the BOSS-CMASS sample~\citep{Manera:2012sc}. The~HOD catalogues have been used to study the impact of MG in several galaxy statistics (e.g., \cite{Alam2021desiMG}). The authors of \cite{Devi:2019swk} populated the ELEPHANT simulations with galaxies using the SHAM approach to study the environmental dependence of the galaxy luminosity function in MG models. The~authors matched the observed SDSS $r$-band luminosity function \citep{Dragomir2018} with the ELEPHANT's $V_{\rm max}$ distribution of halos and~sub-halos.

More recently, the authors of \cite{Arnold2019:MGArepo} and \cite{Hernandez2021:MGArepo} implemented the $f(R)$ Hu-Sawicki and nDGP gravity models into the {\sc Arepo} code \citep{Arepo} to run the SHYBONE (simulating hydrodynamics beyond Einstein) simulations. The~authors employed the {\sc IllustrisTNG} model of galaxy formation \citep{Pillepich:2017jle,Weinberger2017:IllustrisTNG} to simulate the evolution of gases, stars, black holes and dark matter in MG models, i.e.,~the interplay between baryonic feedback processes and MG. Due to the high computational cost of these calculations, the~SHYBONE simulations were limited to small volumes despite their high resolution. The~simulations evolved $2\times 512^3$ dark matter and gas particles in volumes of $(25\,h^{-1}\rm{Mpc})^3$ and $(62\,h^{-1}\rm{Mpc})^3$ from $z=127$ to the present epoch. The~SHYBONE simulations have been used to explore the impact of MG on the high-redshift distribution of neutral hydrogen \citep{Leo:2019ada} and on the SZ effect \citep{Mitchell2021:SZ}. The authors of \cite{Mitchell:2021ter} extended the SHYBONE simulations to larger volumes ($L=302\,h^{-1}\rm{Mpc}$) by retuning the {\sc IllustrisTNG} parameters with the aim of studying the gas content in galaxy clusters in MG~cosmologies.

On the other hand, current and future galaxy surveys require simulations that cover large cosmological volumes with many realizations to estimate the errors of the different galaxy statistics accurately. More recently, the authors of \cite{Hernandez-Aguayo:MGglam} and~\cite{Ruan:MGglam} adapted the particle--mesh GLAM code \citep{Klypin2018:GLAM} to run fast full $N$-body simulations of different MG models, including the $f(R)$ and nDGP models. The~authors showed that MG-GLAM is able to run hundreds of large-scale simulations ($L=512\,h^{-1}\rm{Mpc}$) with $1024^3$ particles using reasonable computational resources. These simulations will assist ongoing and future galaxy surveys (e.g.,~DESI and Euclid, see \S\ref{sec:survey}) by providing large-scale galaxy mocks of MG models to carry out RSD and BAO analyses. For~example,the authors of \cite{Ruan:2021rqv} presented a study of the small-scale halo RSDs in MG cosmologies using MG-GLAM simulations as a previous step to modelling redshift-space galaxy~clustering.

\subsection{Simulations for Radio~Sources}

Simulating {LIM} (in particular 21 cm at high redshift) requires Gpc-scale simulation boxes to ensure the statistical modelling of ionized regions and absorption systems. At~the same time, it also requires high resolution to guarantee the interplay of complex physical processes, such as ionized regimes and feedback mechanisms. Due to these challenges, LIM simulations so far have been performed by either post-processing dark matter-only simulations or using a semi-numerical~approach.

Neutral hydrogen HI assignment schema can be added to the dark matter-only $N$-body simulations.
The authors of \cite{Carucci2017:LIM21cmSim} applied the HI distribution model~\citep{Bagla2010:LIM21cm,Carucci2015:LIM21cm,Carucci2017:LIM21cm} and assigned HI masses to each halo by assuming HI to be confined within the halo with a mass proportional to that of the halo mass. The~HI cloud location is converted from real to redshift space (see Section~\ref{subsubsec:vel_observable}). The~HI density is obtained from the HI cloud redshift space position via the {cloud-in-cell} algorithm. This hybrid $N$-body approach can be extended for different DE models~\citep{Alimi2012:DEUS}. In the same paper, the~authors also applied the hybrid approach to non-standard dark matter simulations of axions and late-forming dark matter models~\citep{Corasaniti2017:DM}. In~addition to $N$-body simulations, particle--mesh-based methods (e.g.,~FastPM; \cite{Feng2016:FastPM}) can also be post-processed to study LIM~\citep{Modi2019:HVsim} via different HI-halo (or HI-stellar mass) connections. The~HI-halo connection can be extended to other lines emitted by galaxies such as  
CO and CII~\citep{Moradinezhad2022:LIM}.

Alternative to post-processing dark matter-only simulations, a~semi-numerical approach was proposed to predict the high-redshift 21 cm signal (21 cm FAST \cite{Mesinger2011:21CMFAST}). The~semi-numerical code combines PT for evolving the density and velocity gradient field, excursion-set formalism~\citep{Bond1991:ExcursionSet} for identifying HII regions, and~analytic prescriptions to spin temperature fields. These quantities are used to obtain the brightness temperature for the 21 cm line. The authors of \cite{Heneka2018J:MG21cm} modified 21 cm FAST to incorporate structure growth in the general parametrizations of MG models. The~modification results in a change in the 21 cm brightness temperature via non-linear HI density contrast, velocity gradient field, and~ionization. 


For radio continuum sources, there are two main tracers: star-forming galaxies (SFG) and active galactic nuclei (AGN). The~widely used cosmological radio continuum simulations include the observation-based semi-empirical simulation~(SKAD; \cite{Wilman_2008}) and the $N$-body-based semi-analytic simulation (T-RECS; \cite{Bonaldi_2018}). To date, there are no continuum radio simulations with MG effects as there are a few fundamental challenges in understanding the sources. One difficulty is understanding the supermassive black holes to the level that we can distinguish between radio-active AGN, radio-quiet AGN, and~radio-SFG. To date, we can only tell them apart based on morphological structures, e.g.,~jets and lobes. These morphological properties are, however, not easily accessible even with inputs from other photometric or spectroscopic surveys~\citep{Magliocchetti2022:radio}. In~addition, understanding AGN populations and the physical mechanisms behind them require high-resolution simulations. These leave the radio continuum simulations more space to explore in the upcoming radio~surveys.


\subsection{Simulation-Based~Inference}
\label{subsec:sbi}
In Section~\ref{subsec:observable_models} we presented different PT-based methods to model two-point statistics. However, these models are limited to producing accurate predictions on larger scales. For~instance, the~latest applications of PT methods to data restricted their analysis to use scales $k < 0.2$ \citep{Philcox2022:BOSSPkBk,DAmico2022:BOSSbispec} for both the power spectrum and bispectrum analyses. Unlocking the information content of smaller scales in cosmological analysis, where measurements from galaxy surveys are most accurate, requires precise $N$-body simulations that capture the non-linear growth produced by gravitational evolution. This is especially relevant for MG models, where non-linearities are enhanced by screening~mechanisms.

Moreover, modelling summary statistics beyond two-point functions analytically may not be possible. Most of the summary statistics discussed in the previous sections that could potentially detect deviations from GR lack accurate analytical predictions and can only be computed through $N$-body simulations. Additionally, corrections that  account for observational systematics, such as redshift-dependent completeness, survey  mask and geometry, and~fibre collisions, have only been demonstrated to work for two-point statistics. Their effect on higher-order summaries is unknown. Using $N$-body simulations, these different systematics can be readily included in the forward model \citep{https://doi.org/10.48550/arxiv.2211.02068, https://doi.org/10.48550/arxiv.2211.00660}.

Solving the inverse problem, estimating the posterior over the cosmological and MG parameters given the observed summary statistic, would require the order of $\mathcal{O}(10^6)$ $N$-body simulations to perform Bayesian inference with Markov Chain Monte Carlo. This is currently computationally intractable due to how expensive $N$-body simulations are, particularly for MG theories. Therefore, most studies rely on modelling the dependence of a summary statistic on cosmology with surrogate models that are trained on a small set of $\mathcal{O}(100)$ $N$-body simulations. The~surrogate models are orders of magnitude faster than the original $N$-body simulations and can then be used to sample the posteriors of the parameters of interest. The~most commonly used surrogate models are Gaussian processes \citep{Zhai_2019} and neural networks \citep{Kobayashi_2020}.

For MG models, full $N$-body simulation suites with varying parameters only exist for $f(R)$ and nDGP gravity models~\citep{Arnold_2022}, where the authors demonstrate how a Gaussian process emulator can be used to accurately predict the matter power spectrum in these MG models. The~same simulations have also been used to obtain MG constraints from lensing observables~\citep{https://doi.org/10.48550/arxiv.2211.05779}. 

Simulation-based models can also directly model the likelihood, or~posterior, of~a given summary statistic by using flexible density estimators such as Gaussian mixture densities or normalizing flows. An~example of their application to cosmology can be found in \citep{https://doi.org/10.48550/arxiv.2211.00660}. These models will be necessary to obtain unbiased constraints on cosmology when the likelihood of a summary statistic is not Gaussian~distributed.

Although we have discussed promising summary statistics that could detect deviations from GR, there is no guarantee that all the summaries combined would exhaust the information content of clustering and lensing datasets. Promising alternative venues to maximally constrain gravity models through $N$-body simulations are: field-level inference and learning optimal summary statistics with machine~learning.

\section{Cosmological Surveys of Our~Universe}
\label{sec:survey}

In order to learn about DE and gravity on cosmological scales, 
we want to map the distribution of matter in our Universe, baryonic and dark,  
as best as we possibly can. 
Surveying the Universe can be performed in all wavelengths of the electromagnetic 
spectrum, from~radio to X-rays, using photons from the~sky.  


In this section, we review past, current, and~future relevant
{photon surveys} that map the distribution of matter in our Universe. 
These surveys are all essential to  our understanding of DE, potentially providing evidence that gravity on larger scales deviates from GR.

\subsection{Photometric~Surveys}
\label{subsec:photo_survey}

Photometric surveys are traditional method for systematically observing the cosmos. A~photometric survey comprises a telescope, equipped with a camera and a set of filters, used to observe large portions of the sky with numerous exposures, without~pointing at a specific object or location. {Photometric surveys provide valuable insights into the intrinsic physical properties of galaxies and their evolutionary processes. Additionally, these surveys contribute to constraining cosmological models via, e.g.,~weak gravitational lensing measurements. Photometric surveys are also commonly utilized for the purpose of identifying objects of interest that warrant further spectroscopic investigation (see, e.g., review \cite{Newman2022:PhotozReview}).}
    
There are plenty applications of photometric surveys. Particularly in cosmology, they are necessary for the discovery of type Ia supernovae, and~for the study of galaxy clustering and weak gravitational lensing. On~the one hand, type-Ia supernovae surveys require a high time cadence of observations, a~few days typically, and~precise flux measurements. On~the other hand, WL surveys require deep imaging to discover large number of galaxies and precise measuring of their shapes.
The study of large-scale structures with galaxies can be performed with the angular information provided by the images. However, radial information is limited due to the number of available photometric~bands. 
    
Main parameters defining a photometric survey include: area of sky observed, bands, magnitude limit, telescope main mirror size, field-of-view, number of galaxies, and~redshift range. Photometric redshift estimates have uncertainties of the order of a few percent on {$\sigma_z\!\sim\! 0.05$} typically.

Early photometric surveys include the automatic plate measuring machine (APM;~\cite{Maddox1988:APM}), the~Edinburgh--Durham Southern Galaxy Catalogue (EDSGC; \cite{Collins1995:EDSGC}), the~Classifying Objects by Medium-Band Observations in 17 Filters project (COMBO-17; \cite{Wolf2003:COMBO17}),
the Canada--Hawaii--France Telescope Legacy Survey (CFHTLS; \cite{Ilbert2006:CFHTLS}), the~Spitzer Wide-Area Infrared Extragalactic Legacy Survey (SWIRE; \cite{RowanRobinson2008:SWIRE}), and the Cosmic Evolution Survey (COSMOS;~\cite{Mobasher2007:COSMOS}).
{In addition, there are also the Wide-Field Infrared Survey Explorer~\citep[][]{Wright2010:WISE} and the Two Micron All Sky Survey (2MASS; \cite{Skrutskie2006:2MASS}) operating in the infrared band.}

Current and upcoming photometric surveys typically have redshift uncertainty {$\sigma_z\!\lesssim \!0.01$}. There are three ongoing WL surveys: the Kilo-Degree Survey (KiDS; \cite{deJong2013:KiDS}), the~Dark Energy Survey (DES; \cite{Abbott2016:des-overview}), and~the Hyper Suprime-Cam Survey (HSC; \cite{Aihara2018:HSC}) covering an area of 1500 deg$^2$ with $\sim\! 10^8$ galaxies. {The KiDS Data Release 1 to 3 covered an area of 450 deg$^2$ with 14.7 million galaxies~\citep{deJong2017:KiDSdr3} and the final Data Release 4 KiDS-1000 covered an area of 1350 deg$^2$ with $\sim\! 3\times 10^7$ galaxies \citep{Kuijken2019:KiDS1000}.
The DES Year 1 covered 1321 deg$^2$ with 26 million objects~\citep{DrlicaWagner2018:desY1} and the DES Year 3 covered 5000 deg$^2$ with $\sim\! 3\times 10^8$ objects~\citep{SevillaNoarbe2021:desY3}.}
    
In the future, stage 4 surveys will probe a larger area and more galaxies. The~{Vera C. Rubin Observatory} (Rubin for short, previously also known as LSST; \cite{LSSTCollaboration2009:Science}) will carry out an imaging survey of the Southern Galactic Cap. It will cover a sky area of 20,000 deg$^2$ and reach the $r$-band magnitude limit of $r\!\sim\! 27$. The Euclid satellite mission~\citep{Laureijs2012:Euclid} will conduct a photometric survey covering 15,000 deg$^2$ with $10^9$ galaxies for the redshift range $0<z<2$. The Nancy Grace Roman Space Telescope (Roman for short; formerly the Wide-Field Infrared Survey Telescope or WFIRST \cite{Spergel2013:WFIRST}) will provide an imaging survey of 2000 deg$^2$ with $5\times 10^8$ galaxies within the redshift range of $1<z<3$.

\subsection{Spectroscopic~Surveys} 
\label{subsec:spec_survey}
Spectroscopic surveys {allow us to} map the distribution of galaxies in 3D, where the radial information is recorded from precise redshift measurements. As~the name states, a~spectroscopic survey {is usually} equipped with a spectrometer, decomposing the light of objects into thin wavelength bands. A~photometric survey is commonly the first step in building a spectroscopic survey, requiring a prior list of pointings, such as fibre-fed surveys. Target lists for such spectroscopic surveys are defined based on the angular positions, fluxes, and~colours, which are passed to obtain the spectra.
Emission and/or absorption features are easily seen in the observed spectra; they are the main ingredient for redshift {measurements}, the~main goal of a spectroscopic~survey. 
    
The main parameters defining the power of a spectroscopic survey are the area of the observed sky, the~number density of objects and their redshift range, and~the spectral resolution. Spectroscopic surveys typically have redshift uncertainties of $\Delta z/z\sim 10^{-3}$.

A large variety of cosmological spectroscopic surveys have been and will be explored, including deep surveys on small areas, interesting for galaxy population studies, to~shallower surveys on large areas, ideal for statistical measurements of the density field of tracers. Early galaxy spectroscopic surveys include the 2dF Galaxy Survey (2dFGS; \cite{Colless1999:2dFGS}), 6dF Galaxy Survey (6dFGS; \cite{Jones2004:6dFGS}), WiggleZ Dark Energy Survey (WiggleZ for short; \cite{Drinkwater2010:wigglez}), and VIMOS Public Extragalactic Redshift Survey (VIPERS; \cite{Guzzo2014:VIPERS}). Cosmological programs with Sloan Digital Sky Surveys (SDSS; \cite{Eisenstein2011:SDSSIII,Blanton2017:SDSSIV}), the~Baryon Oscillation Spectroscopic Survey (BOSS;~\cite{Dawson2013:BOSS}) and the extended BOSS (eBOSS; \cite{Dawson2016:eBOSS}) have provided the largest 3D map to date. BOSS covered 10,000 deg$^2$ with $\sim\! 1.5\times 10^6$ objects within $0<z<0.7$, and eBOSS covered 7500 deg$^2$ with $\sim\!10^6$ objects for an extended redshift range $0.6<z<3.5$ with more tracer types.
    
The Dark Energy Spectroscopic Survey~\cite{DESI2016:Part1} is a currently ongoing 5-year survey. By~the end of the survey it will deliver $\sim\! 30\times 10^6$ spectra and cover 14,000 deg$^2$. In~the near future, both the Euclid and Roman surveys will deliver spectroscopic data in addition to imaging data. Euclid will provide $\sim\!5\times 10^7$ spectra for $0.7<z<1.8$, and Roman will provide $\sim\!2\times 10^7$ spectra for $1<z<3$. The~Subaru Prime Focus Spectrograph (PFS;~\cite{Takada2014:PFS}) will be a spectroscopic survey covering 2000 deg$^2$ with $\sim\! 10^7$ emission line OII galaxies for $0.8<z<2.4$. The~4-metre Multi-Object Spectroscopic Telescope (4MOST;~\cite{deJong2019:4MOST}) will deliver $\sim\! 2.5 \times 10^6$ spectra for over 25,000 deg$^2$ for $0.15<z<3.5$.

\begin{figure}[H]
    \includegraphics[width=0.7\textwidth]{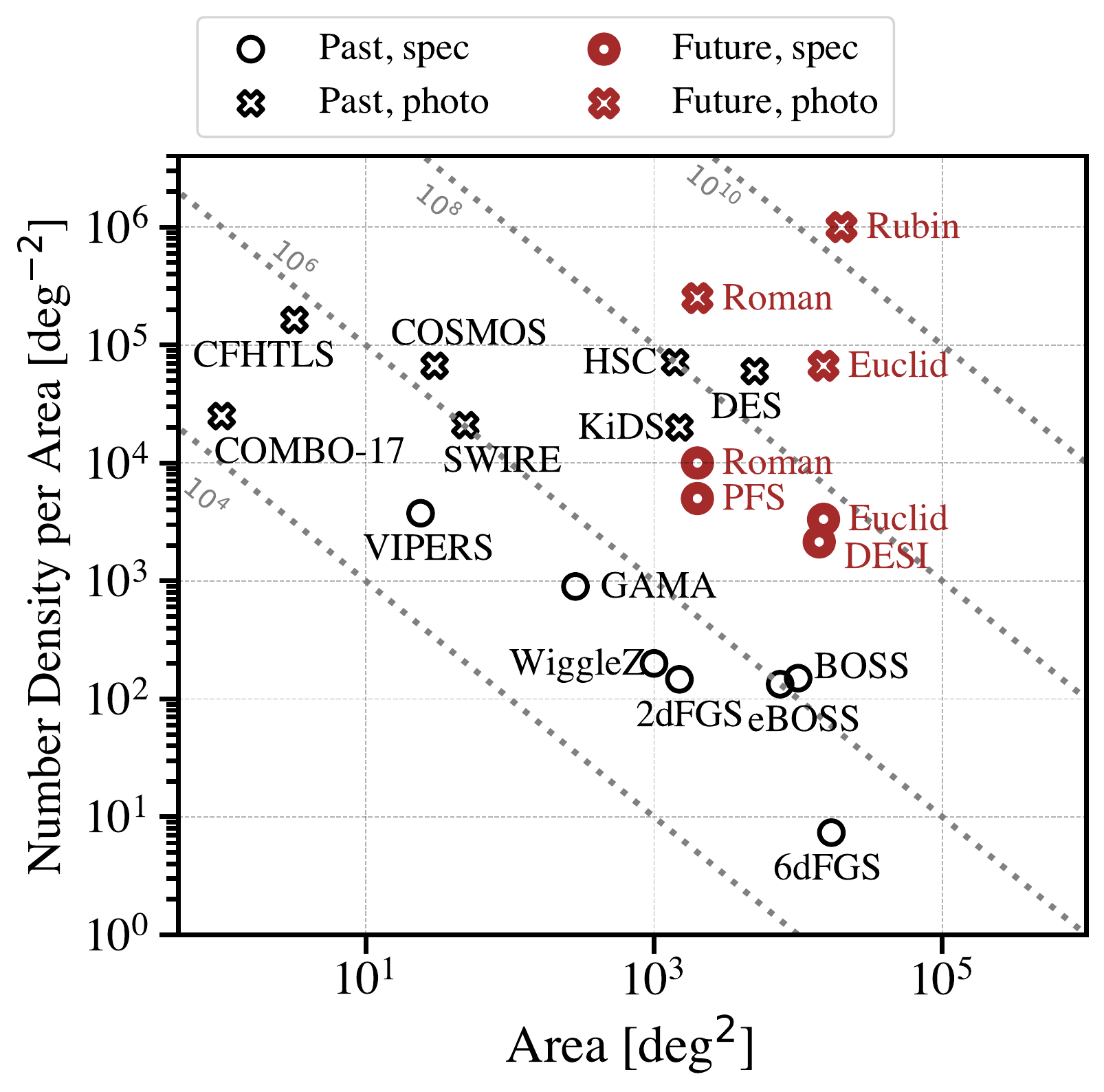}
    \caption{Landscape of past (black) and current/future galaxy surveys (red) for galaxy number density per area as a function of the survey area. Photometric surveys are crosses and spectroscopic surveys are open circles. The grey dotted lines corresponds to $10^4$, $10^6$, $10^8$, and~$10^{10}$ galaxies. Overall, the current/future surveys cover a larger area with a higher number density per~area.}
    \label{fig:gal_survey_landscape}
\end{figure}

Figure~\ref{fig:gal_survey_landscape} shows the landscape of the galaxy surveys with photometric redshifts (cross) and spectroscopic surveys (circle) for galaxy number density per area as a function of the survey area. Overall, the~future surveys (red) occupy the upper right corner, covering a larger survey area with a higher galaxy number density per area relative to past surveys (black).

\unskip

\subsection{Radio Surveys: Continuum and Intensity~Mapping}\label{subsec:radio_surveys}
Radio surveys are a promising tool to test the Universe at unexplored redshifts. Radio continuum at relatively lower wavelengths and HI intensity mapping at high wavelengths are the two main probes in the radio band. {A} radio continuum survey provides a high-angular resolution of radio galaxies but a low resolution in redshifts. The~HI intensity mapping offers a high-radial resolution but a low resolution in the angular direction. {The radio continuum and intensity mapping} are complementary to each other. In~addition, the~HI galaxy redshift surveys are similar to what can be obtained from an optical spectroscopic survey where the galaxy 3D coordinates are provided.

\begin{figure}[H]
    \includegraphics[width=\textwidth]{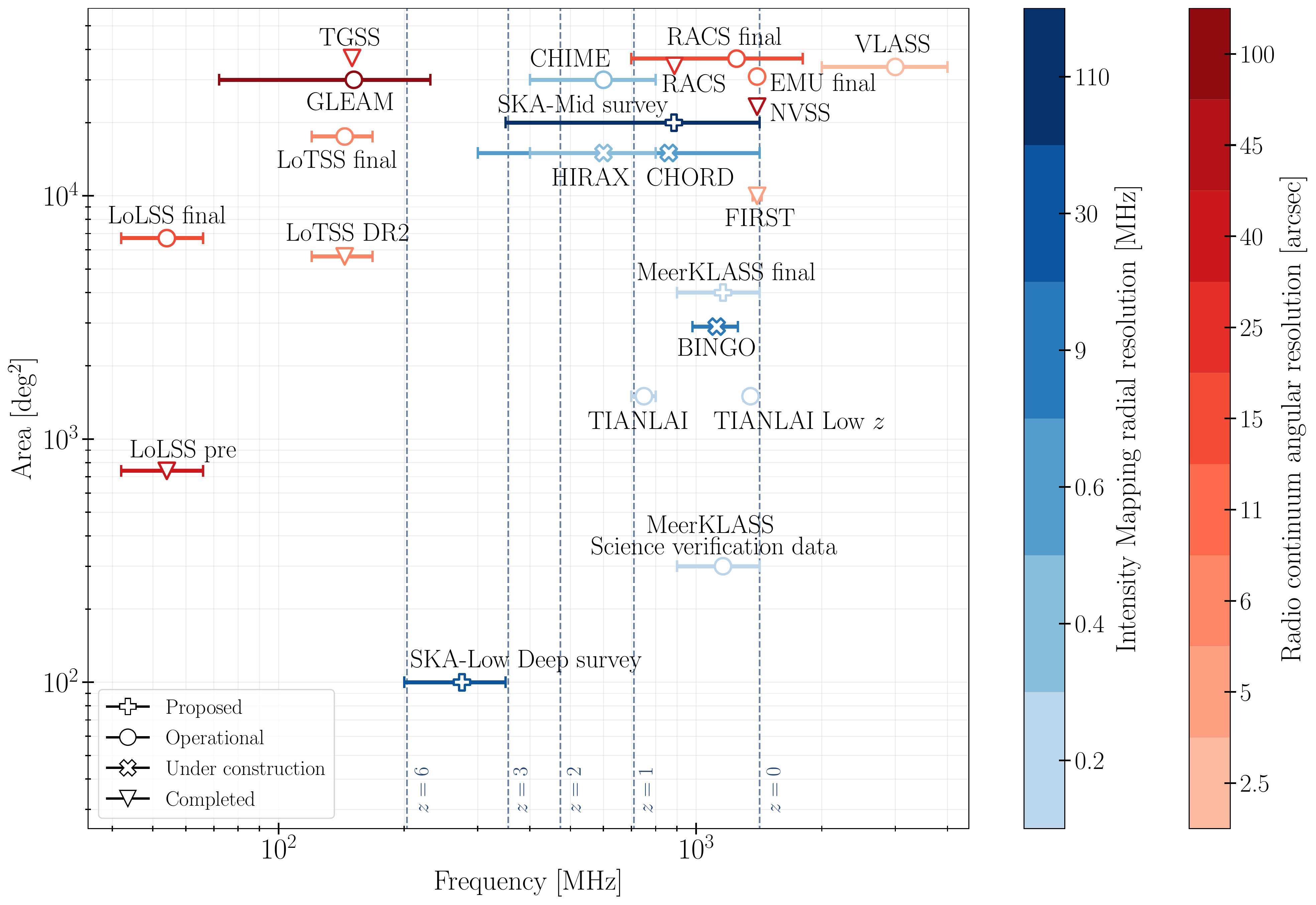}
    \caption{A summary of radio surveys. The~warm {red} colours denote the radio continuum maps while the cold blue colours represent the {21 cm intensity mapping}. The~error bars show the frequency ranges for each of the instruments or surveys. While the advantage of radio surveys is their high-angular resolutions, the~surveys that {measure the 21 cm signal} possess good high-radial resolution, i.e., the frequency and its corresponding redshifts (dashed blue vertical lines). We mark the two kinds of resolutions as different colour gradients.}
    \label{fig:radio_surveys}
\end{figure}

The main parameters defining a radio survey are the frequency range and resolution, which translates into the accessible volume in the redshift. We can categorize them mainly by the area covered, angular resolution, and~used technique, e.g.,~interferometry~\citep{Kovetz:2017agg} or single-dish approach~\citep{Battye:2012tg}. Several instruments are either currently recording data, under~construction, or~being planned. There will be a large amount of radio data available in the coming decades. We summarize the main radio continuum surveys and HI intensity mappings in Figure~\ref{fig:radio_surveys}. We gather the results and review the status of the most relevant radio surveys to forecast the constraining power of such observations on DE/MG~theories.

Large-scale structure studies with radio continuum sources require surveys with wide areas and high-resolutions in either angular or radial directions (reach sub-arcminute level).
In the past decade, surveys such as Faint Images of the Radio Sky at 20 cm (FIRST; \cite{Becker1995:FIRST}), NRAO VLA Sky Survey at 1.4 GHz (NVSS; \cite{Condon1998:NRAO}), and~TIFR GMRT Sky Survey (TGSS-ADR; \cite{Intema2017:TGSSADR}) have improved our understanding of extragalactic radio sources. In~addition, there are the LOFAR LBA Sky Survey (LoLSS; \cite{lolss}), Rapid ASKAP Continuum Survey (RACS; \cite{racs}), Karl G. Jansky Very Large Array Sky Survey (VLASS;~\cite{vlass}) and Galactic and Extragalactic All-Sky MWA
survey (GLEAM; \cite{gleam}). Currently, the~Low-Frequency Array (LOFAR; \cite{vanHaarlem2013:LOFAR}) is the only high-angular resolution (6 arcseconds) and high-sensitivity radio telescope with ultra-low frequency. The~LOFAR two-metre Sky Survey has published two data releases (LoTSS DR1, DR2; \cite{Shimwell2022:LoTSS2}), and~there is the additional LOFAR Low-Band Antenna (LBA) Sky Survey (LoLSS; \cite{deGasperin2021:LOFARLBA1}), which aims to cover the entire northern sky and providing ultra-low-frequency information for $\sim\!10^5$ radio sources. In~the future, there {are plans for} an evolutionary map of the Universe (EMU; \cite{Norris2011:EMU}) that {will} almost cover the whole southern sky, the~precursor of which {is} the EMU pilot~\citep{norris2021evolutionary} which released data 2022, as~well as the Westerbork Observations of the Deep Apertif Northern Sky Survey (WODAN; \cite{Rottgering2011:WODAN}), which covers the northern sky. Together they will cover the whole sky, thus aiding large-scale structure studies. The~Band 1 and 2 of the SKA1-Mid~\citep{Bacon:2018} will provide a continuum WL survey and a wide continuum galaxy survey in the redshift range $z =$ 0.35--3. 

The cosmological principle, which states that the Universe is isotropic and homogeneous, has been challenged by the tension in the cosmic dipole. The~dipole measurements from, e.g.,~the radio continuum surveys NGSS and TGSS, are 2--5 times larger than expected~\cite{bengaly2018probing}. The~results will also potentially favour some of the MG models that lead to inhomogeneity at different scales, with~the near-full-sky area of the radio continuum survey in the future, and~the homogeneity of the Universe will be tested with much higher confidence. Moreover, with~the spectroscopic redshift of the WEAVE-LOFAR survey~\cite{Smith2016:WeaveLofar}, we will have 3D radio galaxy catalogues for the LOFAR survey, both with high-angular and -radial resolution, allowing us to perform higher-order statistics and more local tests in the future with broader cosmic volumes than the optical~surveys.

At present, a~plethora of experiments targeting LIM are currently running, under~construction, or~being proposed. For~21cm IM, the~main instruments are CHIME~\citep{Bandura:2014gwa,CHIMEdetection}, HIRAX~\citep{Newburgh:2016mwi}, LOFAR~\citep{vanHaarlem2013:LOFAR}, GBT~\citep{Masui2013,Wolz2022}, FAST~\citep{Hu:2019okh}, BINGO~\citep{Costa:2021jsk,Wuensche:2021dcx}, CHORD~\citep{chord:2019}, TIANLAI~\citep{Wu:2020jwm,Perdereau:2022ksl}, SKAO~\citep{Bacon:2018} and its precursor MeerKAT~\citep{Santos:2017,Wang:2021,Cunnington:2022uzo}. For~a review of the state-of-the-art ongoing and proposed IM surveys, we refer the reader to~\citep{Kovetz:2017agg}.

{Currently, studies are mostly focused on forecasting the constraining power of future observations since there have yet been no LIM detections.}  For 21cm {intensity mapping}, a variety of scenarios were taken into account.
Some of them focused on the EoR epoch (e.g.,~\cite{Wyithe:2007rq,Chang:2007xk,Heneka2018J:MG21cm}). For~the late-time Universe, 21cm IM is expected to improve the constraints on the background evolution~\cite{Dinda:2018uwm,Wu:2021vfz,Wu:2022jkf}, but~also the perturbations~\cite{masui2010projected}. Several models have been studied, from~phenomenological parametrizations to specific theories. Some studies have tried to forecast the constraints on the $\mu$, $\Sigma$ and $\eta$ MG functions, e.g., \cite{Zhao:2015}, while other focused on EFTofDE~\citep{Berti:2021ccw}. Specific studied models are $f(R)$~\citep{Hall2013:LIM21,Brax2013:LIM21}, generalised scalar--tensor theories~\citep{Bull:2020cpe}, early DE~\citep{CosmicVisions21cm:2018rfq,Karkare:2018sar}, interacting DE~\citep{Xu:2017rfo,Zhang:2021yof}, among~others~\citep{Pourtsidou:2015, Carucci2017:LIM21cmSim}. Recently, several studies have focused on the cross-correlation between IM and other LSS probes, such as WL, GWs, and~galaxy \mbox{clustering \cite{Dash:2020,Scelfo:2022lsx,Casas:2022vik,Abidi:2022zyd}.} The~constraining power on DE of other lines is still mostly uncharted~\cite{Scott:2022fev}. 

The main take-home message of the intense work carried out by the community is that LIM observations are expected to improve significantly the constraining power on beyond-GR models, due to their ability to sample wide ranges of redshift. To~illustrate this, we collect the available forecasts on $\mu$ and $\Sigma$ in {Figure}
~\ref{fig:mu0Sigma0_compiled}.

\subsection{Cosmic Microwave Background~Surveys}
\label{subsec:cmb_survey}
CMB experiments measure the temperature fluctuation and polarization of photons emitted from the last scattering surface. The~anisotropies in the CMB temperature can be categorized into effects that occur before or at the surface of the last scattering surface (primary anisotropies) and integrated effects between the last scattering surface and observers (secondary anisotropies). The~main parameters defining a CMB survey are sky area, angular resolution (beam), frequency channels, sensitivity {to} temperature fluctuation, and~white noise level at different frequency channels. {CMB experiments have different systematics than the LSS surveys due to their distinct observational methods and the nature of the signals they capture. CMB experiments provide insights into the early Universe and complement the LSS probes.}

Main CMB experiments in the past include, {the Cosmic Background Explorer (COBE; \cite{Boggess1992:COBE}) as a full-sky survey with an angular resolution of $7^\circ$. Its two full-sky successors are the Wilkinson Microwave Anisotropy Probe (WMAP; \cite{Bennett2013:WMAP}) with a resolution of $0.3^\circ$ and the Planck satellite~\cite{Planck2016:I} with a resolution of $10\, {\rm arcmin}$}. The~Atacama Cosmology Telescope (ACT; \cite{DeBernardis2016:ACT,Louis2017:ACT}) covered an area of 1000 deg$^2$ with a resolution of $\sim\!1\, {\rm arcmin}$, the~South Pole Telescope (SPT; \cite{Benson2014:SPT,Henning2018:SPT}) covered 500 deg$^2$ with a resolution of $\sim\!1\, {\rm arcmin}$, BICEP/Keck~\mbox{\cite{Keating2003:BICEP, Ogburn2012:BICEP2, Grayson2016:BICEP3},} {and the Cosmology Large-Angular Scale Surveyor (CLASS; \cite{Essinger-Hileman2014:CLASS}) recorded a 75\% sky coverage and a resolution of $<\!1.5^{\circ}$}. 
In the future, the~Stage 4 CMB observatory (CMB-S4; \cite{Abazajian2016:CMBS4}) will cover 8000 deg$^2$ and achieve a resolution of $<3\, {\rm arcmin}$, while the Simons Observatory \cite{Ade2019:Simons} will cover 15,000 deg$^2$ with a resolution of $\sim\!1.5\, {\rm arcmin}$. 

\subsection{{Insights into MG from Current~Surveys}}
\label{subsec:constrain_current_survey}

\begin{figure}[H]
    \includegraphics[width=0.7\textwidth]{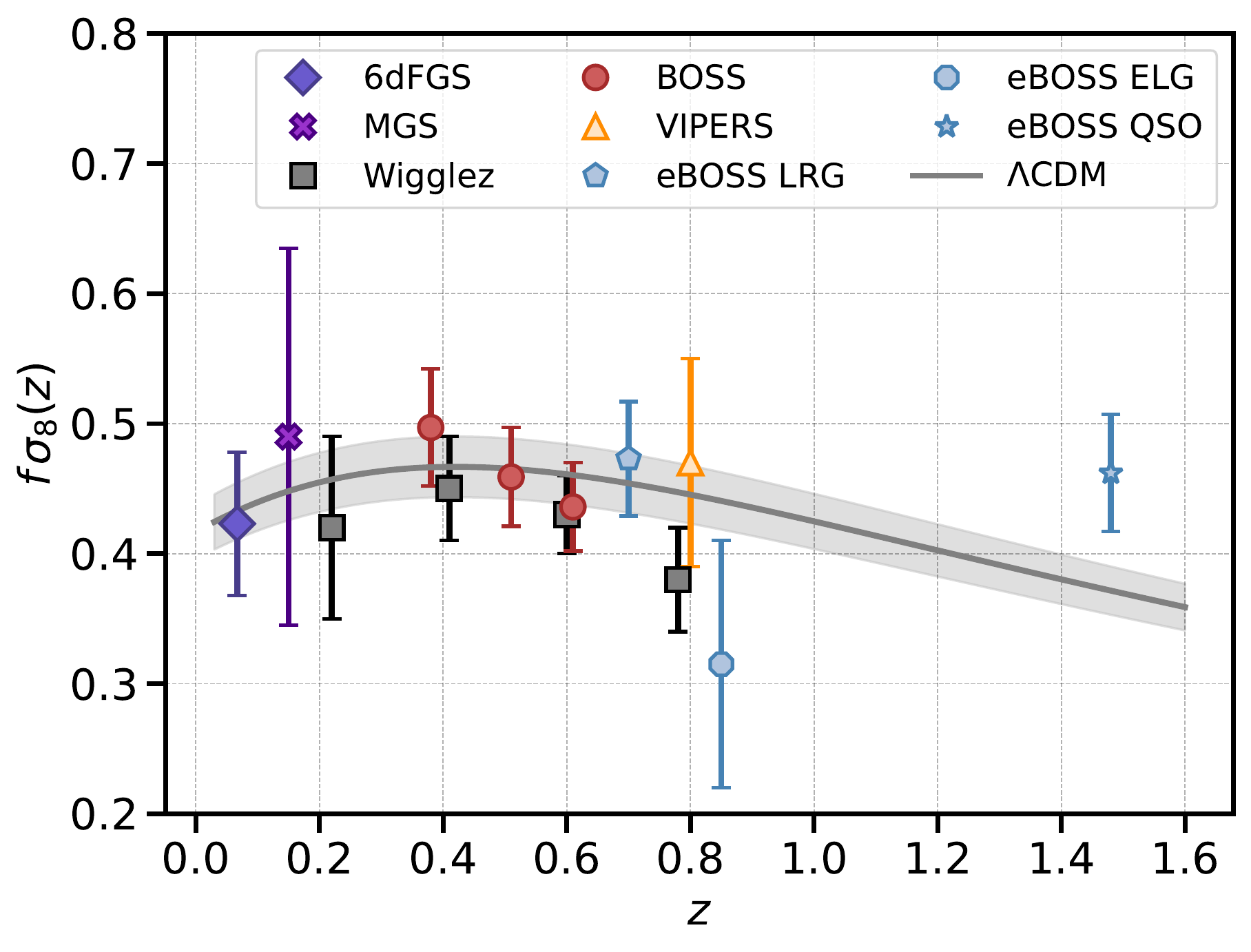}
    \caption{A compiled diagram for $f\sigma_8(z)$ as a function of redshift from different galaxy spectroscopic surveys. The~solid grey line denotes the $\Lambda$CDM with best fitting parameters from Planck15~\citep{Planck2016:XIIIcosmoparam} with a 5\% error bar. The~data points are from the following galaxy spectroscopic surveys utilizing the large-scale RSD information: 6dFGS~\citep{Beutler2012:6dFGs}, WiggleZ~\citep{Blake2011:WiggleZ}, MGS sample~\citep{Howlett2015:MGS}, VIPERS~\citep{delaTorre2013:VIPERS}, BOSS LRG sample~\citep{BOSSCollaboration2017}, eBOSS LRG~\citep{Bautista2021:eBOSSLRG,GilMarin2020:eBOSSLRG}, eBOSS ELG~\citep{Tamone2020:eBOSSELG,deMattia2021:eBOSSELG}, and eBOSS QSO samples~\citep{Hou2021:eBOSSQSO,Neveux2020:eBOSSQSO}.}
    \label{fig:fs8_compiled}
\end{figure}

Since MG can affect the growth of cosmic structure, the~combination of the $f(z)\sigma_8(z)$ (Section~\ref{para:rsd}), inferred from the anisotropic clustering of the galaxy auto-correlation, can be used test for potential deviation from GR (e.g., see the parametrization of $f\sigma_8$ employed in \cite{BOSSCollaboration2017}).
Figure~\ref{fig:fs8_compiled} shows a compiled measurements from the past galaxy spectroscopic surveys, including 6dFGS~\citep{Beutler2012:6dFGs}, WiggleZ~\citep{Blake2011:WiggleZ}, MGS sample~\citep{Howlett2015:MGS}, VIPERS~\citep{delaTorre2013:VIPERS}, BOSS LRG sample~\citep{BOSSCollaboration2017},~eBOSS LRG~\citep{Bautista2021:eBOSSLRG,GilMarin2020:eBOSSLRG}, eBOSS ELG~\citep{Tamone2020:eBOSSELG,deMattia2021:eBOSSELG}, and eBOSS QSO samples~\citep{Hou2021:eBOSSQSO,Neveux2020:eBOSSQSO}. The~$\Lambda$CDM cosmology using the best fitting value from the Planck data~\citep{Planck2016:XIIIcosmoparam} with a $5\%$ error bar is shown as the grey shaded area. The~precision on the current growth rate constraints in terms of $f\sigma_8(z)$ are up to 8\%. Future surveys with increased volume and the number of galaxies will largely improve upon the statistical~precision.

\begin{figure}[H]
    \includegraphics[width=0.6\textwidth]{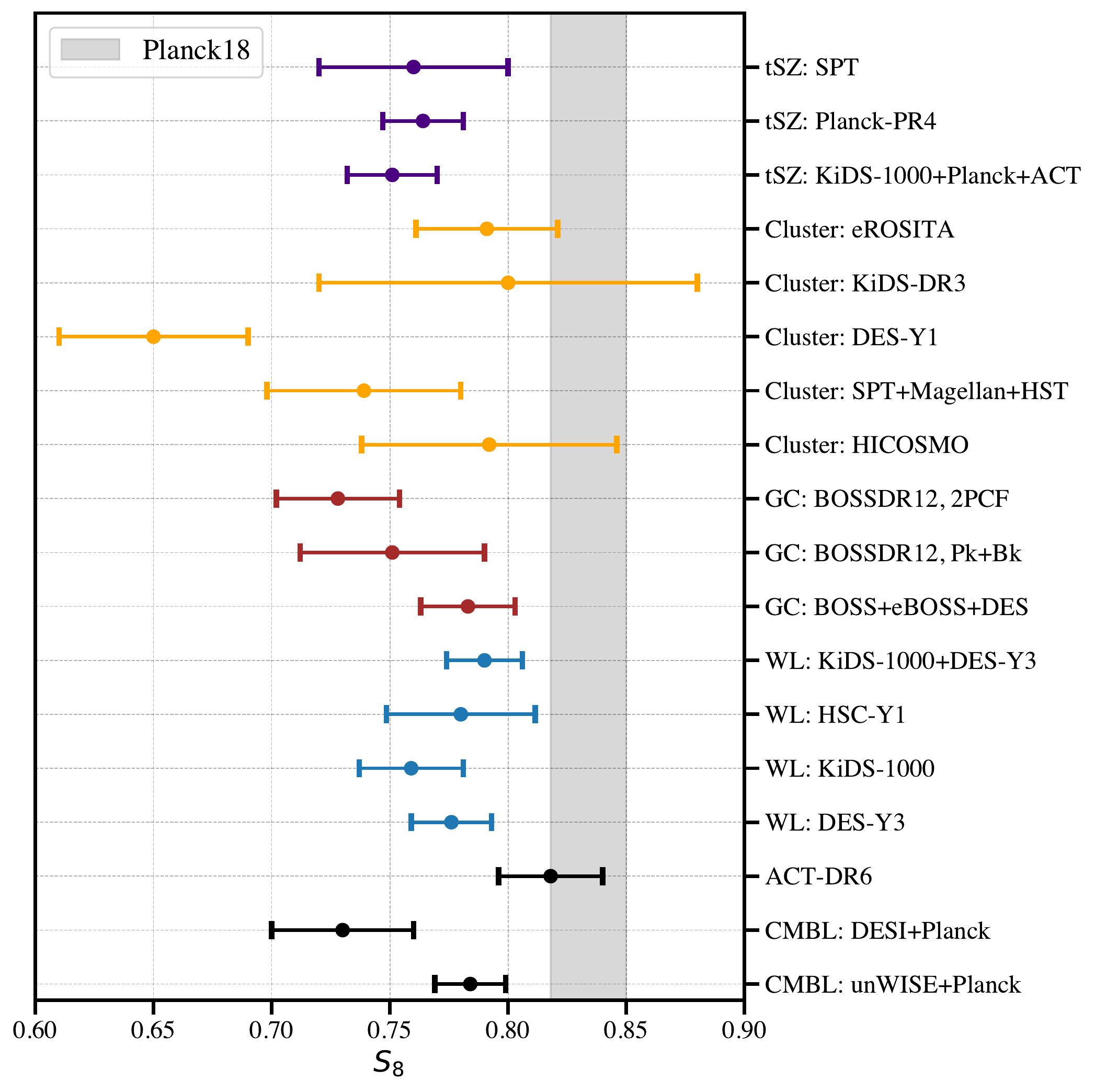}
    \caption{A compiled diagram for $S_8=\sigma_8\sqrt{\Omega_{\rm m}/0.3}$ for CMB lensing (black)~\citep{Krolewski2021:unWISExPlanck,White2022:DESILRGxPlanck,Qu2023:ACTDR6}, galaxy WL data (blue)~\citep{Amon2022-DES-Y3-CS,Asgari2021-KiDS1000-CS,Hikage2019:HSC,DESKiDS2023:JointWL}, galaxy clustering (red)~\citep{Troster2020:BOSS,Semenaite2022:BOSS,Philcox2022:BOSSPkBk}, galaxy clusters  (yellow)~\citep{Schellenberger2017:XrayHICOSMO,Bocquet2019:SPTcluster,Abbott2020:DESY1Cluster,Chiu2022:eROSITA,Lesci2022:KiDScluster}, and~tSZ (purple)~\citep{Douspis2022:SPTtSZ,Tanimura2022:PlanckPR4tSZ,Troster2022:KiDStszACT}. The~grey band is from Planck using TT+TE+EE+lowE~\citep{Planck2020:VI}. An~intriguing $2\sigma$ level tension {in several low-redshift experiments exists when compared to the primary CMB data}.}
    \label{fig:S8_compiled}
\end{figure}

While the constraints on the product of $f(z)$ and $\sigma_{8}(z)$ are consistent with the standard GR scenario, there {is} a potential tension in $\sigma_8(z)$~\citep{Abbott2019:desY1,Riess2019:LMCH0}.
Figure~\ref{fig:S8_compiled} shows a compiled measurement of $S_8$ (see Equation~(\ref{eqn:S8})). We select a few representative astrophysical probes in later stages, including WL, CMB lensing, galaxy clustering, galaxy clusters, and~tSZ. They are compared to the CMB from Planck. The~grey band represents Planck18 constraints. The~CMB lensing data are in black~\citep{Krolewski2021:unWISExPlanck,White2022:DESILRGxPlanck,Qu2023:ACTDR6}, the~galaxy WL data are in blue~\citep{Abbott2022:desY3,Heymans2021:KiDS-1000,Hikage2019:HSC}, while the galaxy clustering \citep{Troster2020:BOSS,Semenaite2022:BOSS,Philcox2022:BOSSPkBk} in red, galaxy \mbox{clusters~\citep{Schellenberger2017:XrayHICOSMO,Bocquet2019:SPTcluster,Abbott2020:DESY1Cluster,Chiu2022:eROSITA,Lesci2022:KiDScluster} in yellow,} and~tSZ~\citep{Douspis2022:SPTtSZ,Tanimura2022:PlanckPR4tSZ,Troster2022:KiDStszACT} are in purple. 
{Extensive research has been conducted on systematics related to various probes, aiming to understand the possible systematic-induced tension between the early-time and multiple late-time measurements (see, e.g., review~\cite{Abdalla2022:review}). For~example, the~CMB estimates are reliant on models via parameters related to amplitude, such as the sum of neutrino mass and the optical depth; they can impact the derived value of $S_8$~\citep{DiValentino2018:tensionS8cmb}. Additionally, there are also anomalies in the Planck lensing amplitude $A_{\rm lens}$~\citep{Planck2016:XLVIsysHFI} and (see also~\mbox{\cite{DiValentino2018:tensionS8cmb,Henning2018:SPT,ACT2020:DR4,Planck2016:XVCMBlens}).} However, these considerations do not fully explain the tension observed. In~the case of WL, notable attention is given to factors such as intrinsic alignment, non-linear matter power spectrum modelling, photo-z estimation, baryonic effects, and~so on, while individually these factors are unlikely to be the sole cause of the $S_8$ tension. The authors of \cite{DESKiDS2023:JointWL} demonstrated that the cumulative impact of these various factors could alleviate the tension. Moreover, further investigation is needed to explore the mass calibration for galaxy cluster counts~\citep{Pratt2019:ClusterMass}.} In addition to the systematics, parameter interpretation (e.g.,~using $\mpch$ unit~\citep{Sanchez2020:hUnit}) cannot be fully excluded. This mismatch in $S_8$ could also be correlated with the tension in the Hubble parameter $H_0$ (see review \cite{Knox2020:H0hunter} and model comparison in \cite{Schoneberg2022:H0Olympics}). 

{In the presence of an intriguing $\sim!2\sigma$ tension between CMB measurements and several late-time measurements could potentially indicate deviations from the standard $\Lambda$CDM model, provided that systematic errors can be entirely ruled out. Tentative candidates are: axion monodromy inflation \citep{Silverstein2008:monodromy,McAllister2010:monodromy,Meerburg2014:monodromy}, sterile neutrinos \citep{Battye2014:CMBLensNeutrino,Bohringer2016:ClusterNeutrino}, alternative DE models~\citep{DiValentino2020:DarkTension,DiValentino2020:InteractDEH0,Camera2019:DEmodel,Davari2020:DEmodel}, and~MG models~\citep{Planck2016:XIV_DEMG,Valentino2016:planckMG,SolaPeracaula2019:BD,Sola2020:BD}. However, if~this discrepancy is attributed to exotic physics, the~proposed model must reconcile the effects in both early and late stages and successfully pass scrutiny from cosmological probes.}

\begin{figure}[H]
    \includegraphics[width=0.9\textwidth]{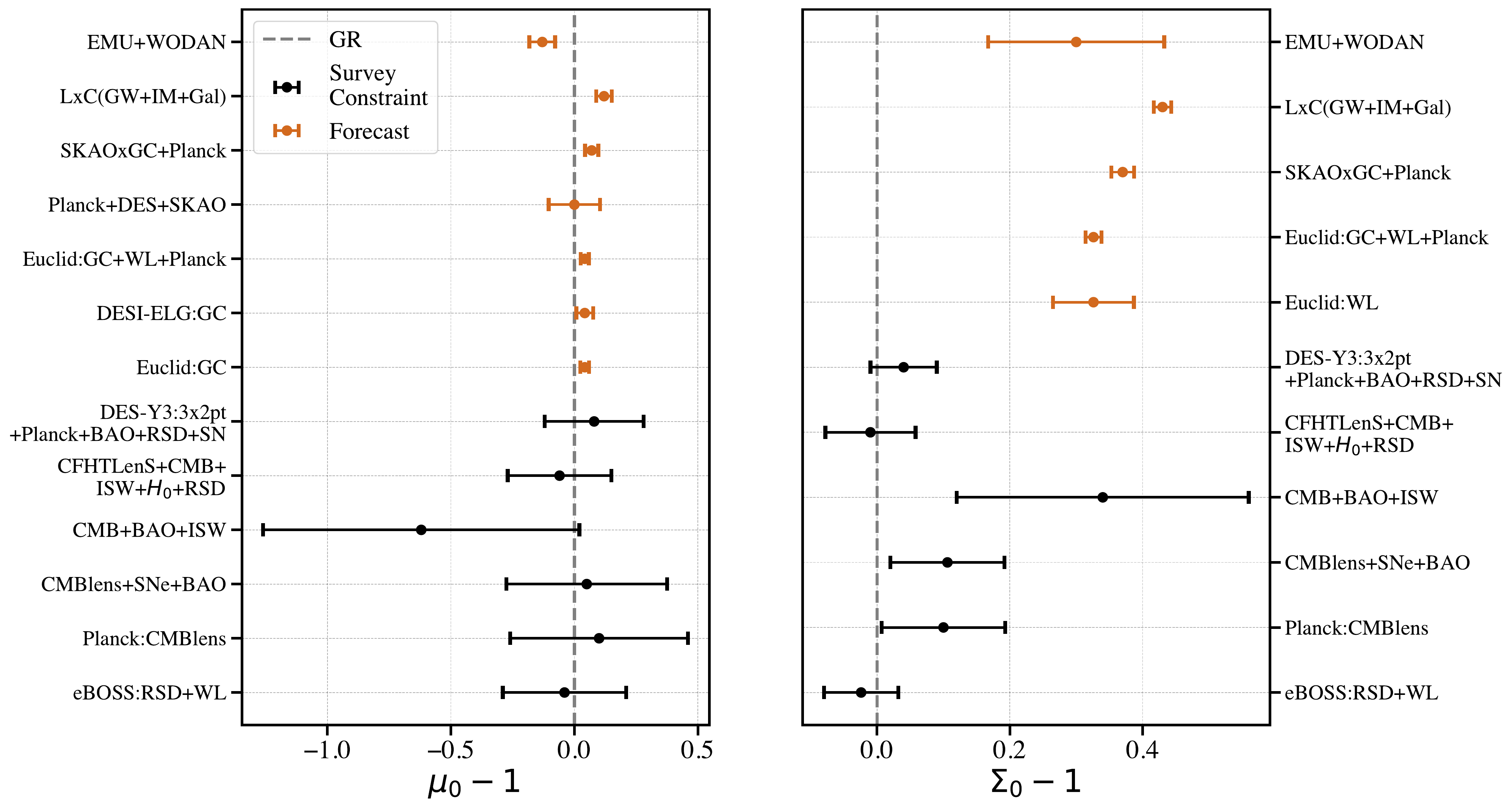}
    \caption{Constraints (black) and forecasts (orange) for $\mu_0=\mu|_{z=0}$ and $\Sigma_0=\Sigma|_{z=0}$.  
    From bottom to top we have RSD+WL~\citep[][]{eBOSScollaboration2021:eboss}, CMB lensing from Planck~\citep{Planck2020:VI}, or~Planck CMB lensing+SupernovaeI+BAO~\citep{Planck2020:VI}, CMB+ISW+RSD together with a $H_0$ prior with or without WL data from CFHTLenS~\citep{CFHTLenS2013:MGRSD}, and DES-Y3+Planck+BAO+RSD+SN~\citep{DES2022:Y3WL}. The~Euclid and DESI-ELG forecast for galaxy clustering and WL only~\citep{Casas:2017eob}, SKAO radio forecast with Planck and DES data~\citep{Zhao2015:SKAradioForecast}, SKAO radio forecast with Planck and galaxy-clustering data~\citep{Casas:2022vik},
    joint forecast using WL and clustering of GW+IM+galaxies~\citep{Scelfo:2022lsx}, and~the radio continuum surveys WODAN+EMU~\citep{Raccanelli2012:Radio}. 
    }
    \label{fig:mu0Sigma0_compiled}
\end{figure}

Figure~\ref{fig:mu0Sigma0_compiled} shows the constraints from the current survey (black) and the forecast (brown) on the $\mu_0=\mu|_{z=0}$ and $\Sigma_0=\Sigma_0|_{z=0}$ functions (see Equations~(\ref{eqn:param_mu}) and (\ref{eqn:param_Sigma}). In~the case of GR, we expect $\mu_0-1 = 0$ and $\Sigma_0-1=0$. The~current constraints~\citep{eBOSScollaboration2021:eboss,Abbott2019:desY1,Planck2020:VI,CFHTLenS2013:MGRSD} are all consistent with GR. Future surveys on galaxy clustering, WL, LIM, CMB, and~GW will greatly improve upon the constraints~\citep{Zhao2015:SKAradioForecast,Casas:2017eob,Casas:2022vik,Scelfo:2022lsx}.

\begin{figure}[H]
    \includegraphics[width=0.55\textwidth]{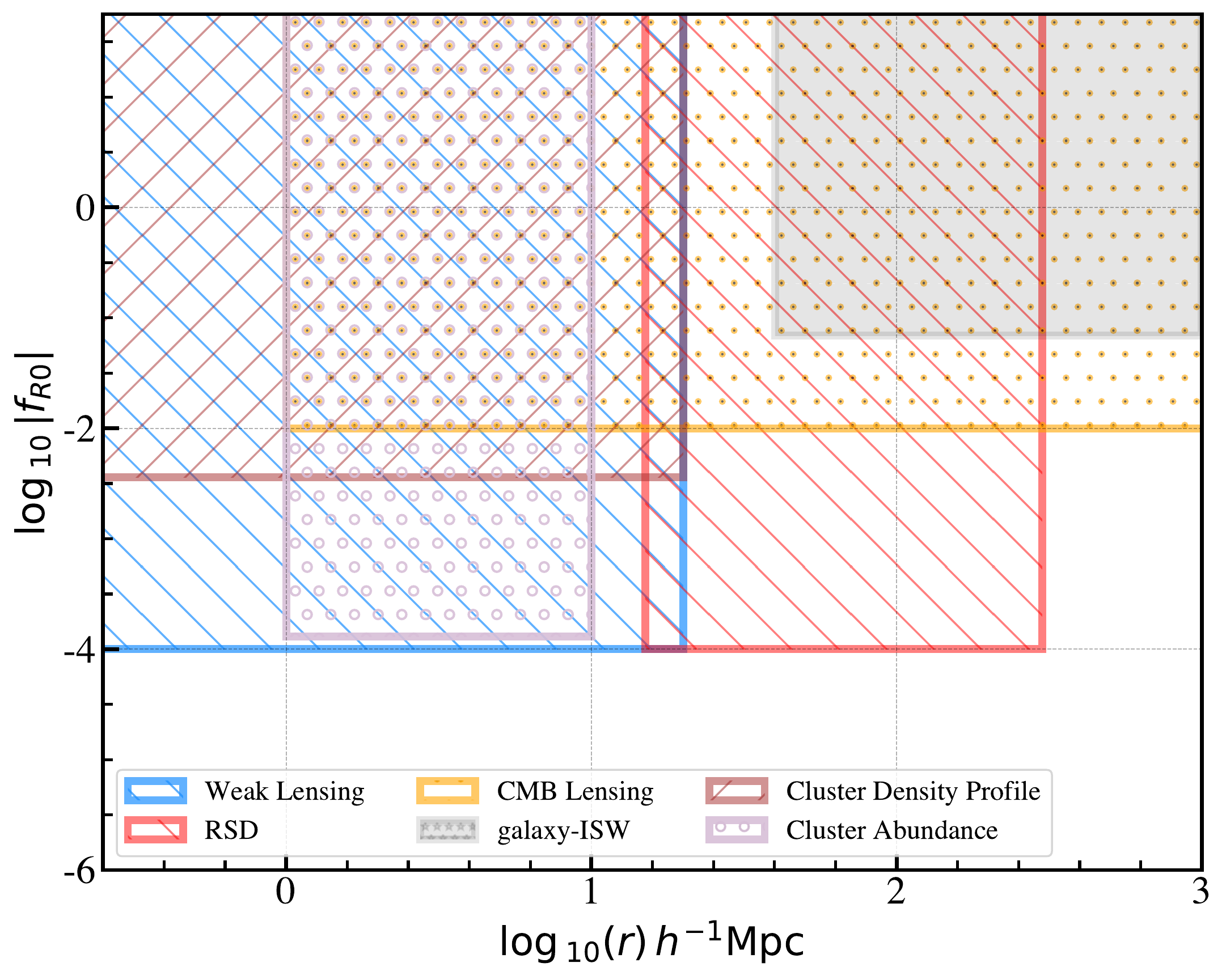}
    \caption{{Constraints} 
 on $f(R)$ from the {past and current} surveys. CMB lensing (SPT, ACT, Planck;\mbox{\cite{Marchini2013:fRCMBlen,Marchini2013:fRCMBlensPlk,Hu2013:fRCMBlens}),} galaxy power spectrum RSD (SDSS; \cite{Yamamoto2010:SDSSLRGfRrsd}), galaxy$-$ISW cross$-$correlation (WMAP, SHOES, SCP, \linebreak  SDSS;\cite{Song2007:ISWfR,Giannantonio2010:ISWfR,Lombriser2012:ISWfR}), cluster abundance (Chandra, MaxBCG;\cite{Schmidt2009:ClusterMG,Ferraro:2010gh,Lombriser2012:ISWfR}), cluster density profile (MaxBCG;\cite{Lombriser2012:ClusterMG}), and~{WL (CFHTLenS; \cite{HarnoisDeraps2015:CFHTLS}).\protect\endnotemark}}
    \label{fig:fR_constraints}
\end{figure}

{Finally, we summarize the current constraints on the $f(R)$ gravity parameter as a function of scale for different probes in Figure~\ref{fig:fR_constraints}. While the current surveys can only establish an lower limit for the present-day value of $f_{R0}$ at approximately $10^{-4}$, future galaxy surveys (Sections~\ref{subsec:photo_survey}--\ref{subsec:radio_surveys}) and CMB experiments (Section~\ref{subsec:cmb_survey}) hold promising potential for enhancing the constraints. Moreover, probes such as 21 cm \citep{Wang2021:21cmMG,Heneka2018J:MG21cm} and underdense regions and 2D underdensities (LSST, Euclid;~\cite{Cautun2018:VoidMG}) can be harnessed to offer further promising avenues.}

 \endnotetext{The authors of \cite{Troster2022:KiDStszACT} found the $f_{R0}$ is unconstrained in the cosmic shear analysis after marginalizing over nuisance or cosmological parameters.}
 
\section{Conclusions and~Discussion}
\label{sec:conclusion}

{In this review, we have presented a concise overview of the gravity tests, with~a particular emphasis on the interplay between theory and data.} We focused on the astrophysical probes sensitive to structure growth at cosmological scales during later stages of our Universe's evolution. {Within this review, we have summarized ideas derived from effective field theories and parametrized frameworks of gravity, which hold promise in providing more robust constraints on gravity in the future, even when extending our investigations into non-linear regimes. The~validity of specific models can be tested by comparing their predictions to observations. In~particular, we focused on the two most representative MG models: $f(R)$ and nDGP. We discussed how gravity under these models can have a different impact on the potential, density, and~velocity fields. As~a consequence, derived quantities and their corresponding summary statistics of large-scale structure and CMB photons are altered when compared to the standard $\Lambda$CDM model.}

{While MG models were initially motivated as more natural explanations for the accelerated expansion, it is worth noting that a broader class of MG models, including Horndeski theories, may encounter challenges in accounting for cosmic acceleration~\citep{Lombriser2017:GWacc}. Notably, the~$f(R)$ and nDGP models, specifically focused on, are encompassed within this broader class. Considering these circumstances, a~fundamental question arises: {Why investigate MG from a cosmological standpoint?}}

{Among the various attempts to modify GR, $f(R)$ and nDGP have remained as viable options} (see Section~\ref{sec:overview_model}). Even though these two models might not necessarily describe the true gravity theory of our Universe, their mechanisms are implemented into simulations (Section~\ref{sec:simulations}), and~most model-specific studies are based on these two models (Section~\ref{sec:methods}).
With a model-specific approach, we can look for distinctive features predicted by a given model; it is easier to search for such features when comparing the predictions to observations, particularly when working at a low signal-to-noise regime. From~this {perspective}, these two models provide explanatory examples of how a given observable could be impacted in the presence of a fifth~force.

Cosmology has been in a precision era for almost two decades, and~future surveys will cover larger areas and produce deeper maps of our Universe. One {hopes that}, by~reducing statistical uncertainties, we can potentially reveal departures from GR-$\Lambda$CDM (Section~\ref{sec:survey}) by nailing down the uncertainties in the parameters {of} the surviving MG models. Following this route, we list a few common challenges across the probes in the MG~studies. 

\begin{itemize}
    \item Model-dependent vs. model-independent: above we mentioned the downsides of the model-specific approach. The~parametrization framework is more general and can quantify broader classes of gravity models. To date, most MG studies within the parametrization framework are only suitable in the linear regime. Only recently have there been studies that generalized the approach to all scales (Section~\ref{subsec:pms_gravity}). However, extracting physical insights from the parametrized approach can be a difficult task.
    \item Analytic templates for observables: although the null hypothesis can be applied to quantify potential deviations from GR, a~first-principle-derived template for a given observable helps to deepen the understanding of structure growth in the presence of a fifth force. However, such templates {typically} rely on the specifics of MG models, and~computing such templates including MG effects is usually challenging. Furthermore, the~templates are often motivated by PT. They are more suitable for summary statistics, while templates (even within the standard GR case) that characterize topological properties {are yet} to be developed. 
    \item Cosmic degeneracies: {massive neutrino and baryonic processes can both have impacts on the small-scale matter power spectrum} (e.g., \cite{Lesgourgues2006:neutrino,vanDaalen2011:galformPmm}). {While various analytic approaches \citep{Mead2015:HMCODE,Schneider2015:baryonPmm,Schneider2019:baryonPmm} have been used to study non-linear effects and baryonic processes, they typically require careful calibration upon hydrodynamical simulations. Meanwhile, the~baryonic processes predicted by hydrodynamical simulations~\mbox{\cite {Schaye2010:OWLS,Dubois2014:HorizonAGN,Peirani2017:HorizonAGN,McCarthy2017:BAHAMAS,Weinberger2017:IllustrisTNG}} are affected by resolution, calibration strategies, and~the choice of sub-grid physics (see review by \cite{Chisari2019}). These variations can result in uncertainties in predicting the statistics of small-scale phenomena}.  
    \item Simulation degeneracies: many simulation-based conclusions rely on post-processing of $N$-body simulations. Various assumptions on the construction of observable catalogues can have non-negligible impacts on the conclusions. Typical examples are degeneracies between the HOD parameters or neglect of observational systematics.
\end{itemize}

This list contains interesting standing-alone problems relevant to MG and generally applied in the context of astrophysics and cosmology. However, there is yet no guarantee that we will find a satisfying answer {as to} the {nature of cosmic acceleration}, even if we solve all the problems listed above. Our Universe is a complex system with a huge number of degrees of freedom, and~we have to admit {that} the fact that so many degrees of freedom can be reduced to only a few within the $\Lambda$CDM framework is crucial. 
The~past effort{s} in studying these representative gravity models taught us how much is left in challenging the phenomenological $\Lambda$.

{These lessons encourage those interested in understanding the nature of cosmic acceleration to continue to explore further fundamental theories.} 
\vspace{6pt} 

\authorcontributions{{~~}}
All authors contribute to the manuscripts. Section 1, JH, CC, and JB; Section 2, MB, CH-A, and JH; Section 3, JH, CC, TT, JZ, CH-A, JB, and MB; Section 4, CH-A, JH, and CC; Section 5, JZ, MB, JH, and JB; Section 6, JH, CC, and JB.

\funding{{~~}}
J.H. has received funding from the European Union’s Horizon 2020 research and innovation program under the Marie Sk\l{}odowska-Curie grant agreement No 101025187.
CH-A acknowledges support from the Excellence Cluster ORIGINS which is funded by the Deutsche Forschungsgemeinschaft (DFG, German Research Foundation) under Germany's Excellence Strategy---EXC-2094-- 390783311.
TT acknowledges funding from the Swiss National Science Foundation under the Ambizione project PZ00P2\_193352. 
The project leading to this publication has received funding from the Excellence Initiative of Aix-Marseille University - A*MIDEX, a French ``Investissements d'Avenir'' program (AMX-20-CE-02 - DARKUNI). JZ is supported by the project "NRW-Cluster for data intensive radio astronomy: Big Bang to Big Data (B3D)“funded through the programme "Profilbildung 2020", an initiative of the Ministry of Culture and Science of the State of North Rhine-Westphalia.The sole responsibility for the content of this publication lies with the authors. MB acknowledges support from the INFN INDARK PD51 grant.

\dataavailability{{~~}} 
No new data were created or analyzed in this study. Data sharing is not applicable to this article.

\acknowledgments{We thank Dominik Schwarz, Anna Bonaldi, David Bacon, Alex Krolewski, Lucas Lombriser, Enrique Paillas, Zachary Slepian, and Hubert Wagner for useful discussions. We especially thank Alex Krolewski for their input on CMB lensing and ISW effects, Enrique Paillas for the density split method, Alessandra Silvestri and Emilio Bellini for the valuable feedback on the theory discussion, Hubert Wagner for dedicated discussions on the topological data analysis, Marta Spinelli for valuable support on technical aspects of IM. We acknowledge Fabian Schmidt for the feedback on the manuscript. We are grateful for the anonymous referees for their detailed comments and suggestions.}  

\conflictsofinterest{{~~}} 
The authors declare no conflict of interest.

\label{lastpage}

\begin{adjustwidth}{-\extralength}{0cm}
\printendnotes[custom]
\reftitle{References}

\PublishersNote{}
\end{adjustwidth}
\end{document}